\documentclass{aa}
\usepackage{psfig}
\usepackage{rotating}
\usepackage{graphicx}
\usepackage{latexsym}
\usepackage{amssymb}
\usepackage{natbib}
\usepackage{txfonts}
\def\etal{et al.}

\def\teff{\ifmmode T_{\rm eff} \else $T_{\mathrm{eff}}$\fi}

\def\ltsima{$\buildrel<\over\sim$}
\def\lsim{\lower.5ex\hbox{\ltsima}}
\newcommand{\hi}{H~{\sc i}}
\newcommand{\hii}{H~{\sc ii}}
\newcommand{\ha}{\ifmmode {\rm H}\alpha \else H$\alpha$\fi}
\newcommand{\hb}{\ifmmode {\rm H}\beta \else H$\beta$\fi}
\newcommand{\lya}{\ifmmode {\rm Ly}\alpha \else Ly$\alpha$\fi}

\def\kms{km s$^{-1}$}

\def\msun{\ifmmode M_{\odot} \else M$_{\odot}$\fi}
\def\msunyr{\ifmmode M_{\odot} {\rm yr}^{-1} \else M$_{\odot}$ yr$^{-1}$\fi}
\def\zsun{\ifmmode Z_{\odot} \else Z$_{\odot}$\fi}
\def\lsun{\ifmmode L_{\odot} \else L$_{\odot}$\fi}

\def\mup{\ifmmode M_{\rm up} \else M$_{\rm up}$\fi}
\def\mlow{\ifmmode M_{\rm low} \else M$_{\rm low}$\fi}

\newcommand{\oh}{\ifmmode 12 + \log({\rm O/H}) \else$12 + \log({\rm
O/H})$\fi}
\def\hyperz{{\em Hyperz}}
\def\flyf{\ifmmode f_{\rm Lyf} \else $f_{\rm Lyf}$\fi}
\def\pz{\ifmmode P(z) \else $P(z)$\fi}
\def\ki2{\ifmmode \chi^2 \else $\chi^2$\fi}
\def\zphot{\ifmmode z_{\rm phot} \else $z_{\rm phot}$\fi}

\newcommand{\xphot}{\ifmmode x_\gamma \else $v_\gamma$\fi}
\newcommand{\xobs}{\ifmmode x_{\rm obs} \else $x_{\rm obs}$\fi}
\newcommand{\xcmf}{\ifmmode x_{\rm CMF} \else $x_{\rm CMF}$\fi}
\newcommand{\vexp}{\ifmmode V_{\rm exp} \else $V_{\rm exp}$\fi}
\newcommand{\vmax}{\ifmmode V_{\rm max} \else $V_{\rm max}$\fi}
\newcommand{\nh}{\ifmmode N_{\rm HI} \else $N_{\rm HI}$\fi}
\newcommand{\dv}{\ifmmode \Delta v({\rm em-abs}) \else $\Delta v({\rm em}-{\rm abs})$\fi}

\begin{document}
\title{3D \lya\ radiation transfer. III. Constraints on gas and stellar
properties of $z \sim 3$ Lyman break galaxies (LBG) and implications
for high-$z$ LBGs and \lya\ emitters}

\author{Anne Verhamme\inst{1}, Daniel Schaerer\inst{1,2}, 
Hakim Atek\inst{3}, Christian Tapken\inst{4}
}
\offprints{daniel.schaerer@obs.unige.ch}
\authorrunning{Verhamme, Schaerer, Atek, Tapken}
\titlerunning{3D \lya\ radiation transfer. III. Gas and stellar properties
of $z \sim 3$ LBGs}

\institute{
Geneva Observatory, University of Geneva,
51, Ch. des Maillettes, CH-1290 Versoix, Switzerland
\and
Laboratoire d'Astrophysique de Toulouse-Tarbes, 
Universit\'e de Toulouse, CNRS,
14 Avenue E. Belin,
F-31400 Toulouse, France
\and
Institut d'Astrophysique de Paris,
98bis boulevard Arago,
FR-75014 Paris, France
\and
Max Planck Institute for Astronomy, 
K\"onigstuhl 17,
D-69117 Heidelberg, Germany 
}
\date{Received date; accepted date}

\abstract{}{
The Aim of our study is to understand the variety of observed \lya\
line profiles and strengths in Lyman Break Galaxies (LBGs) and \lya\
emitters (LAEs), the physical parameters governing them, and hence
to derive constraints on the gas and dust content and stellar
populations of these objects.}    
{
Using our 3D \lya\ radiation transfer code including gas and dust
(Verhamme \etal\ 2006), we fit 11 LBGs from the FORS Deep Field with
redshifts between 2.8 and 5 observed by Tapken \etal\ (2007). A simple
geometry of a spherically expanding shell of \hi\ is adopted.} 
{
The variety of observed \lya\ profiles is successfully reproduced. 
Most objects show outflow velocities of  $\vexp \sim$ 150--200 \kms;
two objects are most likely quasi-static.
The radial \hi\ column density ranges from $\nh \sim 2 \times 10^{19}$ to 
$7 \times 10^{20}$ cm$^{-2}$.
Our \lya\ profile fits yield values of $E(B-V) \sim$ 0.05--0.2 for the 
gas extinction.
We find indications for a dust-to-gas ratio higher than the Galactic value,
and for a substantial scatter.
The escape fraction of \lya\ photons is found to be determined
primarily by the extinction, and a simple fit formula is proposed. In
this case a measurement of EW(\lya)$_{\rm obs}$ can  yield $E(B-V)$, if the
intrinsic \lya\ equivalent width is known  (or assumed).
Intrinsic EW(\lya)$_{\rm int}$ $\sim$ 50--100 \AA\ 
are found for 8/11 objects,
as expected for stellar populations forming constantly over long
periods ($\ga$ 10-100 Myr). 
In three cases we found indications for younger populations. 
Our model results allow us also to understand observed correlations
between EW(\lya)$_{\rm obs}$ and other observables such as FWHM(\lya), $E(B-V)$,
SFR(UV) etc. 

We suggest that most observed trends of \lya, both in LBGs and LAEs,
are driven by variations of \nh\ and the accompanying variation of the
dust content.  Ultimately, the main parameter responsible for these
variations may be the galaxy mass. 
We also show that there is a clear overlap between LBGs and LAEs: at
$z \sim 3$ approximately 20--25 \% of the LBGs of Shapley \etal\
(2003) overlap with $\sim$ 23 \% of the LAEs of Gronwal \etal\ (2007).
Radiation transfer and dust effects should also naturally explain the
increase of the LAE/LBG ratio, and a higher percentage of LBGs with
strong \lya\ emission with increasing redshift.
}{}


\maketitle
\section{Introduction}
\label{s_intro}

\lya\ line radiation, often of the brightest emission lines in distant 
star-forming galaxies, is now frequently observed over a wide redshift
range
\citep[e.g.]{Hu98,Kudritzki00,Malhotra02,Ajiki03,Taniguchi05,Shimasaku06, 
Kashikawa06,Tapk06,Gron07,Ouchi07}.
Numerous narrow-band and other surveys use this line to
search for galaxies at specific cosmic ages, and to hunt for the most
distant objects in the universe \citep[see][]{Willis05,Cuby07,Stark07}.
Furthermore extremely deep ``blind'' spectroscopic exposures have
revealed very faint objects through their \lya\ emission \citep{Rauch07},
illustrating also the discovery potential of future instruments
such as the Multi Unit Spectroscopic Explorer (MUSE) for the Very Large
Telescope (VLT) and extremely large telescopes.

\lya\ measurements are used to infer a number of properties such
as redshift, star formation rates, constraints on the ionisation
of the intergalactic medium and hence on cosmic reionisation,
trace large scale structure at high redshift etc. \citep[see e.g.][ for
an overview, and references therein]{Schaerer07}.
However, given the physics of this generally optically thick resonance line 
quantitative interpretations of \lya\ are often difficult or even ambiguous,
as shown by detailed studies of nearby starbursts carried out
during the last decade \citep{lequeux95,Kunth98,Mas-Hesse03,hayes05,hayes07a,Atek08}. 

With the increased computer power and the availability of \lya\
radiation transfer codes
\citep[cf.][]{Ahn00,Ahn02,Cantalupo05,Hans06,Dijkstra06I,Tasitsiomi06,Verh06}
time is now ripe not only to predict \lya\ in different 
astrophysical situations (e.g.\ galaxy and cosmological simulations)
but especially also to confront observed \lya\ properties (line
profiles, equivalent widths, etc.) of individual galaxies (nearby and distant ones)
with detailed 3D radiation transfer calculations.  This is one of the
aims of our paper.

Two galaxy populations, the well known Lyman Break Galaxies (LBGs) and 
the \lya\ emitters (LAEs), represent currently the largest samples of 
distant galaxies, at least from $z \sim 3$ to 6.5.
Important questions remain unanswered about them, closely related to
their \lya\ emission and absorption.

LBGs show a great diversity of \lya\ profiles and strengths, reaching from 
strong emission, over P-Cygni type profiles, to strong absorption 
\citep[see e.g.][]{Shap03}. Furthermore the \lya\ properties are found to
correlate with other quantities such as the strength of interstellar
lines (IS), extinction, the star formation rate (SFR) etc. 
However, the origin of these variations and correlations remains
largely unknown or contradictory at best.  
For example, \citet{Shap01} suggested, age as the main difference
between LBG groups with different \lya\ strength, where objects with
\lya\ in absorption would be younger and more dusty.  However, it is
not clear why older LBGs would contain less dust, especially since
outflows, supposedly used to expel the dust, are ubiquitous in all
LBGs and since precisely these outflows are the location where the
emergent \lya\ spectrum is determined.
\citet{Ferrara06} proposed that LBGs host short-lived ($30 \pm 5$ Myr) 
starburst episodes, whose outflows -- when observed at different 
evolutionary phases -- would give rise to the observed correlations 
between IS lines and \lya.
However, the ages obtained from SED and spectral fits of LBGs show
older ages and fairly constant star formation histories 
\citep[cf.][]{Elli96,Pett00,Shap01,Papovich01,Pentericci07}.
\citet{Pentericci07} find that $z \sim 4$ LBGs without \lya\ emission
are on average somewhat older ($\sim 410\pm70$ versus $200\pm50$ Myr)
and more massive than those with \lya\ in emission.  However, given
the significant amount of ongoing star formation inferred for both
galaxy subsamples (with/without \lya\ emission) types from their SED
fits, intrinsic \lya\ emission is expected in both subsamples.  The
apparent differences in age and ratio of age over star formation time
scale ($\tau$) can therefore not be a physical cause for the observed
\lya\ variations.
\citet{Erb06}, from an analysis of $z \sim 2$ LBGs, also find that
more massive galaxies have fainter \lya\ (or \lya\ in absorption);
following the arguments of \citet{Mas-Hesse03} they suggest that this
is mostly due to an increased velocity dispersion of the interstellar
medium, indicated by the increased strength of the saturated IS lines,
which would increase the fraction of \lya\ being absorbed. 
However, from radiation transfer modelling we find that the behaviour 
of the \lya\ escape fraction is not linear with the velocity dispersion 
in the ISM \citep[and below]{Verh06}. Other physical parameters may 
play a more important role in the destruction of \lya.    

Important questions remain also concerning the properties of LAEs,
their similarities and differences with respect to LBGs,
and about the overlap between these galaxy populations.
Since a fraction of LBGs show strong enough \lya\ emission to be 
detected with the narrow-band technique mostly used to find LAEs
there must definitely be an overlap.
For example, at  $z \sim 3$ approximately 25 \% of the LBGs of 
Steidel and collaborators \citep{Shap03} have EW(\lya)$_{\rm obs}$ $\ga$ 20 \AA\
(restframe), sufficient to be detected in the LAE survey of \citet{Gron07}.
However, what the properties of LAEs are e.g.\ in terms of
stellar populations (age, star formation histories, $\ldots$), mass,
dust content, outflows, metallicity etc.\ is not yet well established,
although first such analysis have recently become available
\citep[see][]{SP05,Lai07,Lai08,Gron07,Gawiser07,Pirzkal07,Finkelstein07}.
Understanding the nature of LAEs and their relation to LBGs is also
crucial since the contribution of the LAE population to the known
starburst population seems to increase with redshift
\citep{Hu98,Shimasaku06,Nagao07,Ouchi07,Dow07,Reddy07}.

Last, but not least, several types of theoretical models have been
constructed during the last few years aimed at understanding LAE and 
LBG populations, the relation between the two populations,
and to use them as constraints for galaxy formation scenarios,
cosmic reionisation, and other topics
\citep{Thommes05,Ledelliou06,Mori06,Dijkstra07,Kobayashi07,Mao07,Stark07model,
Nagamine08}. Observational constraints on crucial parameters
such as the \lya\ escape fraction from LBGs and LAEs, and other insight
from radiation transfer models are, however, badly needed to reduce 
uncertainties and degeneracies in these modeling approaches.

With these questions about LBGs and LAEs in mind, we have recently
started to model a variety of $z \sim 3$ starbursts with our new \lya\
radiation transfer code \citep[paper I]{Verh06}. First we have studied
the well known $z \sim 2.7$ LBG MS1512-cB58 (cB58 in short), whose
spectrum is dominated by strong \lya\ absorption \citep[paper
II]{SV07}. In this third paper of the series we present an analysis of
11 LBGs observed with FORS2 at the VLT with sufficient spectral
resolution ($R \sim 2000$) to allow detailed \lya\ profile fitting to
constrain their properties.  Taken together, the objects analysed in
paper II and III cover a wide range of \lya\ strengths and also
different morphologies, including absorption dominated \lya\ and \lya\
emission lines with equivalent widths between $\sim$ 6 and 150 \AA\
(restframe). The variety of objects modeled in paper II and III covers
thus in particular the entire range of \lya\ strengths defining the 4
spectral groups of the LBG sample of \citet{Shap03}, the largest
currently available at $z \sim 3$. Furthermore, several of the objects
we model have strong enough \lya\ emission to classify as LAEs,
according to the criteria used in many surveys.
Our analysis represents the first modeling attempt of \lya\ line
profiles of high redshift galaxies with a detailed radiation
transfer code including gas and dust and treating line and continuum 
radiation.

The remainder of the paper is structured as follows.  A description of
the radiation transfer code, the assumptions, and input parameters is
given in Sect.\ \ref{s_model}. In Sect.\ \ref{s_tapken} we model the
\lya\ profiles of the individual objects.
Our main fitting results are discussed and confronted to other observations
in Sect.\ \ref{s_discuss}. Other properties are derived in Sect.\ \ref{s_props}.
In Sect.\ \ref{s_scenario} we finally propose a unifying scenario for LBGs
and LAEs and discuss several implications.
Our main conclusions are summarised in Sect.\ \ref{s_ccl}.

\section{\lya\ radiation transfer modeling}
\label{s_model}

To fit the observations we used our 3D Monte Carlo (MC) radiation transfer 
code {\em MCLya} \citep{Verh06}.
The code solves the transfer of \lya\ line and adjacent continuum photons 
including the detailed processes of \lya\ line scattering, dust scattering,
and dust absorption.
The main assumptions required for the modeling concern the geometry,
the choice of the input parameters, and the input spectrum.
We discuss them now in turn.

\subsection{Geometry}
For simplicity, and given empirical evidence in favour of a fairly
simple geometry in $z \sim 3$ LBGs discussed by \citet{Verh06}, we
adopted first a simple ``super-bubble'' model to attempt to fit the
observed \lya\ line profile.  The assumed geometry is that of an
expanding, spherical, homogeneous, and isothermal shell of neutral
hydrogen surrounding a central starburst emitting a UV continuum plus
\lya\ recombination line radiation from its associated
\hii\ region. We assume that dust and \hi\ are uniformly mixed.

The homogeneity and a large covering factor of the absorbing medium
are supported by observations of LBGs and nearby starbursts. For
example, the outflow of cB58 is located well in front of the stars and
covers them almost completely, since it absorbs almost all the UV
light from the background stars. Indeed, \citet{Sava02} find only a
small residual mean flux above the zero level in the core of the \lya\
absorption feature, whereas it is black for
\citet{Pett02}. \citet{Heck01} estimate an area covering factor for
optically thick gas of 98\% from the residual intensity at the core of
the C~{\sc ii} $\lambda$1335 line.
A somewhat lower covering factor may be indicated for LBGs with strong
\lya\ emission \citep{Shap03}. For simplicity, and in the absence of
further observational constraints, we will assume a covering factor of 
unity.

\subsection{Shell parameters}
As described in paper II, the outflow is modelised by a
spherical, homogeneous, and isothermal shell of neutral hydrogen and
dust centered on a point source. 
Four parameters characterise the physical conditions in the shell:
\begin{itemize}
\item the expansion velocity \vexp,
\item the Doppler parameter $b$,
\item the neutral hydrogen column density \nh,
\item the dust absorption optical depth $\tau_a$.
\end{itemize} 

In principle \vexp\ is constrained by observations, either directly
measured by the blueshift of low ionisation interstellar lines (hereafter LIS)
compared to stellar
lines \citep{Pett02}, or from the shift between absorption
LIS lines and \lya\ in emission, \dv, when the stellar
lines are too faint to be observed \citep{Shap03}. 
Otherwise \vexp\ will be constrained by \lya\ line profile fits.
In \citet{Verh06} we showed that radiation transfer effects lead to
$\dv \approx 3\times\vexp$  in expanding shells with 
$\nh \ga 10^{20}$ cm$^{-2}$. For lower column densities the peak of the 
redshifted \lya\ emission may trace $\sim \vexp$ (leading to $\Delta
v({\rm em-abs}) \sim 2 \vexp$), instead of twice
this value \citep[cf.][and also Fig.\ref{vary_nh}]{Verh06}.
For three of the 11 objects to be modeled here \dv\
has been measured (see Table \ref{t_overview}).

The Doppler parameter $b$, describing the random motions of the
neutral gaz possibly including microturbulence, is kept as a free
parameter.
For indication, $b \sim 13$ \kms\ corresponds to thermal motions for
$T=10^4$K; for the lensed LBG cB58 \citet{Pett02} derived $b \sim 70$ \kms\ 
from fits of LIS lines.

Although presumably the neutral column density and the dust amount are 
physically related, e.g.\ by a given dust-to-gas ratio,
both parameters are kept free in our modeling procedure.
The resulting values will later be compared to available observational
constraints.

We assume that dust and \hi\ are uniformly mixed.
As discussed in \citet{Verh06}, the dust optical depth $\tau_a$
relates to the usual extinction $E(B-V) \approx (0.06 \ldots 0.11) \,
\tau_{a}$, where the numerical coefficient covers
attenuation/extinction laws of \citet{Calz00}, \cite{Seat79} and
similar.  Here we assume $E(B-V) = 0.1 \tau_a$.

\subsection{The intrinsic spectrum in the \lya\ region}

The synthetic stellar spectrum of star-forming galaxies 
close to \lya\ is described in paper II. 
The stellar continuum presents an absorption
feature around \lya\, whose strength varies in time, depending on the
SF history, on the age of the star-forming galaxy, and less on its
metallicity \citep{Scha03,Delg05}. 

The main H and He recombination lines created in the
\hii\ region surrounding the starburst are also predicted by the models of 
\citet[ hereafter S03]{Scha03}: 
for metallicities between 1/50 \zsun\ and solar the strength of \lya\
varies from EW(\lya)$_{\rm int}$ $\sim$ 250--360 \AA\ at early time after the
burst, and declines until zero for a burst whereas it reaches an
equilibrium value of 60--100
\AA\ for objects with a constant star formation rate (SFR) after $\sim
50-100$ Myr (see also Fig.\ \ref{f_s04_sfr}).

In paper II, we showed that the fitting of the observed
star-forming galaxy cB58 depends only very little on the details of
the stellar continuum around \lya.  Since the objects modeled here
show stronger \lya\ emission than cB58, which is dominated by
absorption, neglecting the detailed shape of the stellar continuum is
even more justified here. Therefore we model the input spectrum as a
flat continuum plus a Gaussian emission line described by two
parameters:
\begin{itemize}
\item the intrinsic equivalent width, EW(\lya)$_{\rm int}$, and
\item the intrinsic full width at half maximum, FWHM(\lya)$_{\rm int}$.
\end{itemize}

What ``reasonable'' values should we adopt for these parameters ? 
Our first approach, to reduce the number of free parameters in the
model, was to test if a unique scenario was conceivable, i.e.\ if we
could fit all data with the same intrinsic \lya\ spectrum, the
differences in the observed spectra would then come from radiation
transfer effects in the outflowing medium. 
These objects are likely starburst galaxies with a constant star
formation (SF) history as derived from their UV low/medium-resolution
spectra \citep{Noll04,Mehl06}, so we fix the intrinsic equivalent
width to EW(\lya)$_{\rm int}=60-100$ \AA, as derived from the S03 models. 
We adopted the intrinsic value FWHM(\lya)$_{\rm int}=100$ \kms, as it
is comparable 
to the values measured from the velocity dispersion of H$\alpha$ and CO 
lines in cB58\citep{Tepl00, Bake04}, and the dispersion measured
in 16 starbursts at $z \sim 2$ by \citet{Erb03}.

\begin{table*}[tb]
\caption{Sample of 11 LBG galaxies taken from \citet{Tapk06} with
  their observational constraints: ID (col.\ 1), \lya\ profile type (2),
  systemic redshift from \citet{Noll04} except for FDF1267, where $z$ is 
  from T07 (3), the UV (4) and \lya\ (5) star formation rate, 
  the slope of the UV continuum $\beta$ (6), the velocity shift
  between the LIS lines and \lya, \dv\ (7), the
  observed EW(\lya)$_{\rm obs}$ (8) and FWHM(\lya)$_{\rm obs}$ (9).
  EWs and FWHM are given in the restframe; we here denote 
them by ``observed'' for distinction with ``intrinsic'' or ``theoretical'' 
values to be derived later.}
\begin{tabular}{ccccccccc}
\hline
ID & type & $z$  & SFR$_{UV}$ & SFR$_{\lya}$
& $\beta$  & \dv\ &
EW(\lya)$_{\rm obs}$ & FWHM(\lya)$_{\rm obs}$  \\
 & & & [\msunyr] & [\msunyr] & & [\kms] & [\AA] & [\kms] \\
\hline
\hline
$1267$ & C & $2.788\pm 0.001$ & $1.16\pm 0.25$ & $1.49\pm 0.08$ & & & $129.8\pm 27.41$ & $235\pm 34$ \\
$1337$ & A & $3.403\pm 0.004$ & $27.28\pm 1.15$ & $2.10\pm 0.14$ &$-2.43$& $607$ & $6.69\pm 0.46$ & $597\pm 84$ \\
$2384$ & A & $3.314\pm 0.004$ & $22.74\pm 0.77$ & $10.8\pm 0.27$ & $-0.55$ & &$83.19\pm 3.89$ & $283\pm 47$ \\
$3389$ & A & $4.583\pm 0.006$ & $14.85\pm 2.47$ & $9.20\pm 0.38$ & & &$38.82\pm 10.95$ & $354\pm 70$ \\
$4454$ & A & $3.085\pm 0.004$ & $1.98\pm 0.49$ & $2.25\pm 0.08$ & $-2.42$ & &$74.38\pm 11.84$ & $323\pm 47$ \\
$4691$ & B & $3.304\pm 0.004$ & $17.88\pm 0.75$ & $16.31\pm 0.14$ & $-2.46$ & &$79.44\pm 1.61$ & $840\pm 115$ \\
$5215$ & C & $3.148\pm 0.004$ & $26.20\pm 0.80$ & $9.57\pm 0.21$ & $-1.71$ & &$32.48\pm 1.06$ & $483\pm 90$ \\
$5550$ & A & $3.383\pm 0.004$ & $44.78\pm 1.07$ & $3.27\pm 0.20$ & $-1.81$& $620$ & $6.36\pm 0.40$ & $424\pm 85$ \\
$5812$ & A & $4.995\pm 0.006$ & $5.24\pm 0.79$ & $9.60\pm 0.18$ & & &$153.8\pm 26.6$ & $226\pm 23$ \\
$6557$ & A & $4.682\pm 0.006$ & $13.85\pm 1.39$ & $3.35\pm 0.15$ & & &$30.51\pm 3.04$ & $380\pm 135$ \\
$7539$ & B & $3.287\pm 0.003$ & $29.87\pm 0.78$ & $2.45\pm 0.46$ &$-1.74$ & $80$ & $6.84\pm 0.46$ &$1430\pm 230$\\
\hline
\end{tabular}
\label{t_overview}
\end{table*}

\subsection{Description of the method}

We ran a grid of $\sim500$ models with varying physical conditions in
the shell. The expansion velocity was varied from 0 to 400 \kms\ in
steps of 50 \kms, the neutral column density from $2\times10^{19}$ to
$2\times10^{21}$ cm$^{-2}$, the dust amount from $E(B-V)=0$ to 0.4
($\tau_a=$0, 0.1, 0.5, 1., 2., 3., and 4.), and the Doppler parameter
from $b=10$ to 200 \kms\ (10, 20, 40, 80, and 200 \kms).

In contrast to the four shell parameters, there is no need to run a
new Monte Carlo simulation each time we want to change the input
\lya\ spectrum. This can be made a posteriori without resorting to any
simplifying assumption. 
To do so we run each simulation of the four-dimensional grid with a
flat continuum as input spectrum, e.g.\ the same number of photons
per frequency bin, and we memorise this input frequency for each
photon. Once the simulation is done, we construct output spectra
corresponding to the input frequency bins and we assign a different
weight to each, in order to reconstruct any input spectrum shape
(for example, a flat stellar continuum+a Gaussian centered on \lya, or
a synthetic starburst spectrum as in paper II.

From each calculation we derive the integrated spectrum (\lya\ line
profile) emerging from the expanding shell.  For comparison to the
observations, our synthetic spectra are convolved with a Gaussian with
FWHM=150 \kms\ corresponding to the experimental resolution.
As shown below, moving in the space of these 6 input parameters, we
can reproduce the whole diversity of observed \lya\ spectra, ranging
from double-peaked profiles to broad absorption or asymmetric emission
lines.

Finally, our calculations allow us also to derive the \lya\ escape fraction
$f_e$.
This is computed from 
\begin{equation}
f_e = \frac {\int^{\infty}_{-\infty} f_e(v)\times \phi(v) \, d\nu}
           {\int^{\infty}_{-\infty} \phi(v) \, d\nu}, 
\end{equation}
where $\phi(v)$ is the intrinsic \lya\ line profile computed from the
MC simulation, and $f_e(v)$ is the escape fraction in each input
frequency bin computed from the MC simulation.

\section{Fits of the FORS Deep Field sample}
\label{s_tapken}

\subsection{Description of the FDF sample} 
Our work uses the FORS Deep Field sample presented by 
\citet[hereafter T07]{Tapk07}. 
Therefore, we give a brief overview of their work. T07
present the medium-resolution spectra (R=2000) of 16
high-redshift galaxies. The target selection for the objects of T07
was based on the FDF spectroscopic survey \citep{Noll04}.
The FDF spectroscopic survey aimed at obtaining
low-resolution spectra (R$\approx$200) of intrinsically bright
galaxies with a photometric redshift \citep{bender2001,gabasch2004}
 between $z \approx$ 1 and 5 with a
high signal-to-noise ratio ($\geq$10).  The spectra of 90 galaxies
with redshift between 2 and 5 were analysed and published
electronically by \citep{Noll04}. The deep (up to 10h integration
time with FORS1/FORS2) low-resolution spectra allowed them to derive the
redshift with high accuracy and reliability. 
ad to search for even weak signs of AGN activity of the objects. Based
on the low-resolution spectra, T07 selected starburst
galaxies with strong \lya\ emission and/or with bright UV-restframe
continuum for the follow-up medium-resolution spectroscopy.

These medium-resolution spectra were
obtained with FORS2 at the VLT UT4 using the holographic grisms 1400V
and 1200R. The spectral resolution of both grisms is $R$ $\approx$
2000. The spectral range of the 1400V (1200R) grism is about 4500 to
5800 (5700 to 7300) \AA .  All data were collected in service mode
using one single MXU mask for each grism.  The total integration time
of the 1400V (1200R) observations is 6.25 h (9.45 h).  The data were
reduced using the MIDAS-based FORS pipeline \citep{Noll04}). For more details see T07.  

The spectra of all objects of T07 include the \lya\
profile. However, only eleven \lya\ profiles have a sufficient SNR
($>$10), which allows a detailed comparison with our theoretical
models. The properties of this sample are listed in Table
\ref{t_overview}. Note that except stated otherwise all equivalent widths 
are given in the restframe; we denote them by ``observed''
equivalent widths for distinction with ``intrinsic'' or ``theoretical'' 
values to be derived later.

Eight galaxies have redshifts around z$\approx$3, while 3 galaxies
have redshift with 4.5$< z <$ 5. While only a few \lya\ profiles of
our sample show an absorption component (FDF5550), all our profiles
display an emission component. The equivalent width of the emission
component range between EW(\lya)$_{\rm obs}$ = 6 and 150 \AA . This \lya\
equivalent width is measured using the continuum redwards of the \lya\
emission line (at $\approx$1300 \AA ). Although the majority of LBGs
have \lya\ equivalent widths lower than 20 \AA, 
8/11 of our galaxies have an equivalent width (of the total \lya\
line, including absorption and 
emission) higher than 20 \AA . Therefore 70\% of our sample would be
detected in a typical narrow-band survey, searching for LAEs. 

As described by T07 the profiles show a wide range of
morphologies. For convenience mostly, 
we divide the galaxies in three groups according to
their \lya\ profile: (A) \lya\ emitters with asymmetric profiles:
FDF1337, FDF2384, FDF3389, FDF4454, FDF5550, FDF5812, and FDF6557, 
(B) double-peak profiles FDF4691 and FDF7539, and (C)
asymmetric \lya\ plus a blue bump: FDF5215 and FDF1267.

\subsection{The fitting procedure}

For fits with our synthetic spectra the observed, non-normalised spectra were transformed
to velocity space using the redshift listed in Table \ref{t_overview}.
If necessary $z$ was adjusted within the error bars cited.
We then use the same normalisation as \citet{Tapk07} to determine
EW(\lya)$_{\rm int}$.
Finally, we overlay synthetic spectra on observed ones and estimate fit qualities.
The spectral parts we focus on are location of the peak, the
shape of the peak and the extended wing, knowing that the blue side 
of the spectrum could be affected by the surrounding IGM. 

The parameters of the best fits, as well as derived parameters from
our model like the escape fraction are summarised in Table~\ref{t_summary}. 
Multiple entries correspond to multiple solutions of similar quality.

\begin{table*}
\caption{Summary of the best fits derived from our model of a
  spherical expanding shell surrounding a starburst to reproduce
  a sample of 11 spectra from \citet{Tapk06}. Cols.\ 1 and 2 are the
  object ID and  \lya\ profile type respectively. Cols.\ 3 to 8 give
  the model parameters, col.\ 9 the derived \lya\ escape fraction.}
\begin{tabular}{ccccccccc}
\hline
ID  & type & \vexp  [\kms]& b [\kms] & \nh [cm$^{-2}$] & $\tau_a$ & 
EW(\lya)$_{\rm int}$ [\AA]& FWHM(\lya)$_{\rm int}$ [\kms] & $f_e$ \\
\hline
\hline
$1337$ & A & $200$ & $20$ & $5\times 10^{20}$ & $1.0$ & $55$ & $100$ & $0.12$ \\
$5550$ & A & $200$ & $20$ & $5\times 10^{20}$ & $1.5$ & $65$ & $100$ & $0.05$ \\
$2384$ & A & $150$ & $20$ & $3\times 10^{19}$ & $1.0$ & $170$ & $100$ &$0.16$ \\
$4454$ & A & $150$ & $20$ & $2\times 10^{19}$ & $0.5$ & $100$ & $150$ & $0.42$ \\
$5812$ & A & $150$ & $20$ & $2\times 10^{19}$ & $1.0$ & $280$ & $100$ & $0.16$ \\
$3389$ & A & $150$ & $20$ & $2\times 10^{19}$ & $0.5$ & $50$ & $150$ & $0.45$ \\
$6557$ & A & $150$ & $20$ & $4\times 10^{19}$ & $1.0$ & $70$ & $100$ & $0.17$ \\
$4691$ & B & $10.$ & $20.$ & $8\times 10^{19}$ & $0.0$ & $80$ & $1000$ & $1.0$ \\ 
$7539$ & B & $25.$ & $40.$ & $5\times 10^{20}$ & $0.5$ & $100$ & $100$ & $0.28$ \\
$5215$ & C & $200$ & $20$ & $2\times 10^{15}$ & $0.01$ & $25$ & $700$ & $1.0$ \\
$5215$ & C & $400$ & $20$ & $7\times 10^{20}$ & $1.0$ & $120$ & $100$ &$0.12$ \\
$1267$ & C & $50$ & $20$ & $2\times 10^{19}$ & $0.1$ & $150$ & $300$ &$0.64$ \\
$1267$ & C & $300$ & $20$ & $3\times 10^{20}$ & $2.0$ & $500$ & $100$ & $0.02$ \\
\hline
\hline
\end{tabular}
\label{t_summary}
\end{table*}

\subsection{The asymmetric profiles (group A)}

Among the 11 objects, 7 present the characteristic \lya\
asymmetric emission line (FDF1337, 5550, 2384, 4454, 5812, 3389 and 6557).
This line shape can be understood by radiation 
transfer effects through an expanding medium \citep{Verh06}. Indeed,
\citet{Tapk07} were able to measure a velocity shift between
the interstellar absorption lines and the \lya\ emission line for two
of these objects, FDF1337 and FDF5550, because their UV continuum is
bright. Both of them present a shift of $\sim 600$ \kms\ (cf.\
Table~\ref{t_overview}), a clear sign of outflows,
which is most likely related to 3 times the expansion velocity of the 
shell as shown in \citet{Verh06}. 

\subsubsection{FDF1337 and FDF5550}
\begin{figure*}
\begin{tabular}{cc}
\includegraphics[height=6cm, width=8cm]{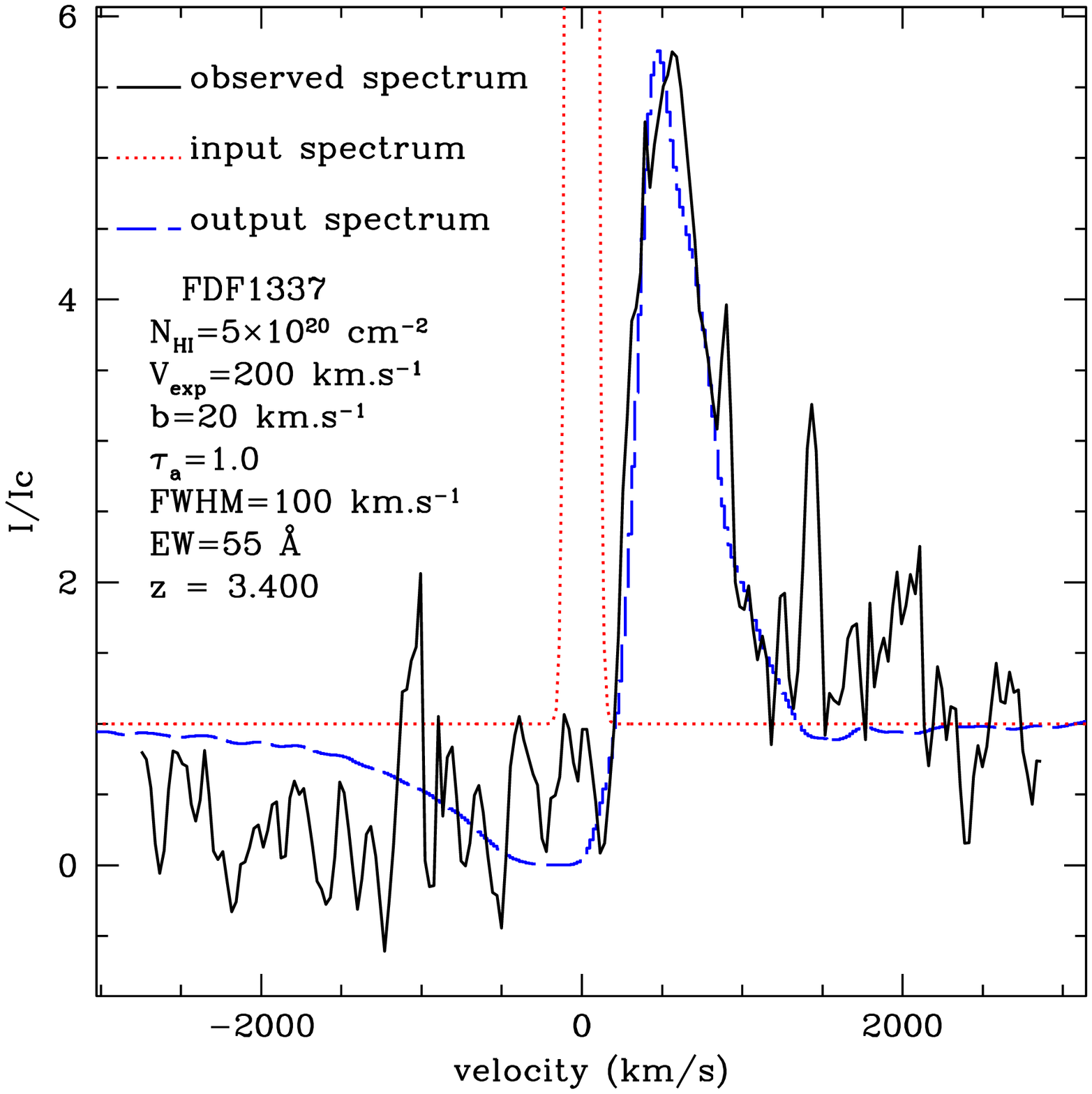} &
\includegraphics[height=6cm, width=8cm]{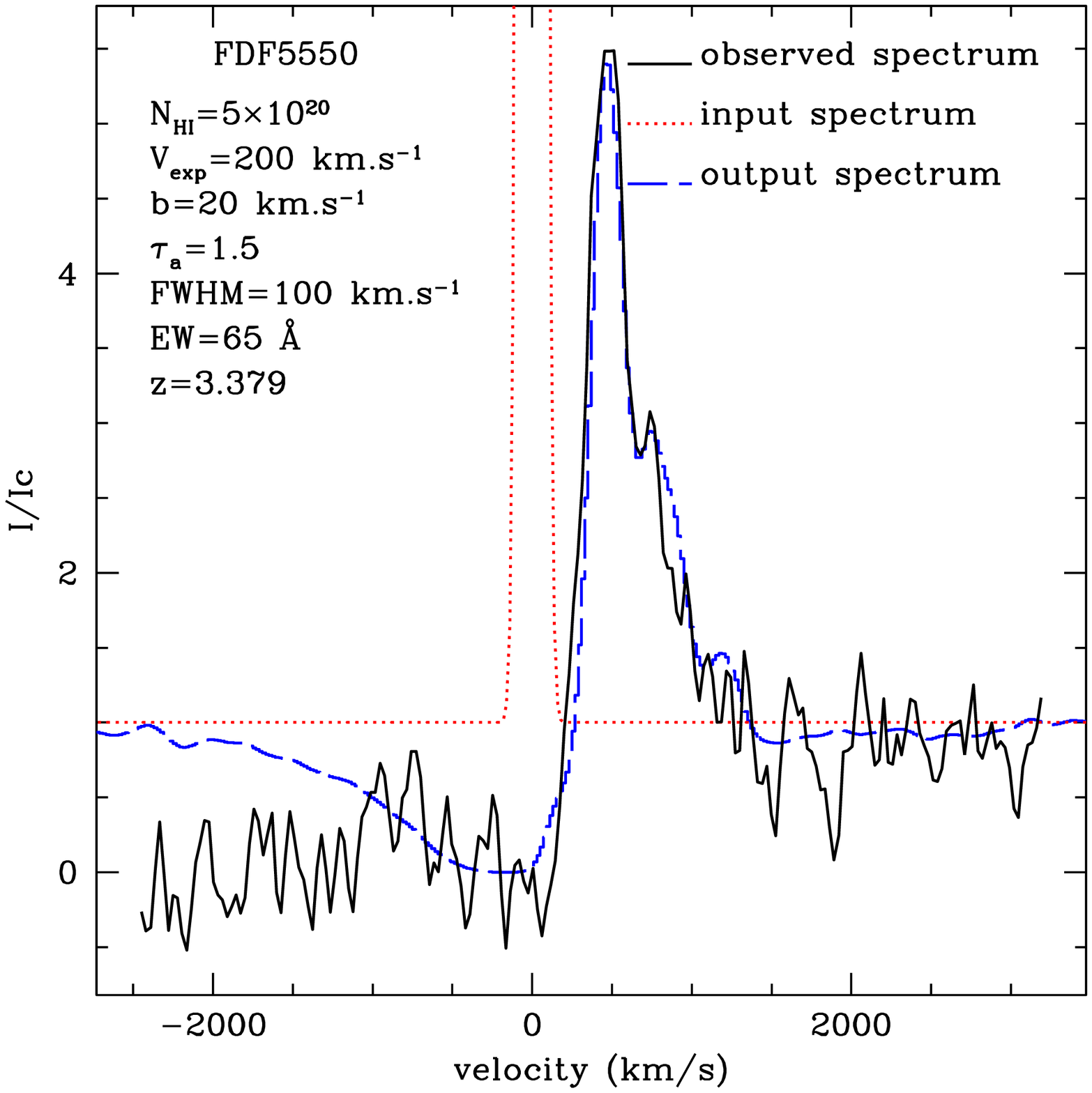} \\
\end{tabular}
\caption{\lya\ line profile fits for FDF1337 ({\em left}), and 
FDF550 ({\em right}), two of the best constrained cases. 
  The observed spectra are shown with the black solid line, model fits as 
  blue dashed curve, the intrinsic (input) profile with the red dotted line.
  All spectra are normalised to unity in the red (positive velocities).
  The fit parameters are indicated in the Figure and in Table
  \ref{t_overview}. 
  The expanding shell model reproduces
  well the faint and broad asymmetric \lya\ emission lines. The secondary
  peak, or the ``bump'' in the red extended wing is also reproduced for
  $b=20$ \kms.
  Note that the intrinsic EW(\lya)$_{\rm int}$ of these objects is larger than the
  observed one by approximately one order of magnitude.}
\label{FDF1337_5550}
\end{figure*}

To model FDF1337 and FDF5550, we have therefore fixed $\vexp = 600/3 = 200$
\kms, and we proceed to adjust the 3 remaining shell parameters. 
Our best fits and
the corresponding parameters are presented in Fig.\ \ref{FDF1337_5550}.    
Note that both spectra are fitted with an intrinsic EW(\lya)$_{\rm int}\sim 60$
\AA, which corresponds to the equilibrium value reached by a galaxy
with constant star formation after 50--100 Myr (cf.\ Fig.\ \ref{f_s04_sfr}).
Even if the observed EW(\lya)$_{\rm obs}$ is an order of magnitude lower, the
intrinsic predicted EW(\lya)$_{\rm int}$ seems to be ``standard'' in this sense.
Dust (we find $\tau_a=1.0$ for 1337 and $1.5$ for 5550, i.e.\ $E(B-V)
\sim 0.1-0.15$) in a high neutral column density ($\nh=5\times10^{20}$
cm$^{-2}$) outflow causes this attenuation.  The derived \lya\ escape
fraction is $f_e=0.052$ for FDF5550 and $f_e=0.121$ for FDF1337. A
rather small Doppler parameter, $b=20$
\kms, is derived compared to cB58 ($b_{\rm cB58}=70$\kms) to reproduce the 
secondary peak on the elongated red wing. Note the extension of this
red wing over $\sim 1500$ \kms. 

\subsubsection{FDF2384, FDF4454, and FDF5812}  
\begin{figure*}
\begin{tabular}{ccc}
\includegraphics[height=6cm, width=6cm]{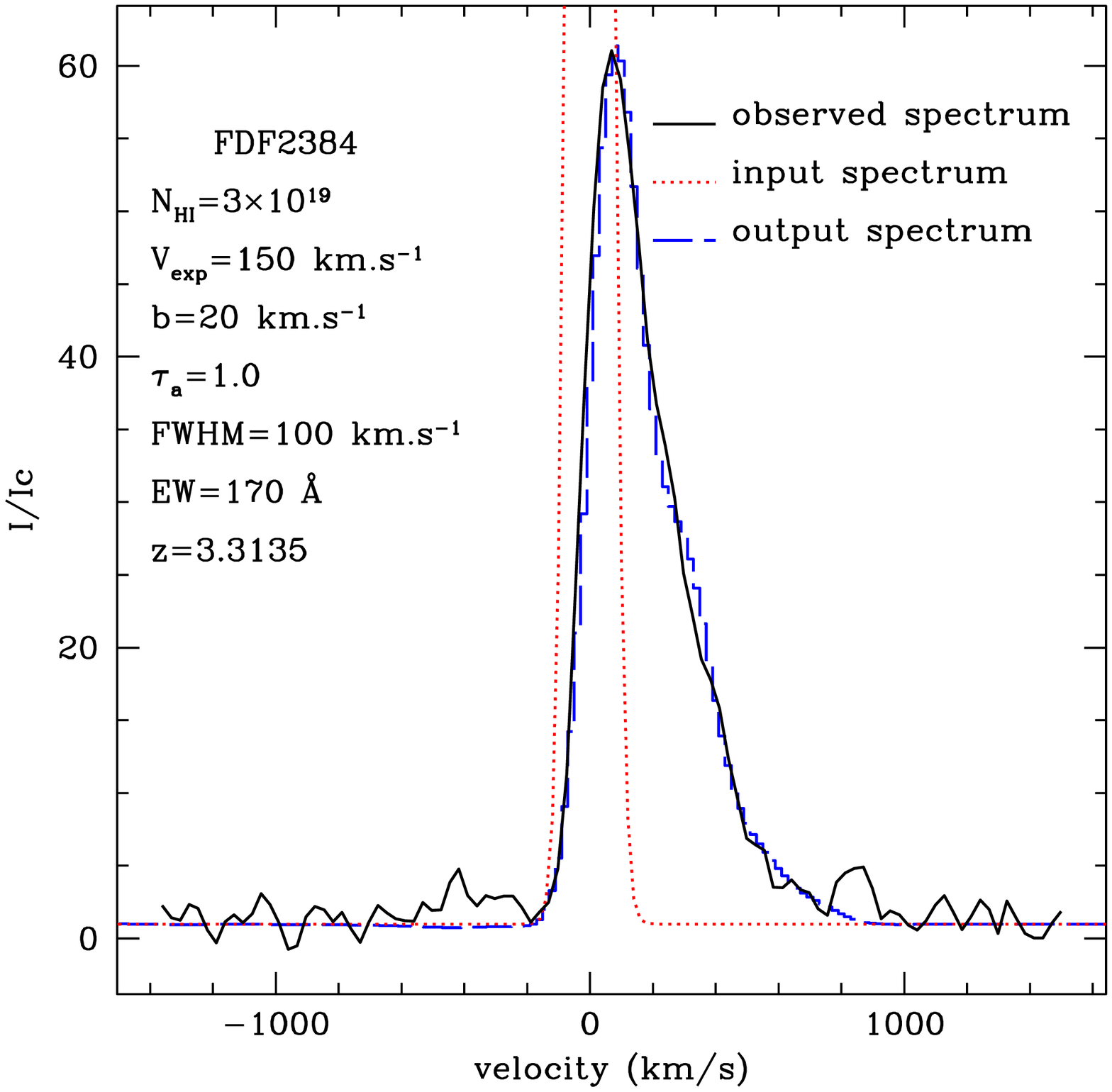} &
\includegraphics[height=6cm, width=6cm]{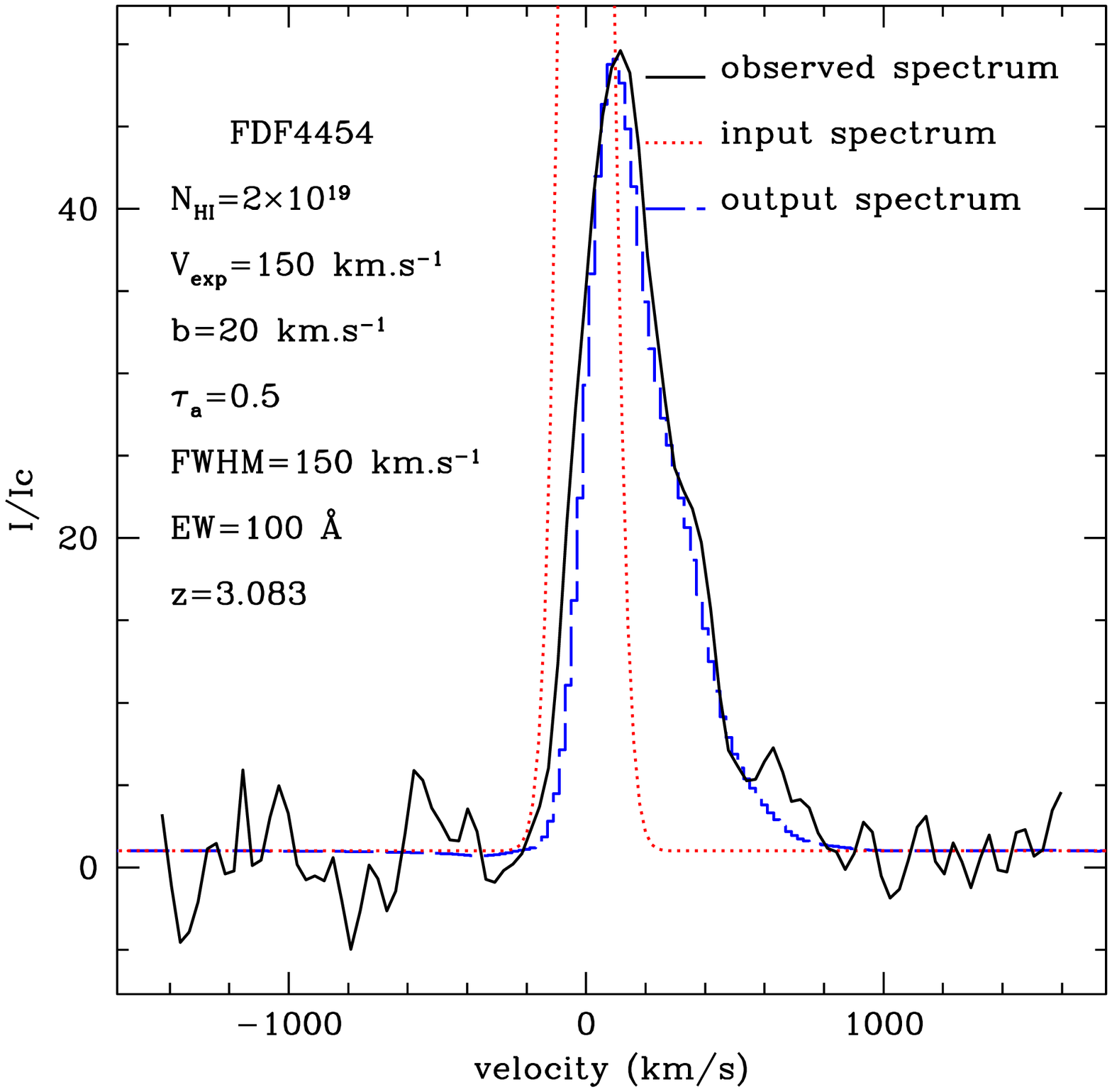} &
\includegraphics[height=6cm, width=6cm]{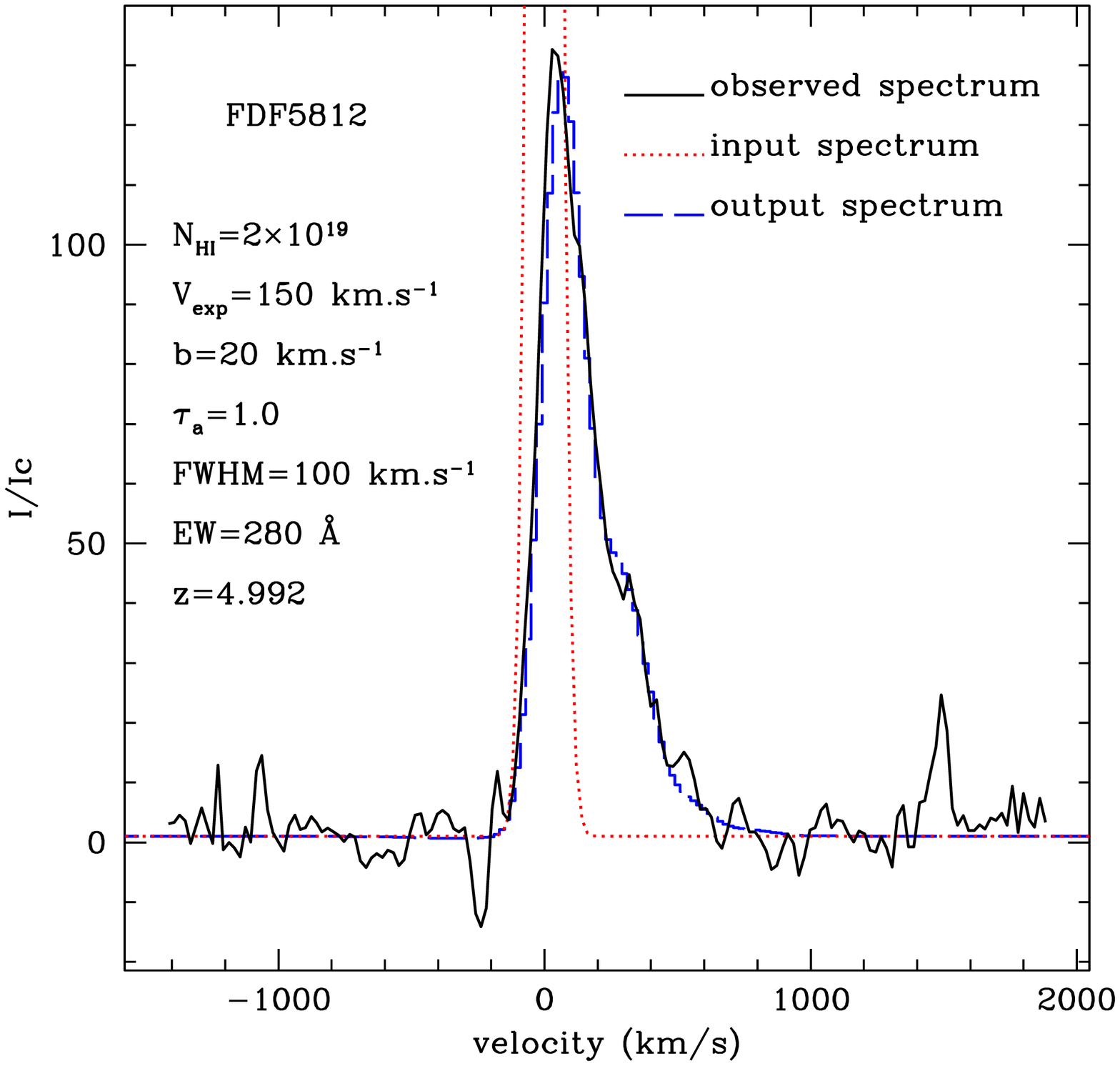} \\
\end{tabular}
\caption{Line profile fits for the three strongest \lya\ 
emitters of the sample, presenting all a narrow asymmetric emission
line: FDF2384 ({\em left}), FDF4454 ({\em middle}), and FDF5812
({\em right}). Same symbols as Fig.\ in \ref{FDF1337_5550}. 
The expansion velocity of the shell is $\vexp\sim150$ \kms, similar
as for the two precedent objects. The dust
content is similar too, but \nh\ is one order of
magnitude lower. The intrinsic \lya\ EW is also larger, particularly
for FDF5812 (EW(\lya)$_{\rm int}=280$ \AA), but these values depend strongly on
the continuum determination, which is quite uncertain for these faint
objects. See text for more details.}
\label{FDF5812}
\end{figure*}

The \lya\ fits for these objects are shown in Fig.~\ref{FDF5812}.
The profiles differ from the former by their
high EW(\lya)$_{\rm obs} > 70$ \AA, and a less extended red wing
($\sim600$ \kms).  
This leads to simulated neutral column densities an order of magnitude
lower and higher escape fractions ($f_e > 0.15$,
cf.\  Table~\ref{t_summary}).

\begin{figure*}
\begin{minipage}[c]{9.5cm}
\begin{tabular}{cccc}
 & $\tau_a=0.5$ & $\tau_a=1.0$ & $\tau_a=2.0$ \\ 
\begin{sideways}\, $\vexp=100$ \kms\end{sideways}&
\includegraphics[height=2.4cm,width=2.4cm]
{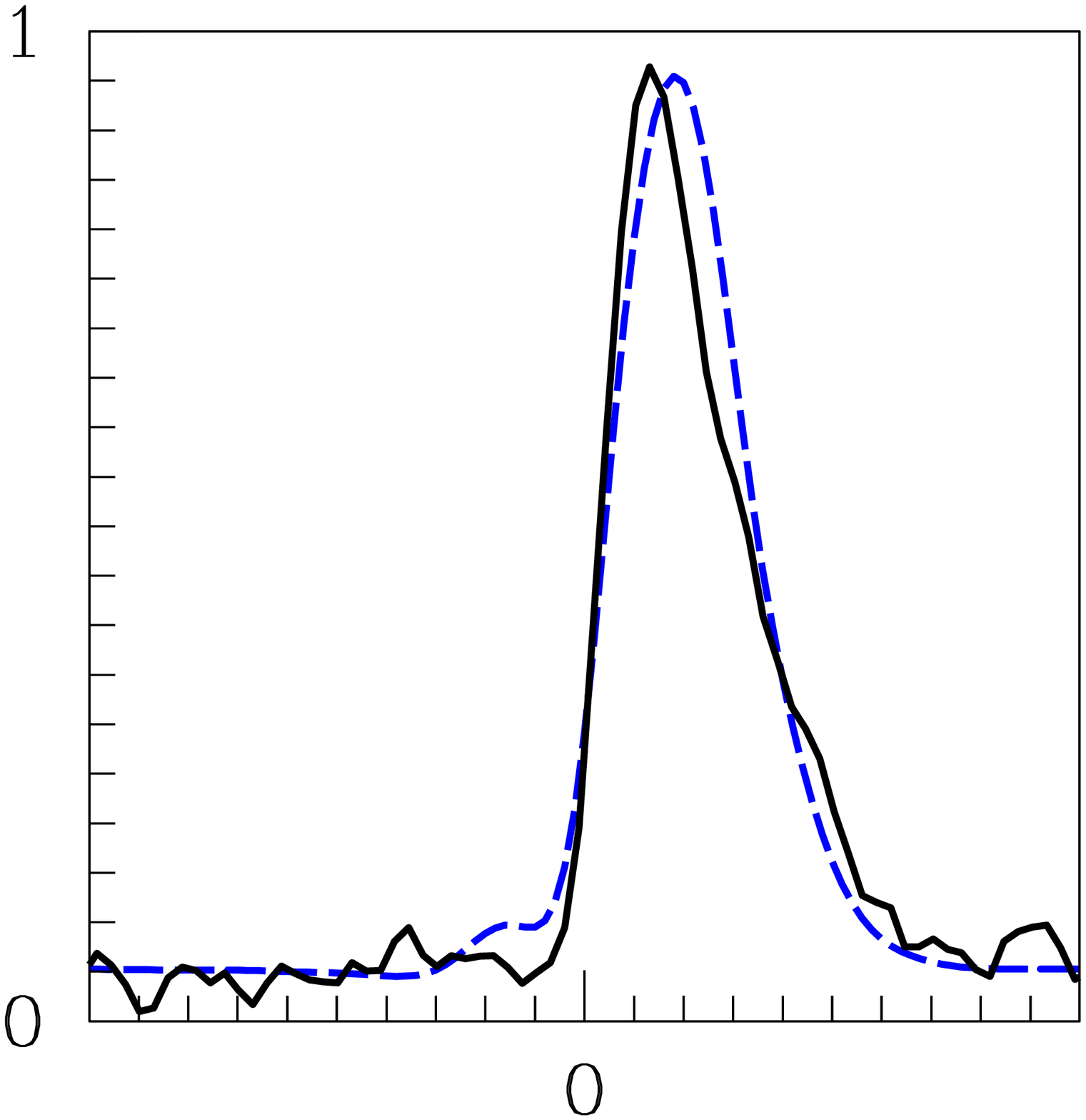} & 
\includegraphics[height=2.4cm,width=2.4cm]
{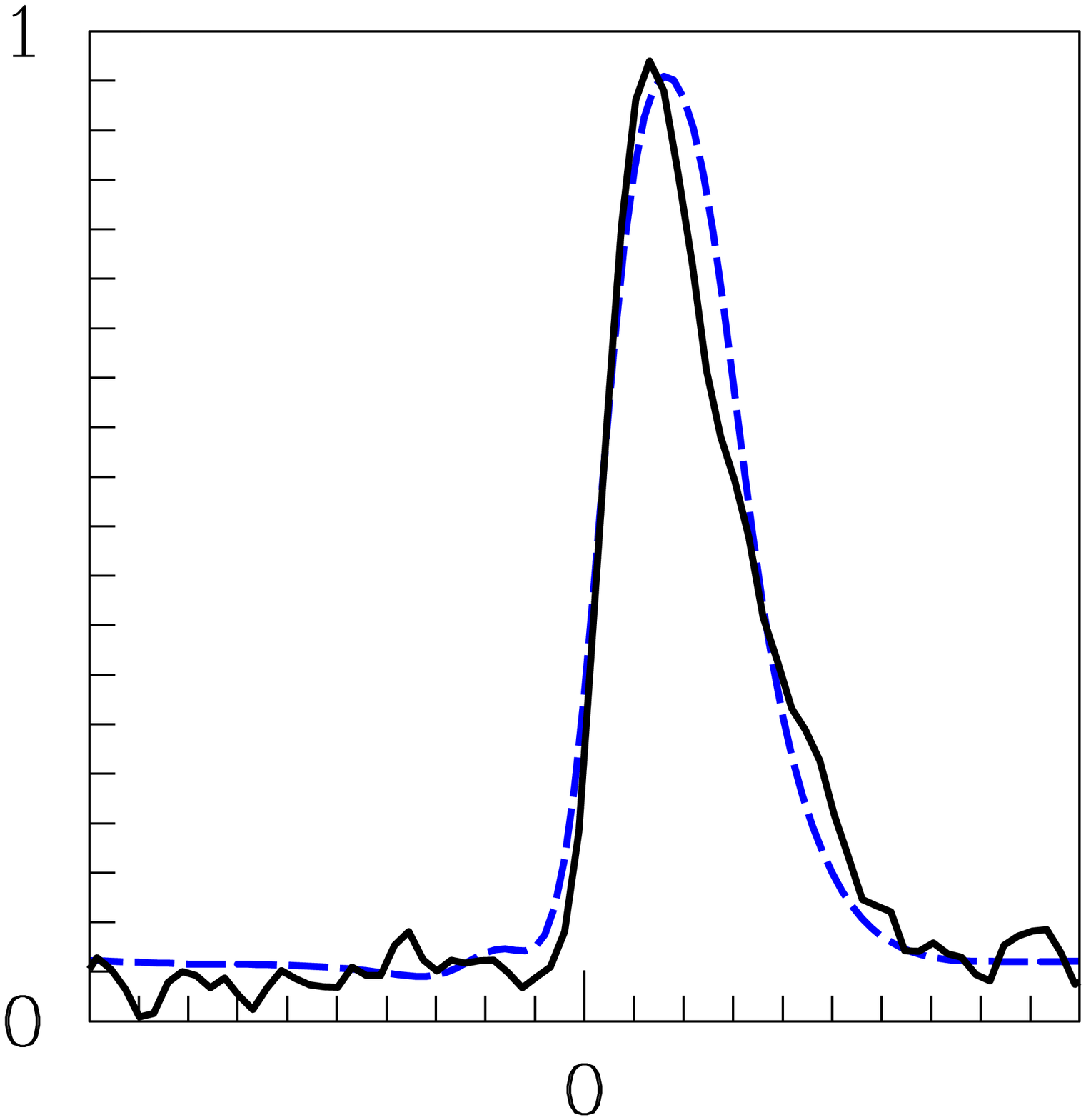} & 
\includegraphics[height=2.4cm,width=2.4cm]
{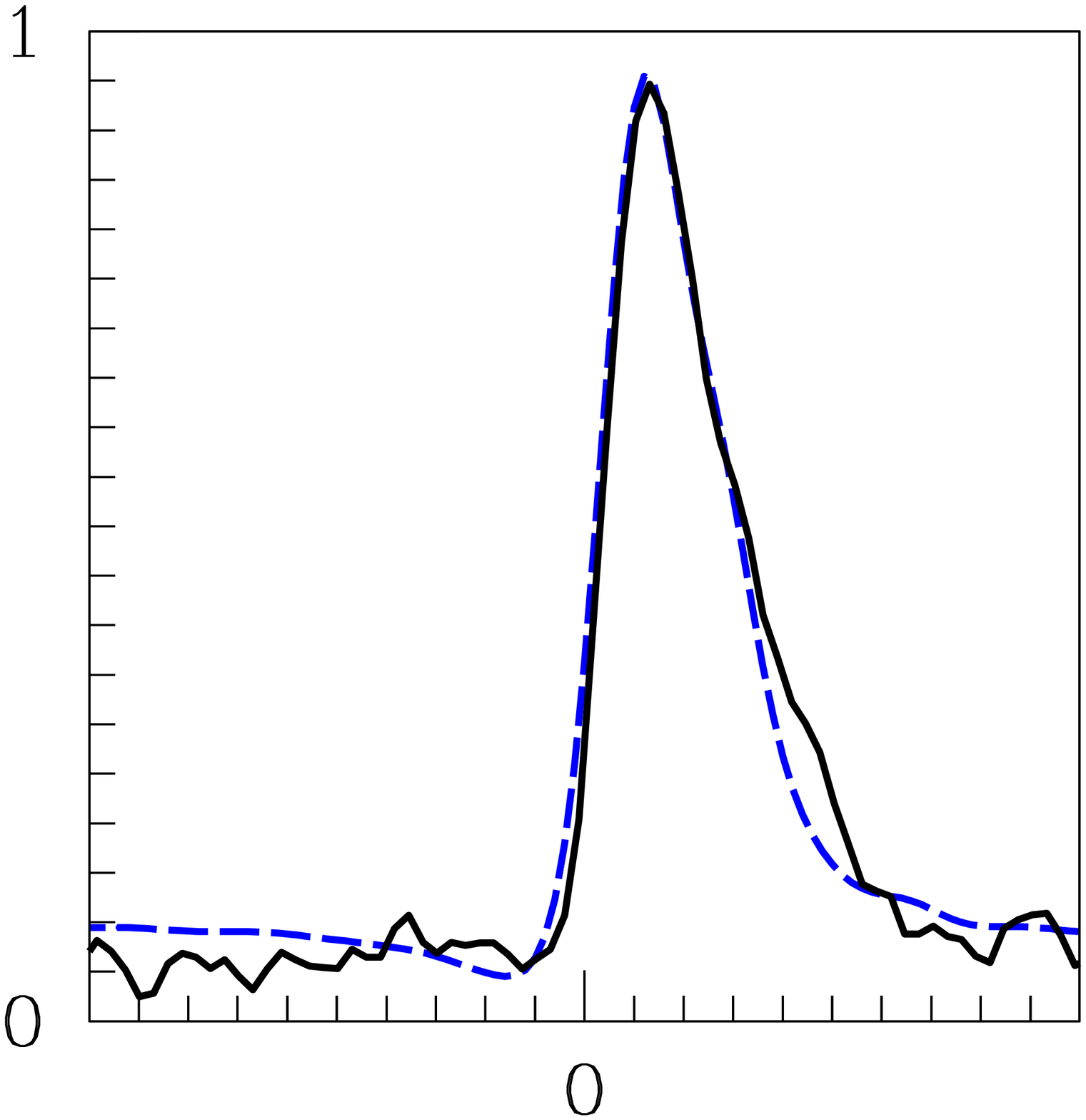} \\
\begin{sideways}\, $\vexp=150$ \kms\end{sideways}&
\includegraphics[height=2.4cm,width=2.4cm]
{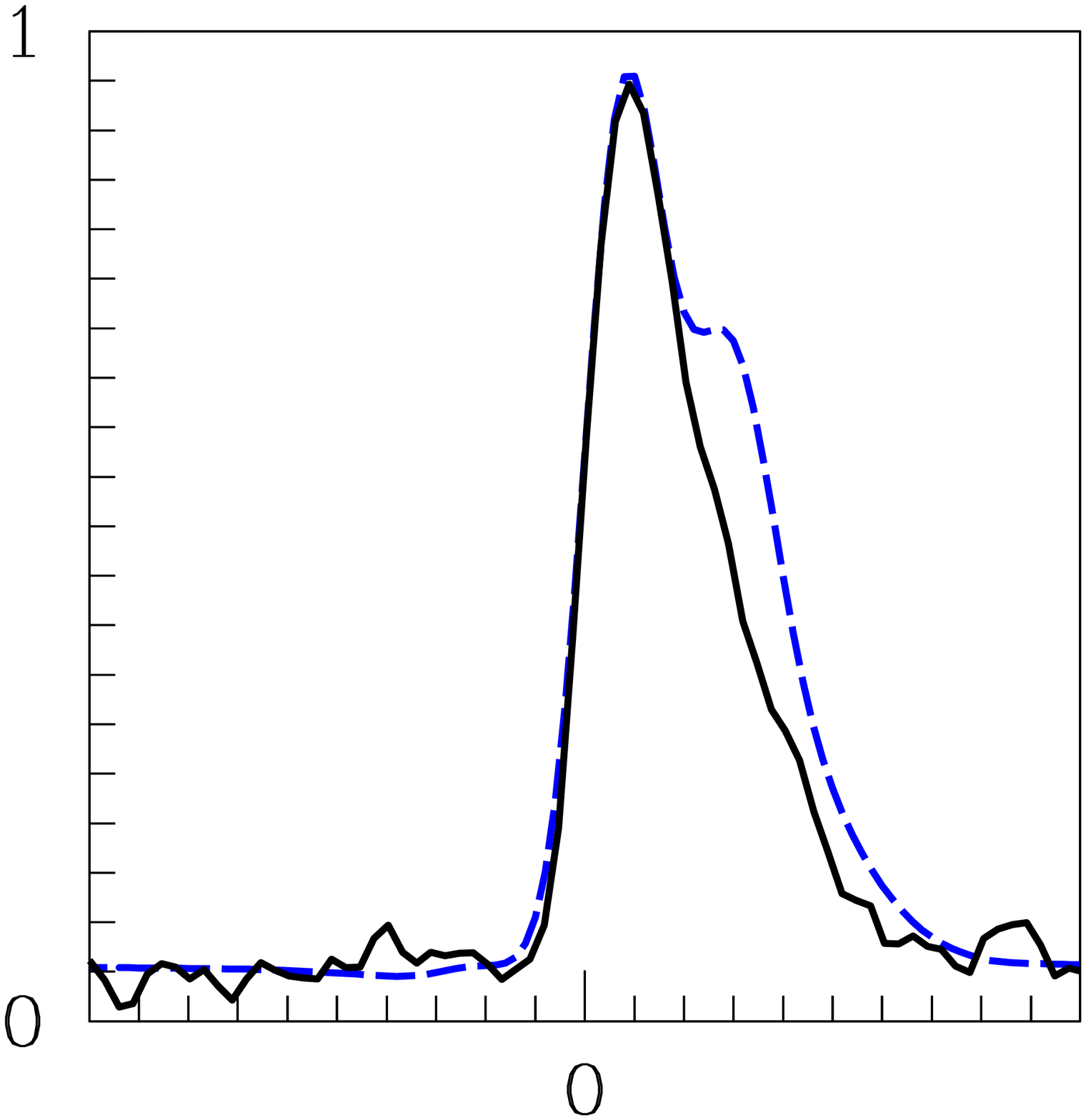} &  
\includegraphics[height=2.4cm,width=2.4cm]
{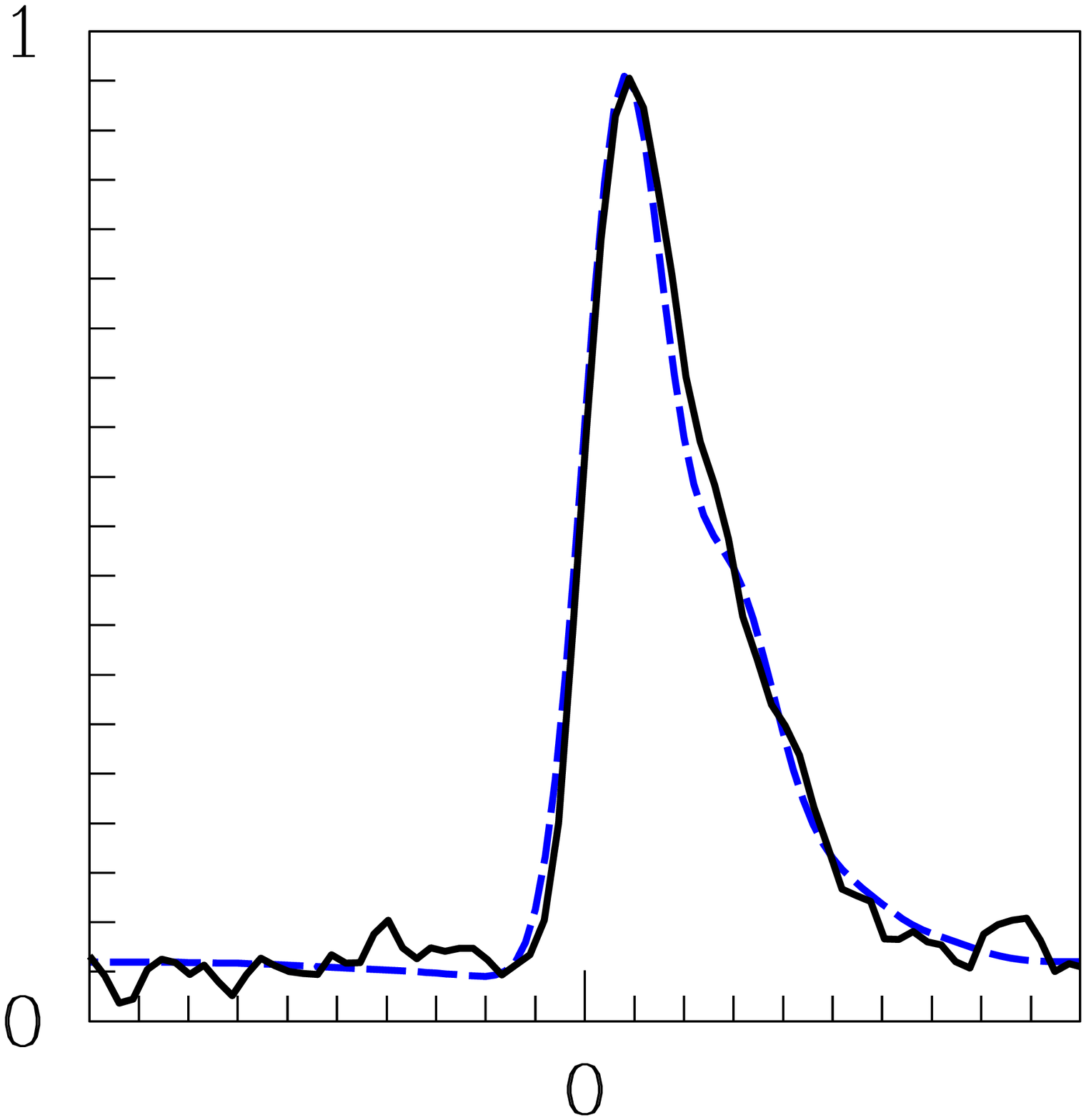} & 
\includegraphics[height=2.4cm,width=2.4cm]
{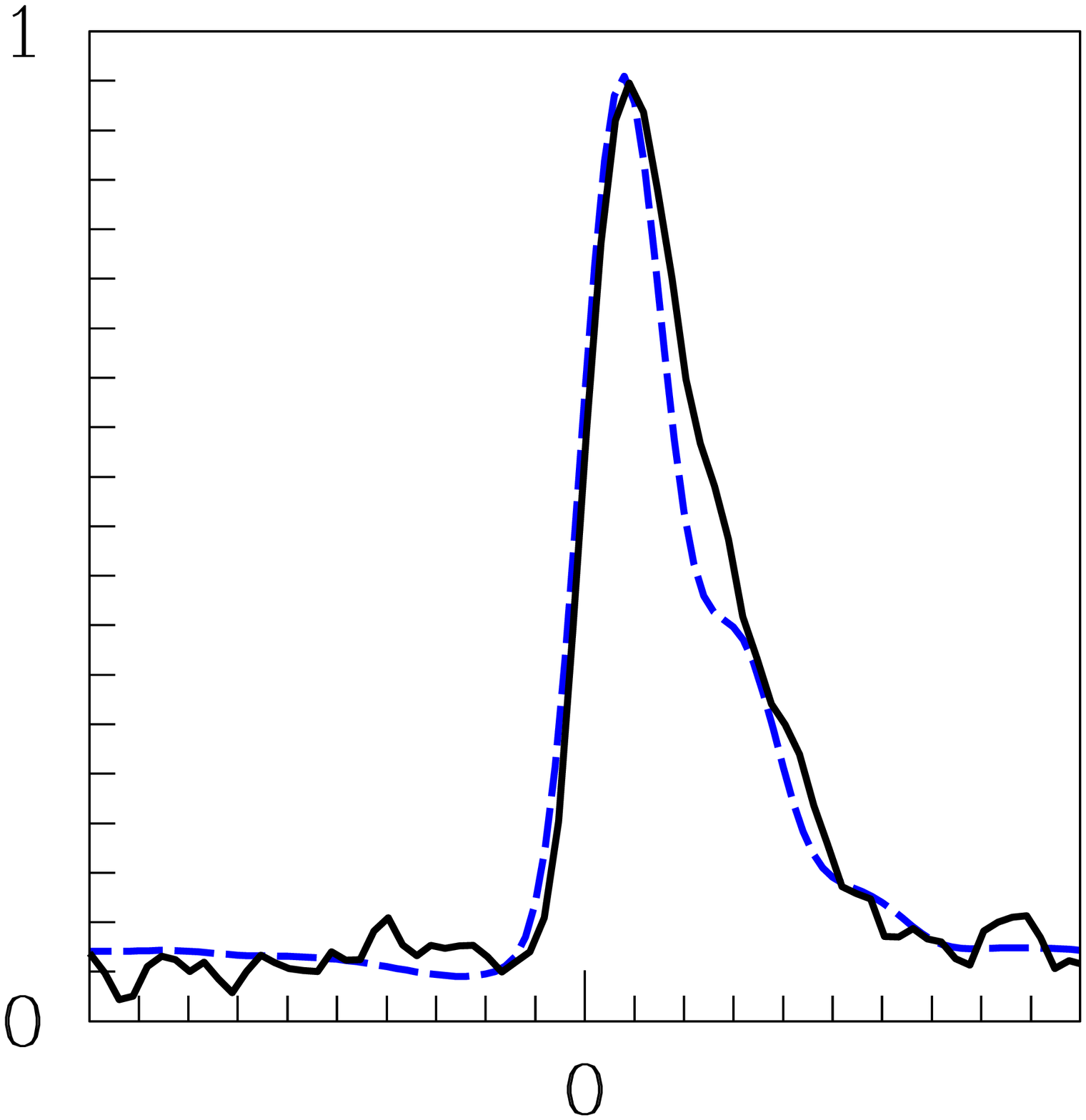} \\
\begin{sideways}\, $\vexp=200$ \kms\end{sideways}&
\includegraphics[height=2.4cm,width=2.4cm]
{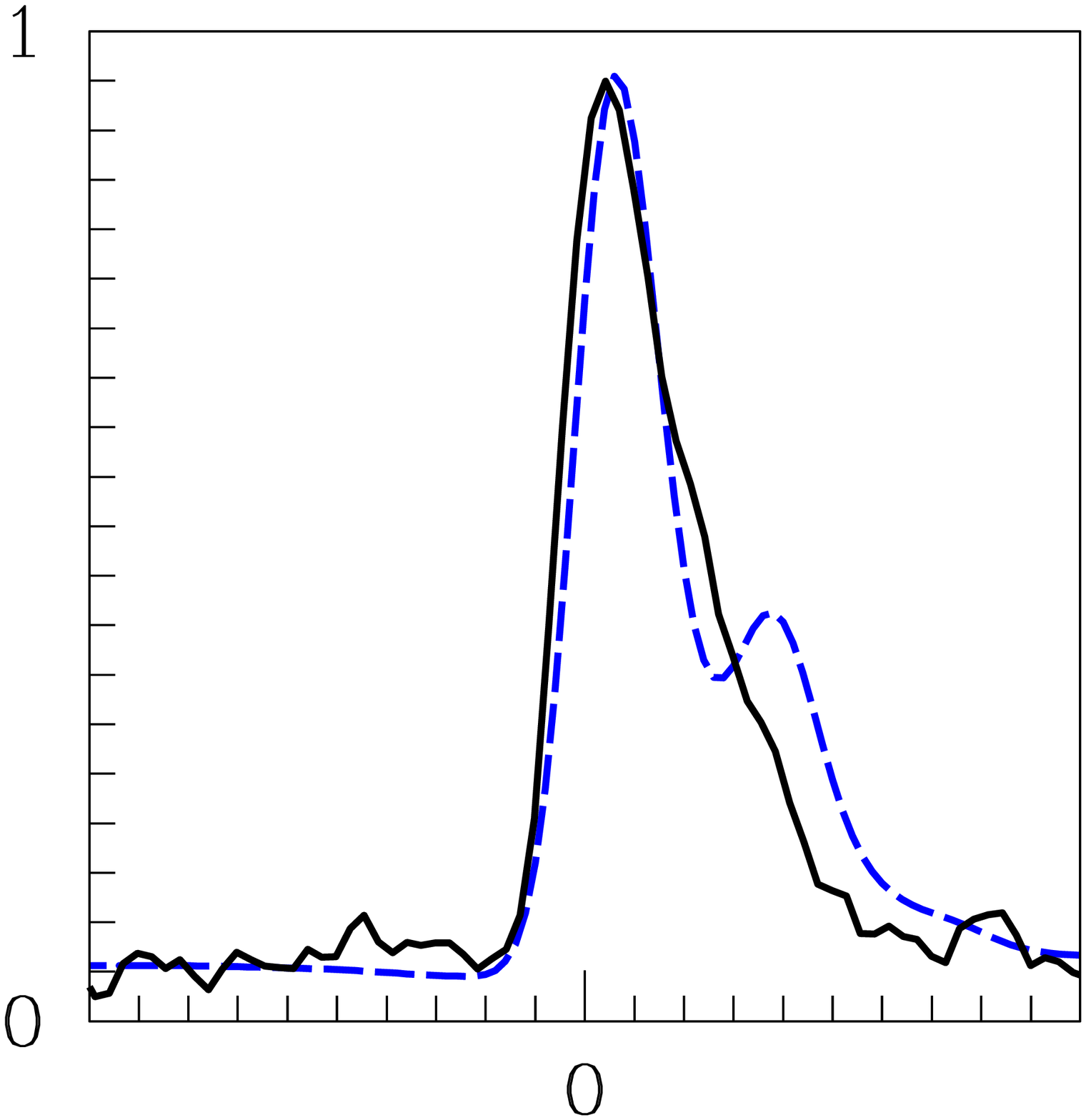} & 
\includegraphics[height=2.4cm,width=2.4cm]
{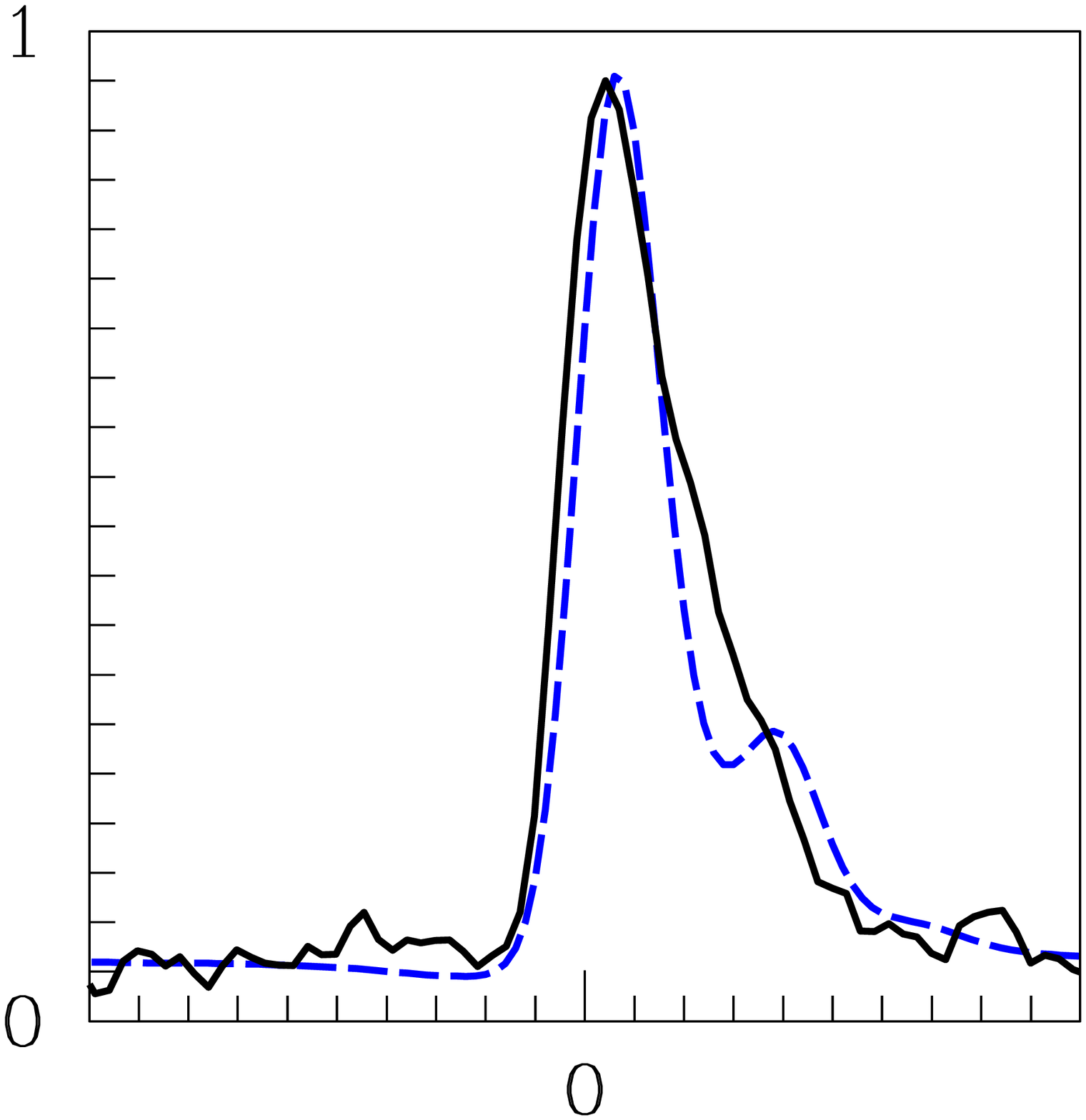} & 
\includegraphics[height=2.4cm,width=2.4cm]
{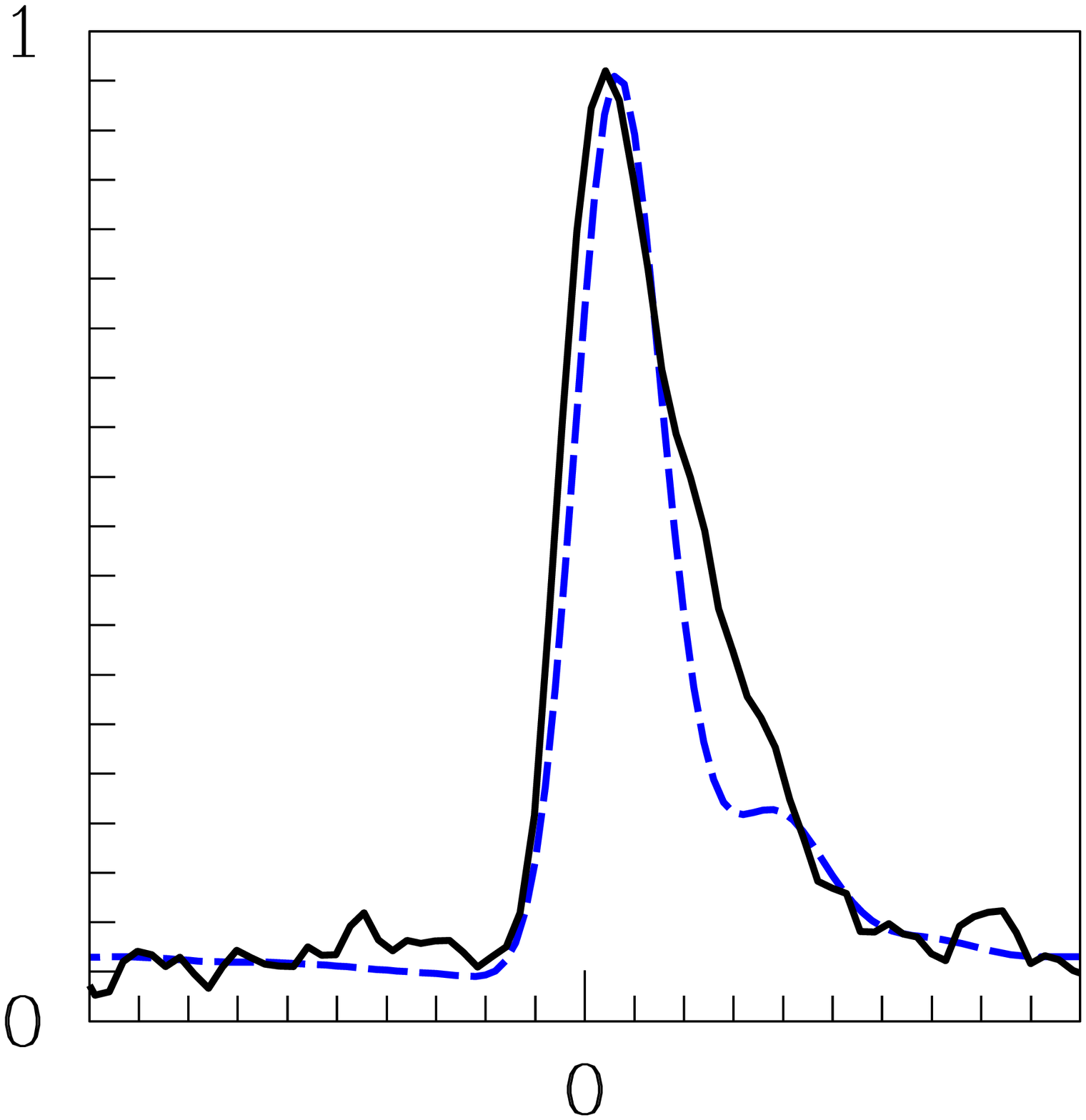} \\ 
\end{tabular} 
\end{minipage}
\begin{minipage}[c]{9.5cm}
\begin{tabular}{cccc}
 & $\tau_a=0.5$ & $\tau_a=1.0$ & $\tau_a=2.0$ \\ 
\begin{sideways}\, $\nh=2\times10^{19}$ cm$^{-2}$\end{sideways}&
\includegraphics[height=2.4cm,width=2.4cm]
{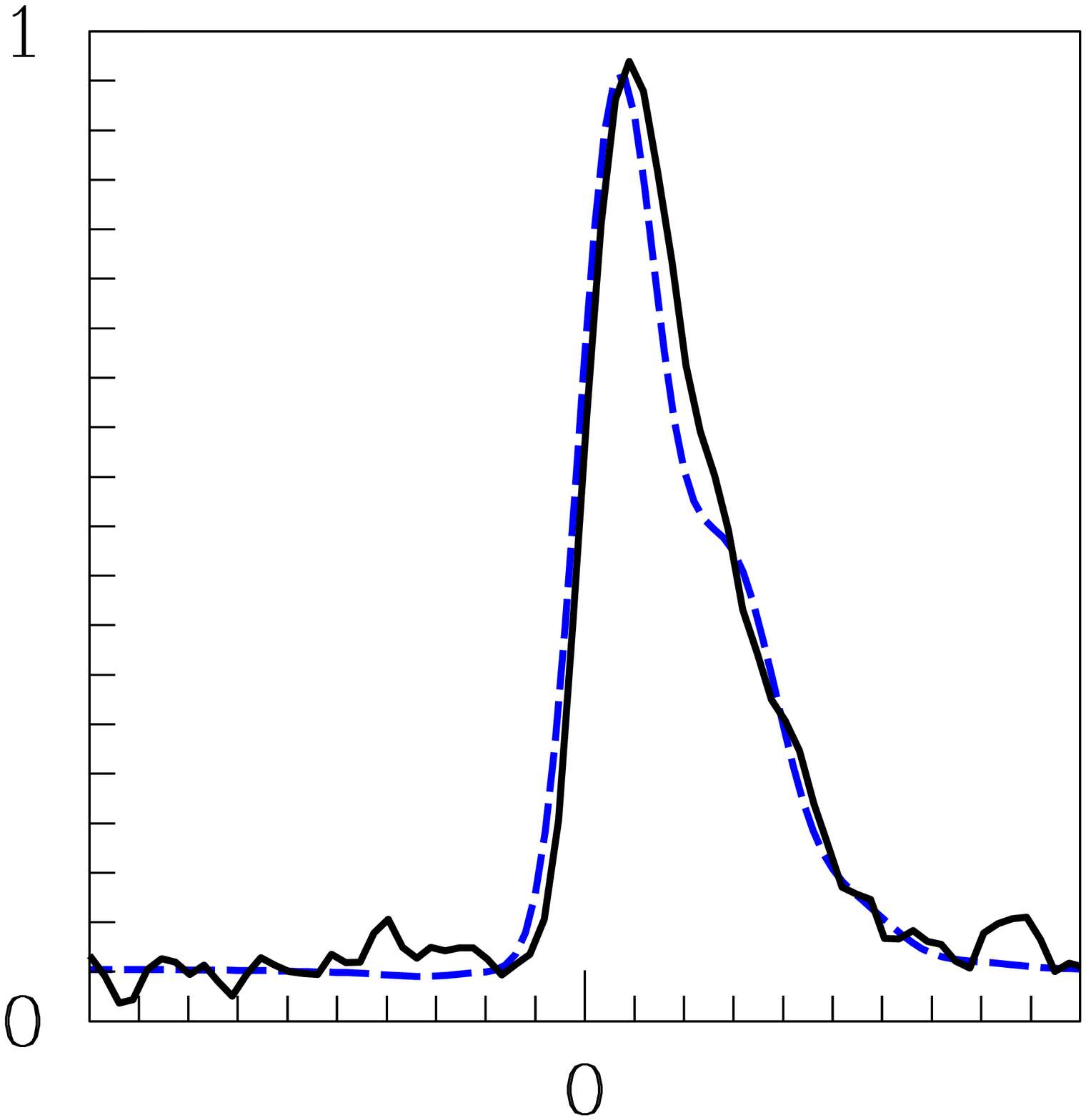} & 
\includegraphics[height=2.4cm,width=2.4cm]
{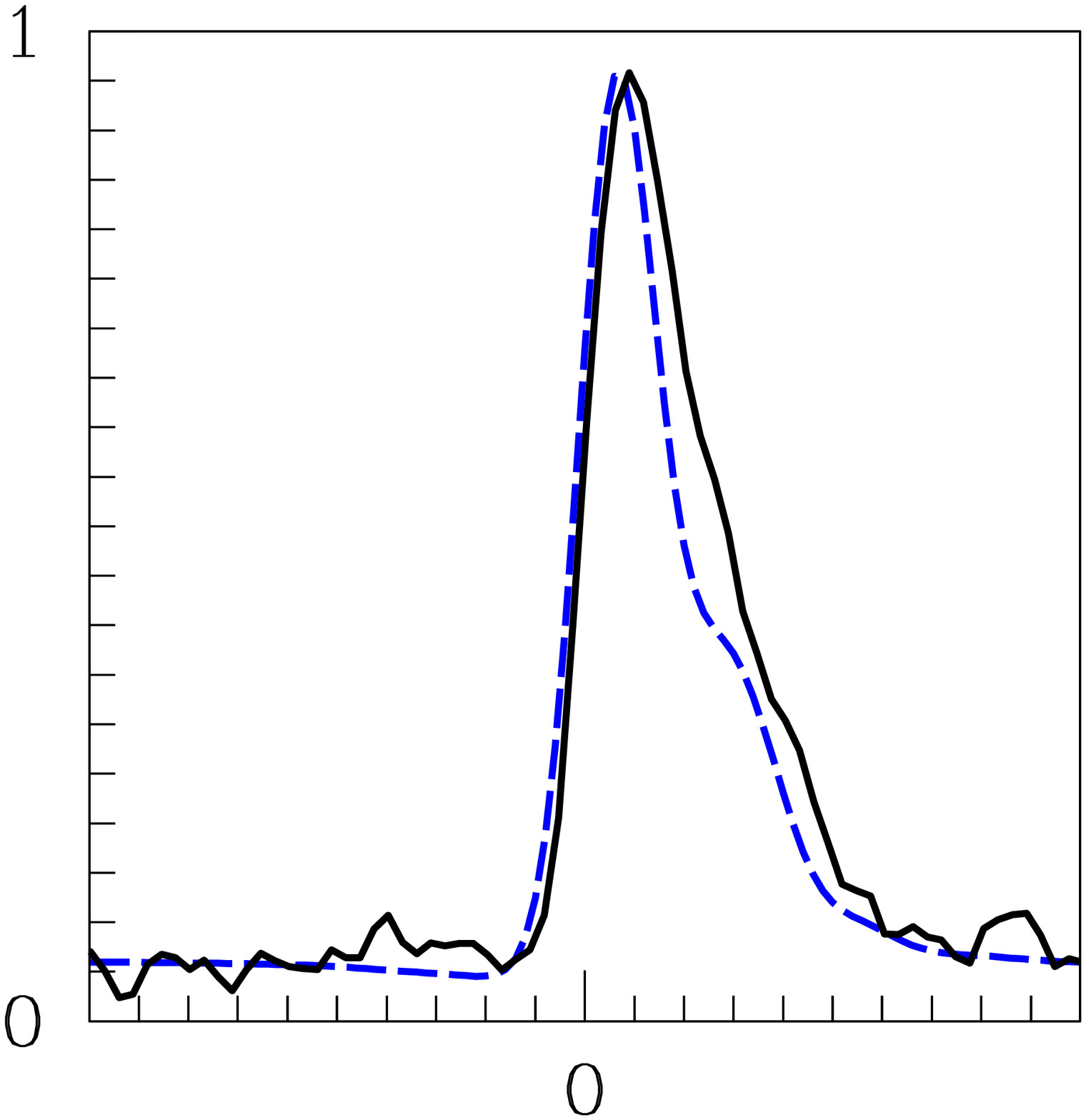} & 
\includegraphics[height=2.4cm,width=2.4cm]
{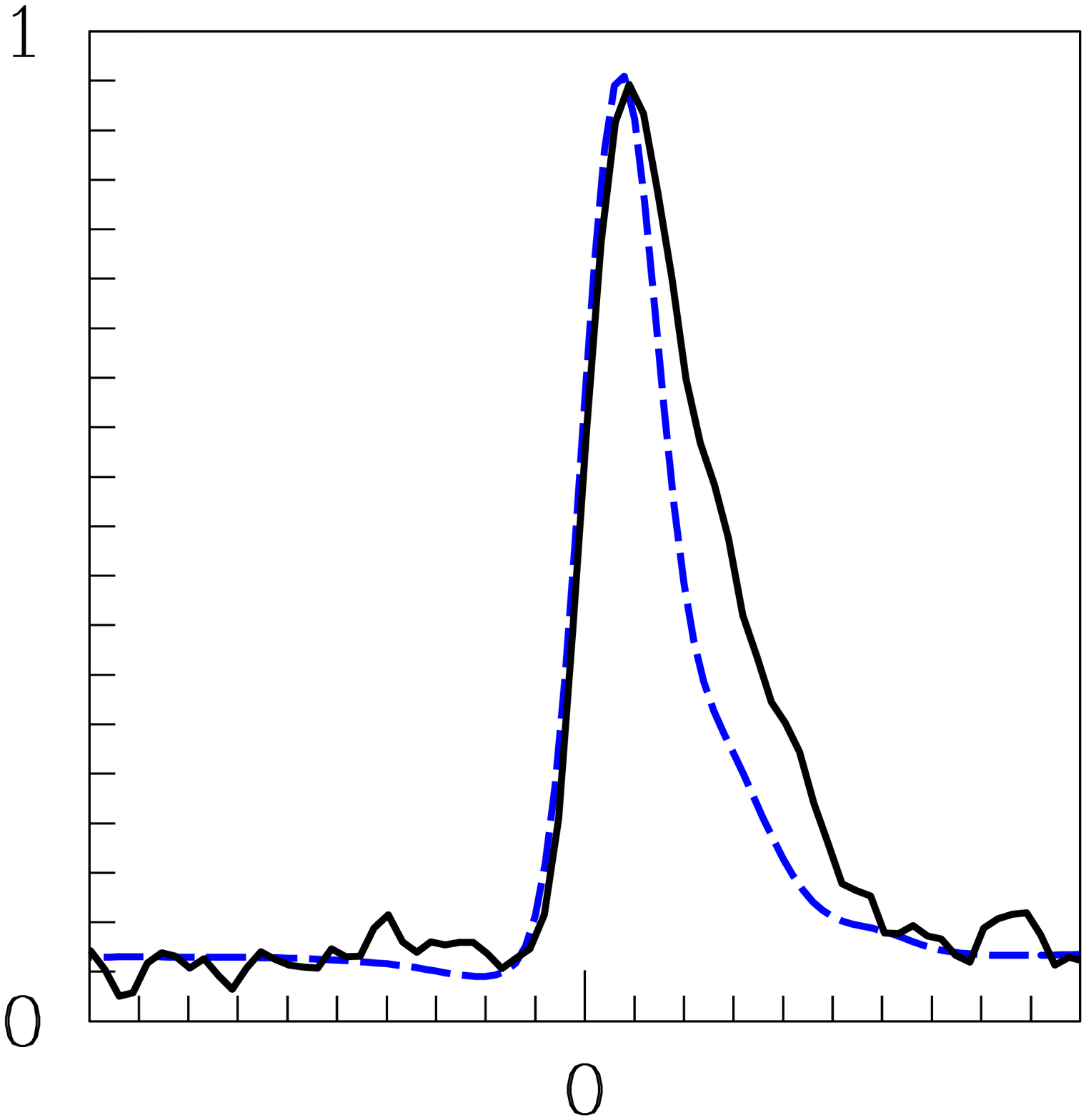} \\
\begin{sideways}\, $\nh=3\times10^{19}$ cm$^{-2}$\end{sideways}&
\includegraphics[height=2.4cm,width=2.4cm]
{fit2384_3N19_V150_d05.eps} &  
\includegraphics[height=2.4cm,width=2.4cm]
{fit2384_3N19_V150_d1.eps} & 
\includegraphics[height=2.4cm,width=2.4cm]
{fit2384_3N19_V150_d2.eps} \\
\begin{sideways}\, $\nh=4\times10^{19}$ cm$^{-2}$\end{sideways}&
\includegraphics[height=2.4cm,width=2.4cm]
{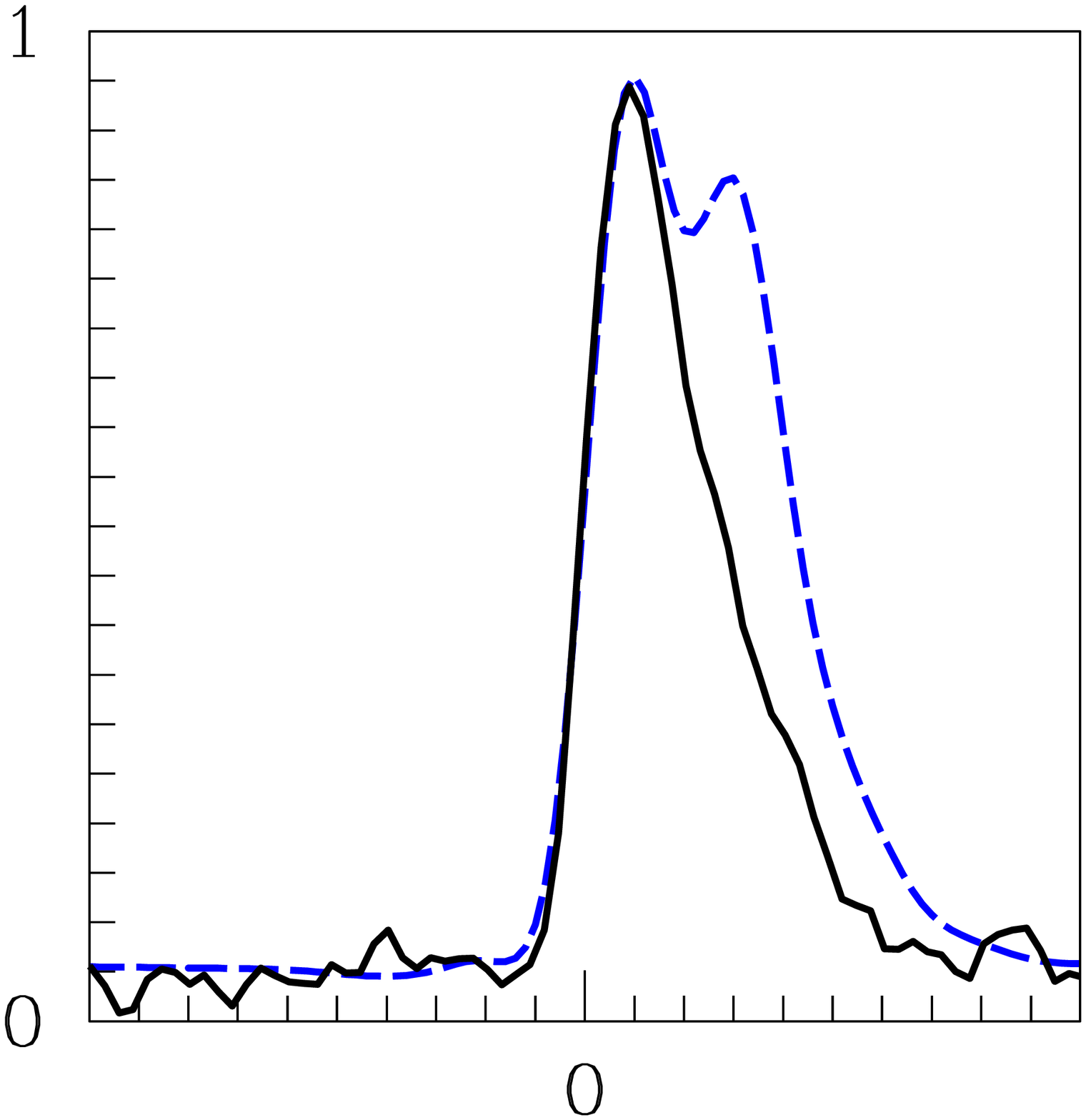} & 
\includegraphics[height=2.4cm,width=2.4cm]
{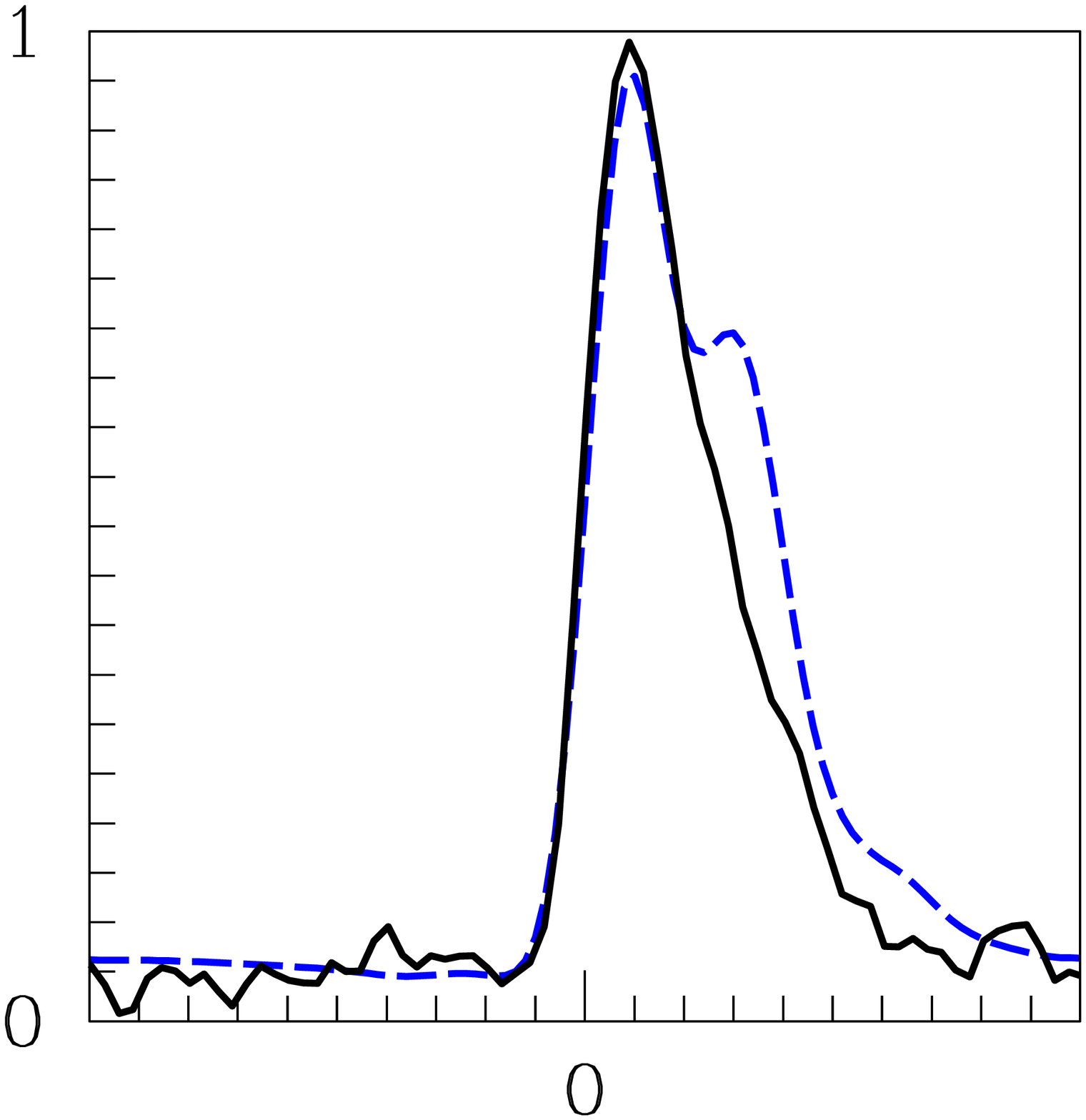} & 
\includegraphics[height=2.4cm,width=2.4cm]
{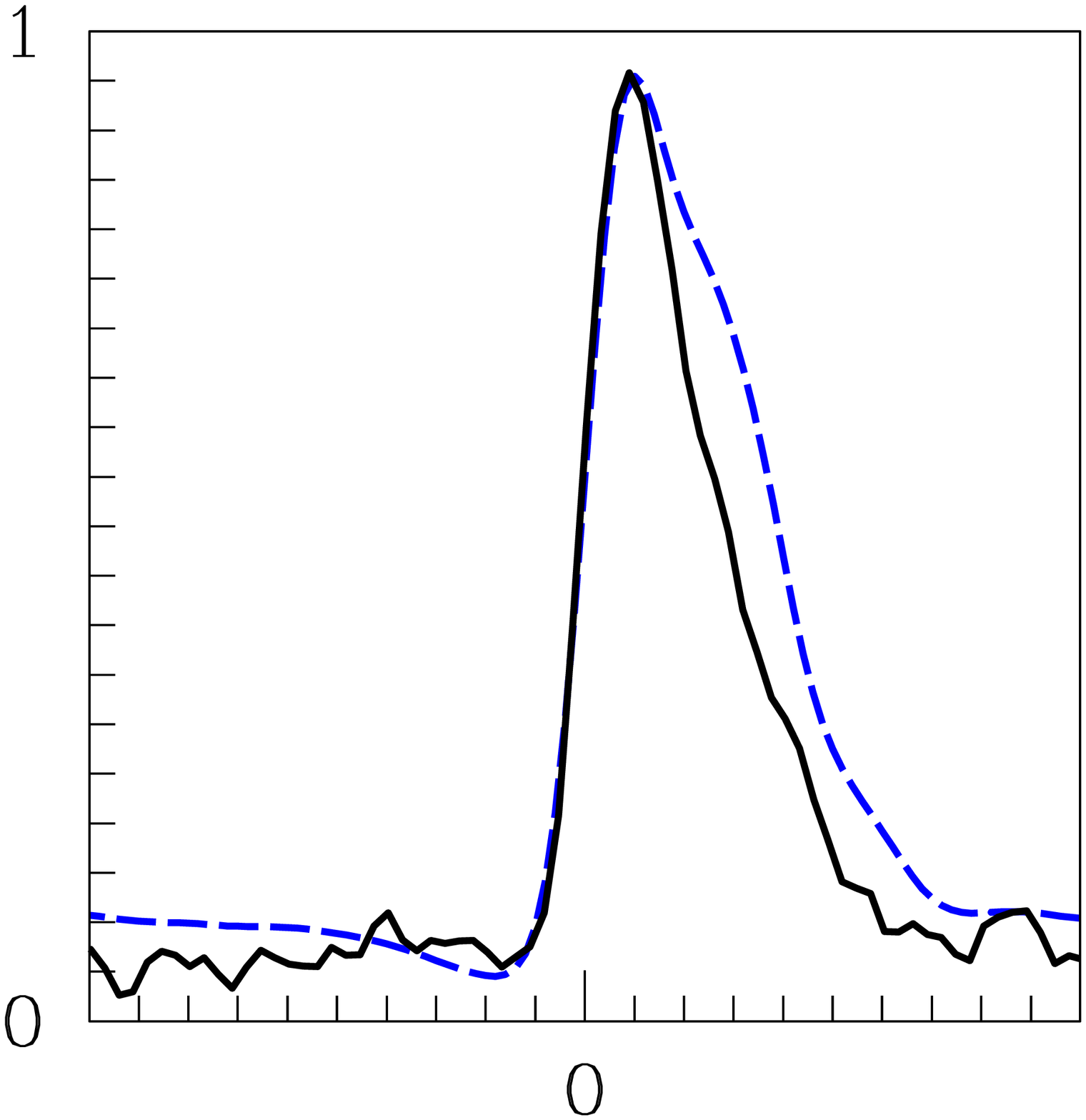} \\ 
\end{tabular}  
\end{minipage}
\caption{Grid of predicted \lya\ line profiles (blue dashed lines)
compared to observed spectral line of FDF2384 to illustrate the 
constraints on the fit parameters \nh, \vexp, and $\tau_a$.
All models have been computed with the same Doppler parameter ($b=20$ \kms),
  and the same input spectrum (a flat continuum+a Gaussian emission
  line with EW$_{\rm int}$=80 \AA and FWHM$_{\rm int}$=100 \kms). 
 {\em Left 3x3 panel:} 
Variations of \vexp\ (from 100 to 200 \kms\ from top to bottom line)
and  $\tau_a$ (from 0.5 to 2.0 from left to right) for fixed $\nh=3\times 10^{19}$ cm$^{-2}$.
{\em Right 3x3 panel:} 
Variations of \nh\ ( from $2\times 10^{19}$ cm$^{-2}$ to
  $\nh=4\times 10^{19}$ cm$^{-2}$) and  $\tau_a$ (from 0.5 to 2.0 from left to right) 
for fixed  $\vexp=150$ \kms.}
\label{mini_grid}
\end{figure*}

To illustrate how well constrained the model parameters are
we will present in detail the fitting of the object FDF2384. 
In Fig.\ \ref{mini_grid} a fraction of the model grid we use is shown
for a fixed value $b=20$ \kms, and for several values of \vexp, \nh\
and $\tau_a$. 
The central profile of each 3x3 grid illustrates the best fitting profile
for FDF2384. The observed profile of FDF2384 is overlayed on each cell.
The overall behaviour of the spectra shown on this Fig.\ can be
summarised as follows.   

When \vexp\ increases, multi-peaks appear on the extended red
wing. Indeed, the location of the second red-peak related to
``backscattered'' photons  --- photons which are reflected by the
 receding shell through the interior --- is at $V_p\sim
2\vexp$, so when \vexp\ increases, the separation between this 
peak and the one at lower velocity (from the first-red-peak due to scatterings
in the forthcoming part of the shell) increases 
(for more details, see \citet{Verh06}). 
 
When \nh\ increases, the ``extension'' of the \lya\ line increases ---
the width between the sharp blue edge and the end of the red extended
wing. Although very clear, this effect is relatively modest here,
since \nh\ changes only by a factor 2. This effect is easily
understood by the increase of the optical depth in the medium, forcing
\lya\ to reach frequencies further from line center to escape.

When $\tau_a$ increases, photons which undergo the highest number of
scatterings will be  destroyed, on average. This enhances the
peaks made of back-scattered photons, whose escape is easier, thanks
to this mechanism, than the escape of diffusing photons. 

The Doppler parameter is fixed to $b=20$\kms, because higher values
broaden the profiles too much, and secondary bumps are smoothed.
Variations of $b$ are discussed in Sect.~\ref{s_degen}.

The overall shape of the \lya\ profile of FDF2384 is smooth, with a
small bump visible 
in the extended red wing (at $V\sim400$ \kms); its position
corresponds to  $\vexp\sim150$ 
\kms. Thanks to the ``extension'' of the line in FDF2384, we can fix 
$\nh \sim (2-4)\times 10^{19}$ cm$^{-2}$. 
Finally, the relative height of the bump compared to the main peak will
determine correlated values for \nh\ and $\tau_a$, and two
combinations are possible: ($\nh=2\times10^{19}$ and
$\tau_a=0.5$) 
provides  a reasonable fit, but the best fit is obtained with
($\nh=3\times10^{19}$ and $\tau_a=1.0$), as shown in
Fig.\ \ref{FDF5812}. 

The derived best fit escape fraction is $f_e\sim0.16$.
There are big differences of the escape fraction, when $\tau_a$
increases from 0.5 to 2, going from $f_e\sim 40\%$ to $f_e<5\%$;
correspondingly the intrinsic strength of the emission line 
varies from EW$_{\rm int}\sim120$ to $170$ \AA.   

We proceed the same way to fit FDF4454 and FDF5812, which appear quite
similar: the expanding shell has the same velocity $\vexp=150$\kms,
the column density in the shell is an order of magnitude lower than in
FDF1337 and FDF5550, but the dust amount, $\tau_a$, remains the same.
Since the intrinsic EW(\lya)$_{\rm int}$ is $\ge$ EW$_{\rm obs}$ and the latter
values are already relatively large, one obtains quite large intrinsic
\lya\ equivalent widths, between $\sim$ 100 and 280 \AA, for
FDF2384, 4454, and 5812. The highest values require fairly young 
ages (see Fig.\ \ref{f_s04_sfr}); however since these objects are
quite faint, their continuum placement may be uncertain and EW$_{\rm int}$
may therefore be overestimated.
The \lya\ escape fraction of FDF4454 is high, $f_e\sim0.42$, because
the shell is less dusty.

\subsubsection{FDF3389 and FDF6557}

\begin{figure*}
\begin{tabular}{cc}
\includegraphics[height=6cm, width=8cm]{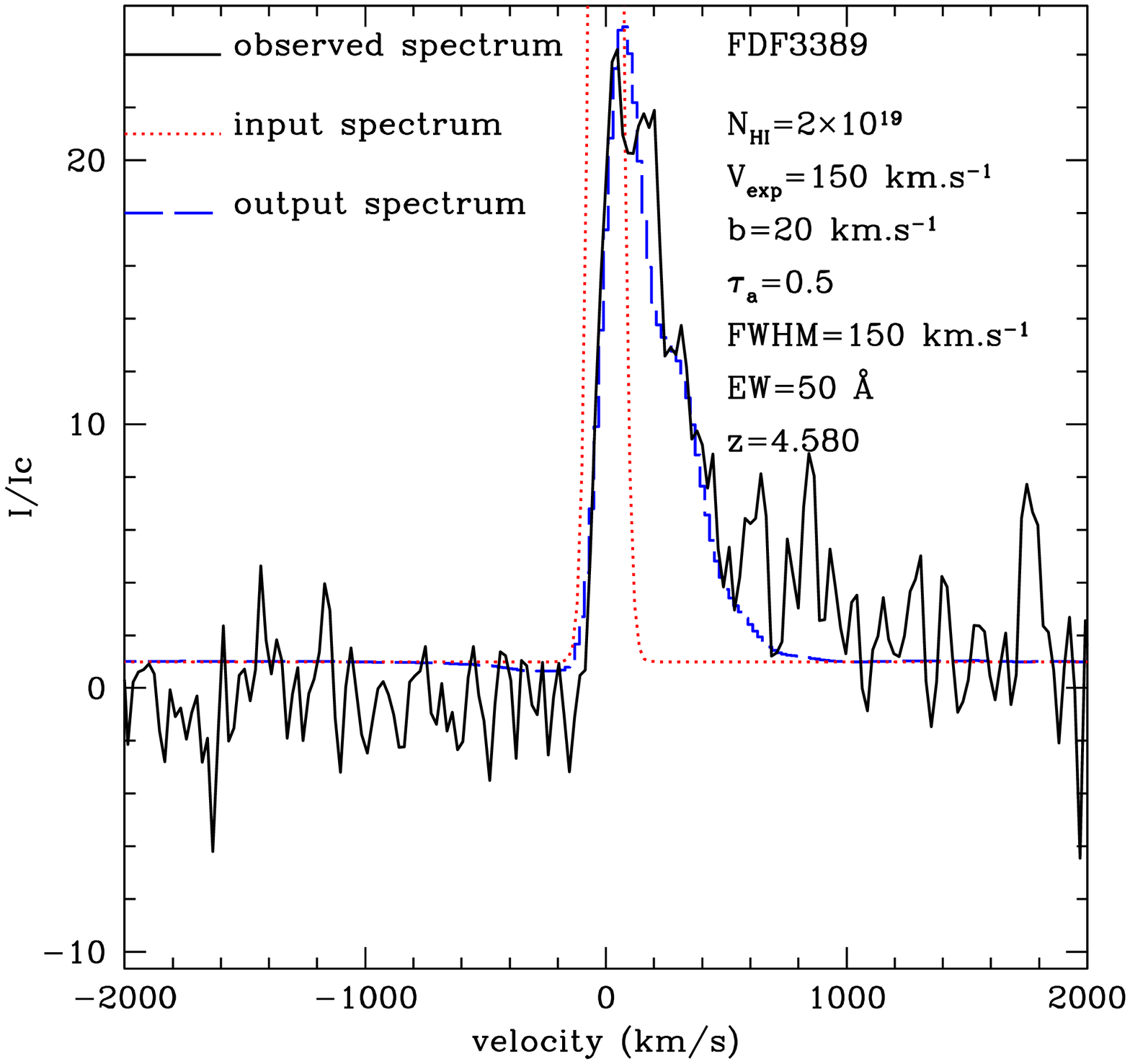} &
\includegraphics[height=6cm, width=8cm]{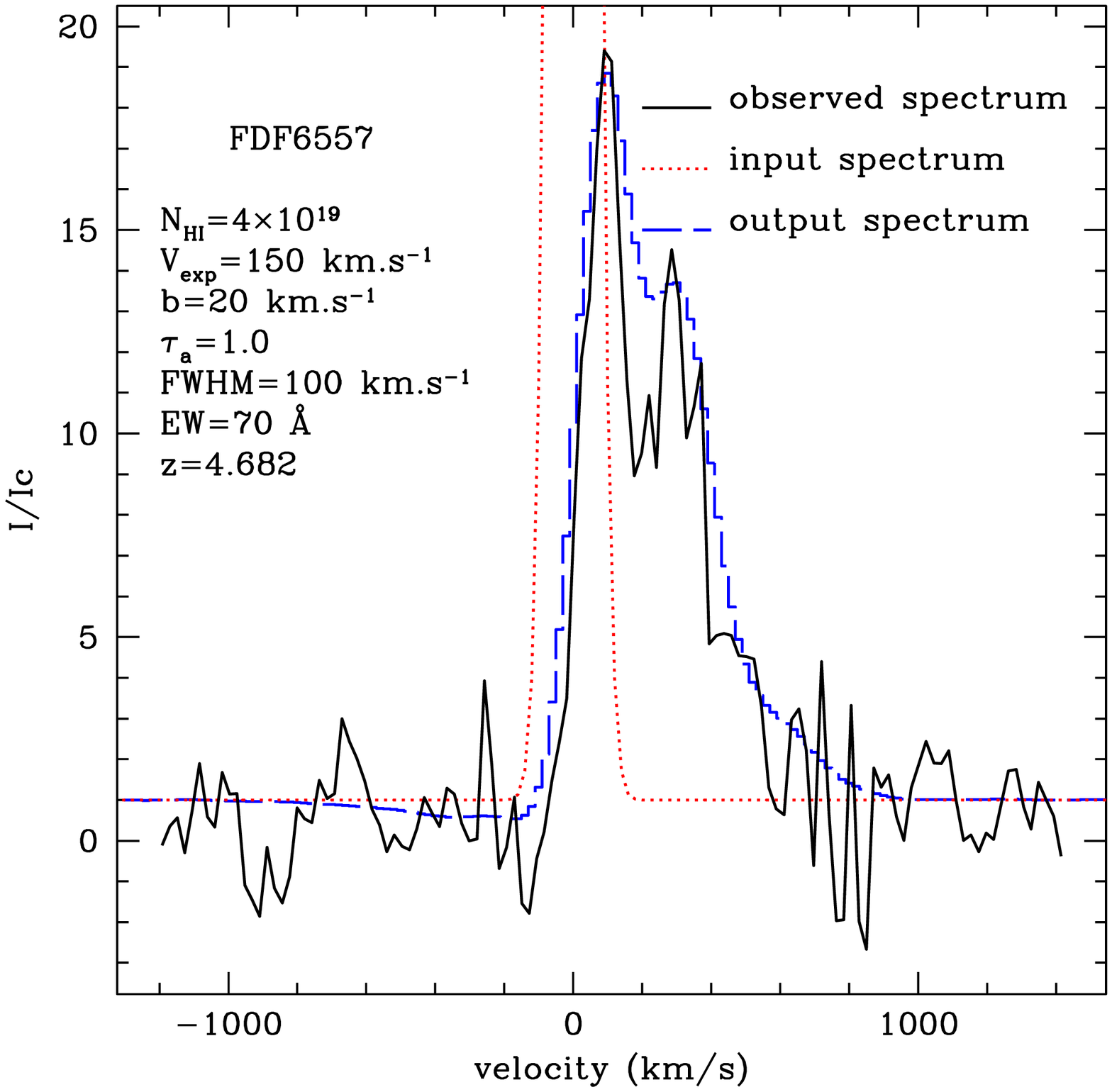} \\
\end{tabular}
\caption{Line profile fits for FDF3389 ({\em left}) and
FDF6557 ({\em right}) showing asymmetric spectra with probable
secondary structures on the red wing. Same symbols as in Fig.\ 
\ref{FDF1337_5550}.  The parameters derived from our fitting
($\vexp=150$ \kms, $\nh\sim 2-4\times10^{19}$ cm$^{-2}$,
$\tau_a=0.5-1$) are similar to the 3 narrow asymmetric cases (cf.\
Fig.\ \ref{FDF5812}) except for lower intrinsic EW, which here show
values compatible with expectations for constant SFR over long
timescales. }
\label{FDF3389}
\end{figure*}

These two objects are both at $z\sim4.5$. They have medium observed
equivalent widths (EW(\lya)$_{\rm obs}=30-40$ \AA),
compared to the former asymmetric profiles, and rather small
extensions ($\sim 600-700$ \kms). We treat them separately, because
their spectra seem more complex than a smooth asymmetric emission
line. They present multi-peaks on the red extension of the line.
However, they are also more noisy, and these secondary features are
only twice the noise amplitude. Nevertheless, we assume that they are
significant, and we derive best-fits, taking these features into
account (see Fig.\ \ref{FDF3389}). 

The parameters
derived from our fitting are close to the 3 narrow asymmetric cases
($\vexp=150$ \kms, $\nh\sim 2-4\times10^{19}$ cm$^{-2}$, $\tau_a=0.5-1$)
except for the intrinsic EW, which have ``canonical'' values --- compatible
with expectations for SFR=const --- again,
because the observed EW are lower for these objects. The escape
fraction is high for FDF3389 ($f_e=0.45$), and lower for FDF6557
($f_e=0.17$) because of a higher column density and so a higher
dust content. 

\subsection{The double-peaked \lya\ profiles (group B)}

Two of the 11 spectra present double-peaked profiles (FDF4691 and
FDF7539). 
Such line morphologies are a natural outcome of  radiation transfer in a static medium \citet{Neuf90}, since in such media 
the \lya\ photons can only escape by diffusing into the red or blue wings, where
the opacity decreases rapidly.

\begin{figure*}[htb]
\begin{tabular}{cc}
\includegraphics[height=6cm, width=8cm]{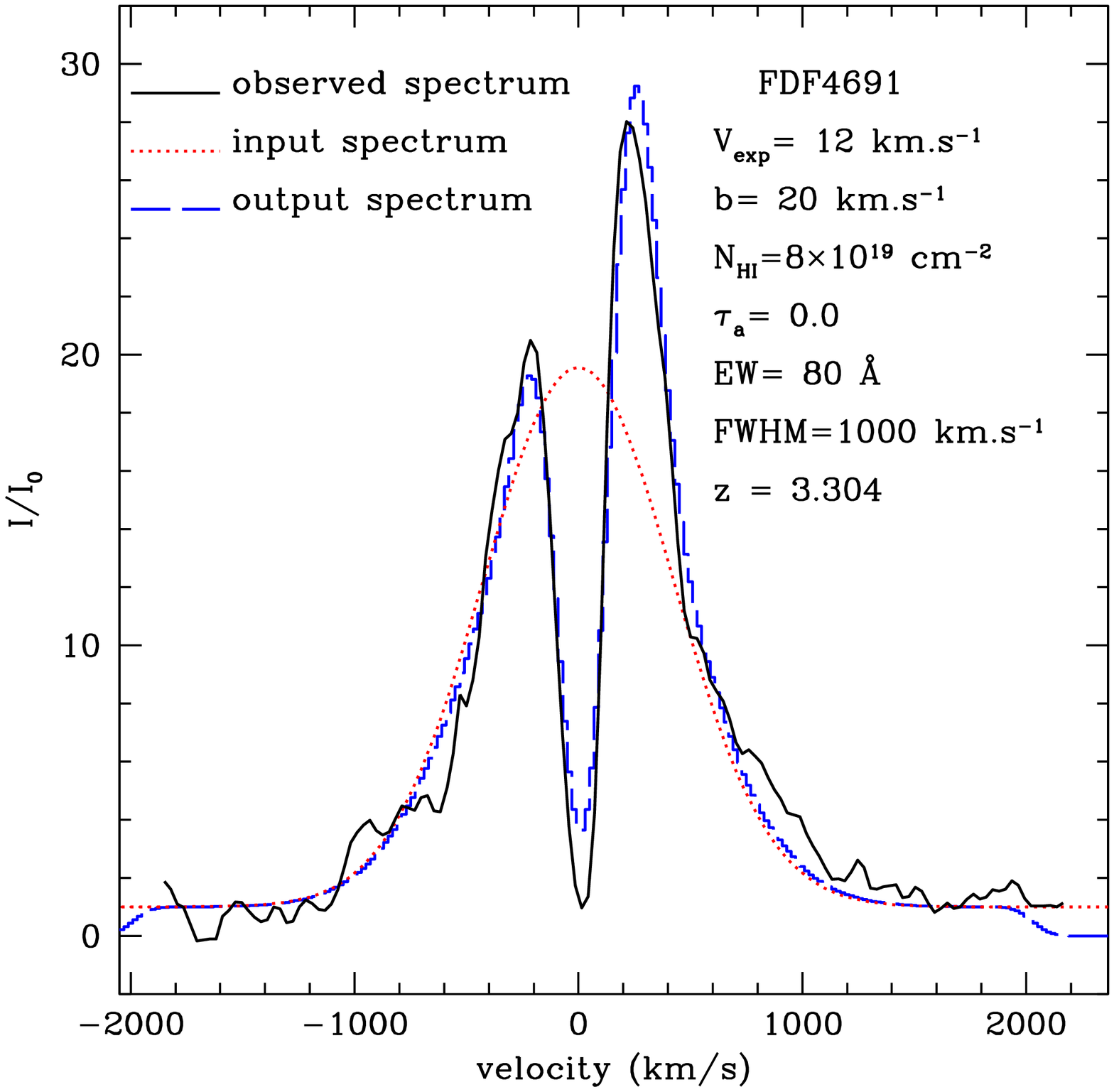} &
\includegraphics[height=6cm, width=8cm]{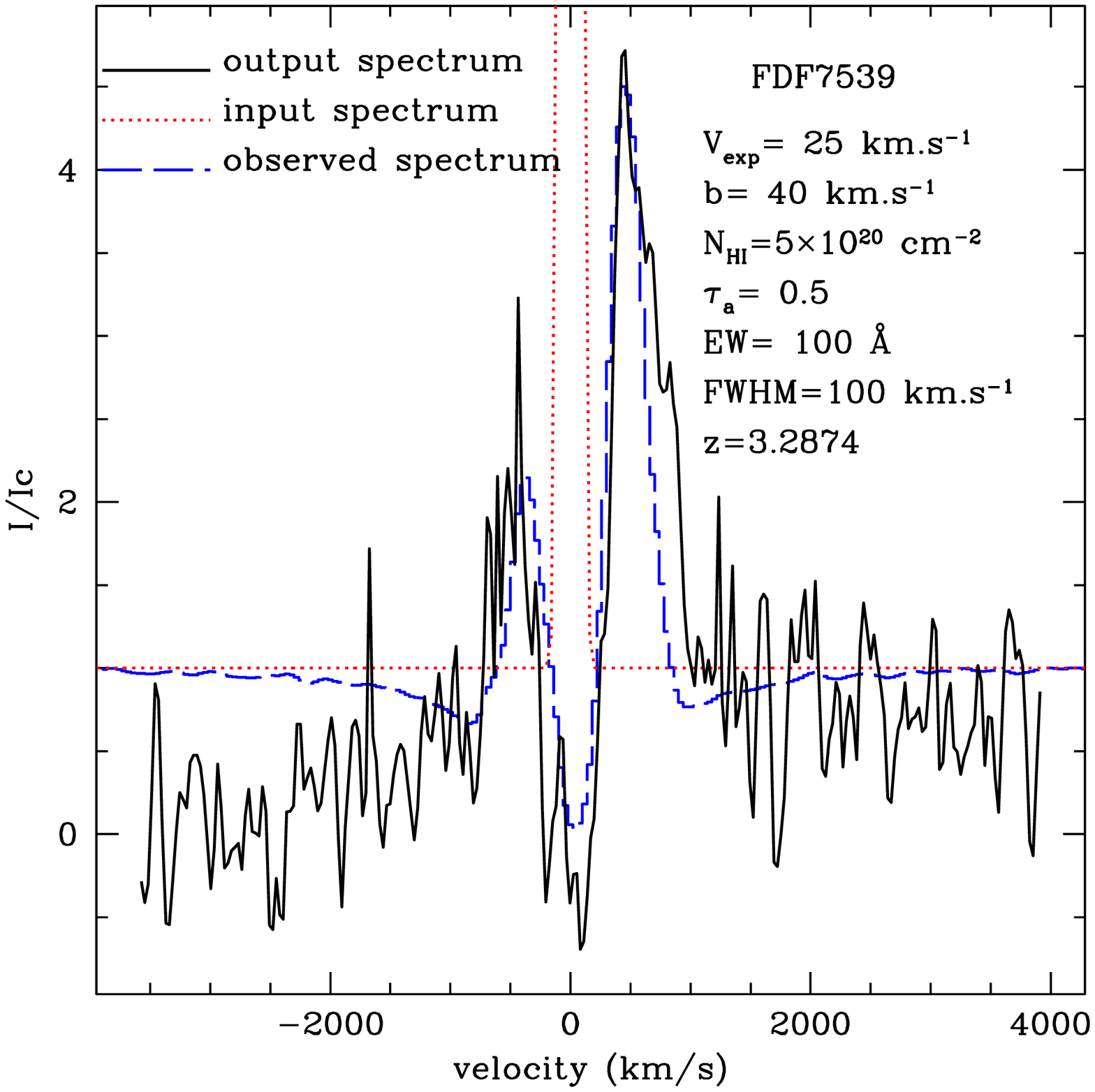} \\
\end{tabular}
\caption{Line profile fits for FDF4691 ({\em left}) and FDF7539 ({\em right}),
  the two double peaked profiles (type B) with static or almost
  static shells ($\vexp=10-25$ \kms). The peak separation and the
  observed EW(\lya)$_{\rm obs}$ are different for these objects, and so
  are the other fitting parameters. FDF4691 is the only object for which a
  very broad input spectrum is derived from the modelling ($FWHM=1000$
  \kms instead of $\sim 100$ for all other objects), maybe a signature
  of a hidden AGN.}
\label{FDF4691_7539}
\end{figure*}

\subsubsection{FDF4691}
This galaxy has a high EW(\lya)$_{\rm obs}\sim80$ \AA.
\citet{Tapk04} fitted this object with a code using a finite element
method \citep{Rich01}, and proposed $\nh=4.\times 10^{17}$ cm$^{-2}$
and $b=60$ \kms\ in an almost static (\vexp=12 \kms) and dust-free
shell as best-fit parameters. The intrinsic spectrum they use is a
Gaussian with FWHM=1000 \kms, and no continuum. 
Using the same parameters we can reproduce their fit.

However, since their code is not suited to high column densities 
we searched for other solutions.
We find a fit of better quality --- the deep gap between the peaks
is better reproduced by a smaller $b$, and the red wing is fitted
with more accuracy starting from an input spectrum with a continuum --- with a
higher column density ($\nh=8\times 10^{19}$ cm$^{-2}$), and consequently 
a smaller $b=20$ \kms\footnote{Indeed, the location
  of the peaks were predicted in static media, and depend on a
  combination of \nh\ and b} 
(see Fig.\ \ref{FDF4691_7539} for the fit and Table~\ref{t_summary} for a
summary of the parameters).  Our best fit is obtained with no dust,
and the fit with $\tau_a=0.1$ is less good than the one with no dust,
from which we estimate an upper limit on the dust content of $\tau_a
\la 0.1$.  
This is consistent with the fact that the \lya\ and UV SFR indicators
derived from observations \citep{Tapk07} are similar.    
To reproduce the very extended wings of the line, the intrinsic
\lya\ emission line has to be very broad. It
is characterised by a very large value of FWHM$_{\rm int}=1000$ \kms,
and a ``st andard'' 
EW(\lya)$_{\rm int}=90$ \AA. Radiation transfer effects are inefficient to
broaden the line in a medium with such a small column density.  
If interpreted as a result of virial motions, such a large
FWHM seems, however, unphysical.
A hidden AGN may be an explanation for the high FWHM, as suggested by
\citet{Tapken05}.

\begin{figure}[tb]
\includegraphics[height=6cm, width=6cm]{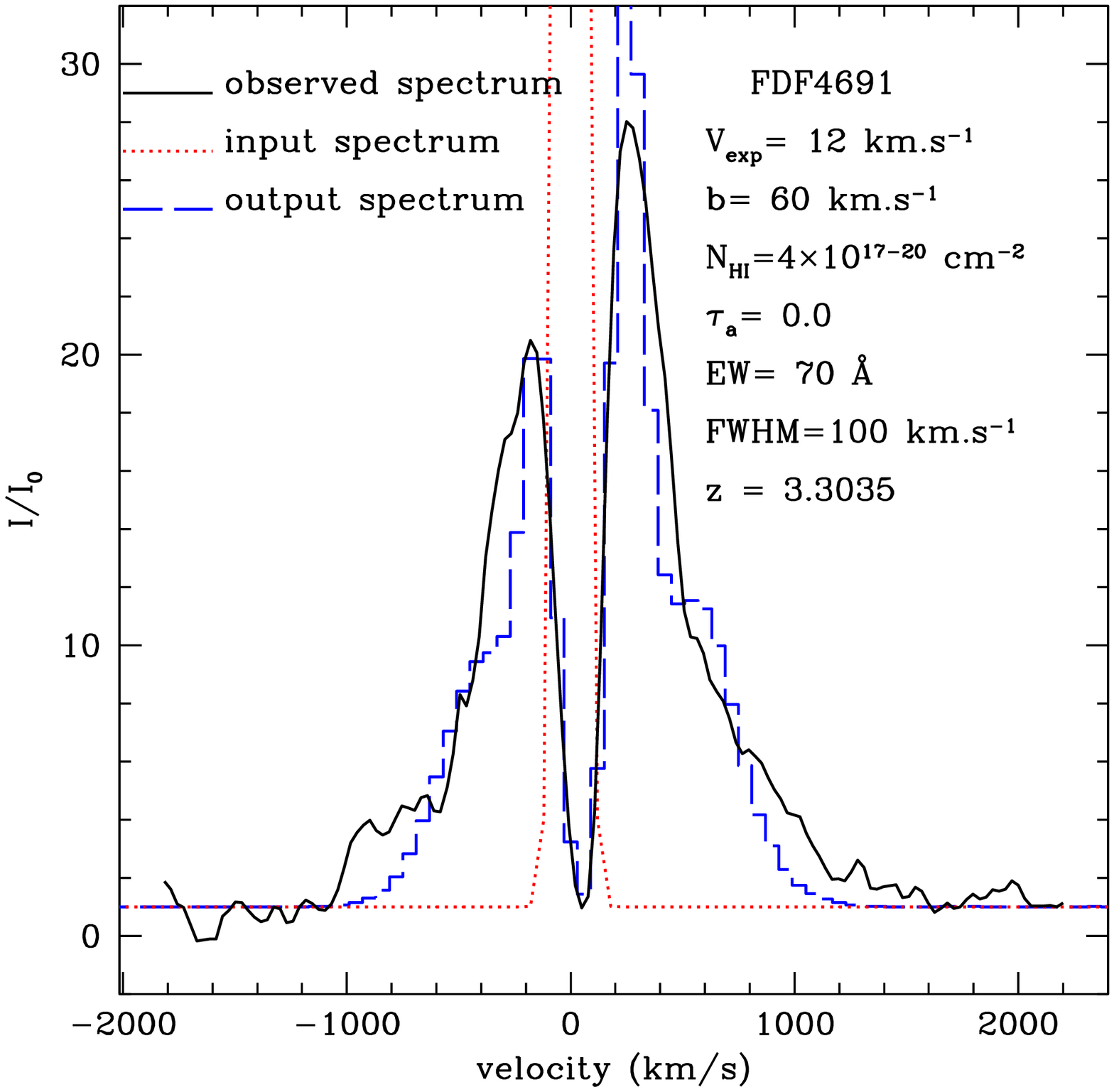} 
\caption{Fit of FDF4691 with a ``two-phase'' model, a low density
  shell with $\nh=4.\times 10^{17}$ cm$^{-2}$ and a high density shell
  with $\nh=4.\times 10^{20}$ cm$^{-2}$. The other parameters of the
  shells are identical: $b=60$ \kms, $\vexp=12$ \kms, no dust,
  EW(\lya)$_{\rm int}=70$ \AA. This allows for a more reasonable
  FWHM(\lya)$_{\rm int}=100$ \kms instead of 1000.}  
\label{FDF4691_2P}
\end{figure}

A solution to reproduce the observed spectrum with a more realistic
intrinsic spectrum (FWHM$_{\rm int}$=100 \kms and EW$_{\rm int}$=80
\AA) is to invoke two 
contributions from two different media: when we sum emergent spectra
from two identical shells except for the column density ($b=20$ \kms,
$\vexp=10$ \kms, no dust, EW(\lya)$_{\rm int}=80$ \AA, and
$nh=4.\times 10^{17}$ cm$^{-2}$ for one and $\nh=4.\times 10^{20}$
cm$^{-2}$ for the other), we are able to reproduce a spectrum with
narrow peaks close to the center and broad wings, starting from a
``standard'' value for the FWHM, FWHM(\lya)$_{\rm int}=100$ instead of
1000 \kms\ (see Fig.~\ref{FDF4691_2P}). This could correspond to a
physical situation where an 
initially thick shell has been stretched until a hole forms, and the
diffuse medium in the hole is still opaque enough to imprint radiative
transfer effects on \lya\ photons. The surfaces of the thick shell and
the hole are of equal size in this first model.  
The parameters listed for FDF4691 in Table \ref{t_summary} are those of the 
homogeneous single shell model discussed above.

\begin{figure*}
\begin{tabular}{cc}
\includegraphics[height=6cm, width=8cm]{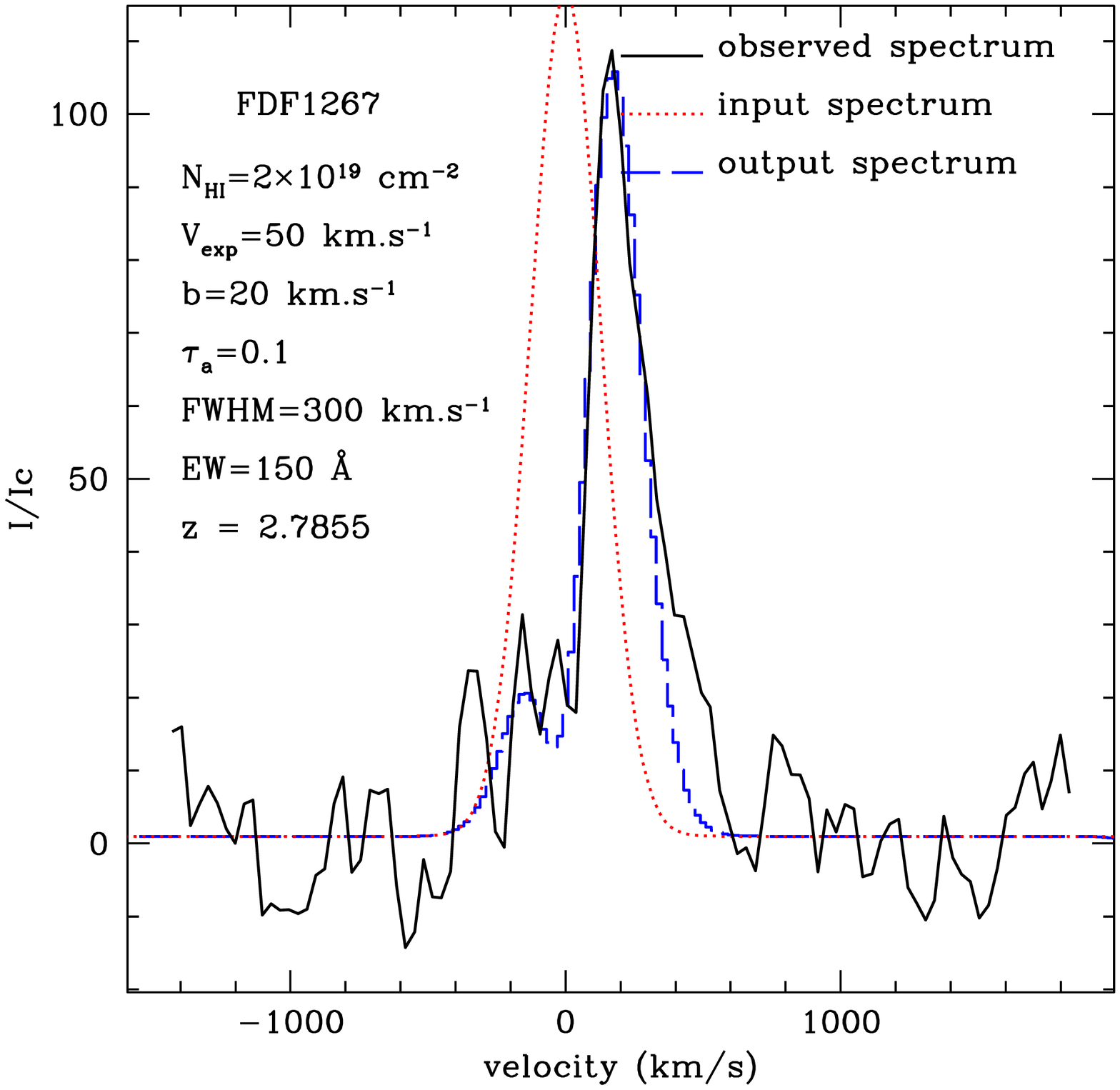}&
\includegraphics[height=6cm, width=8cm]{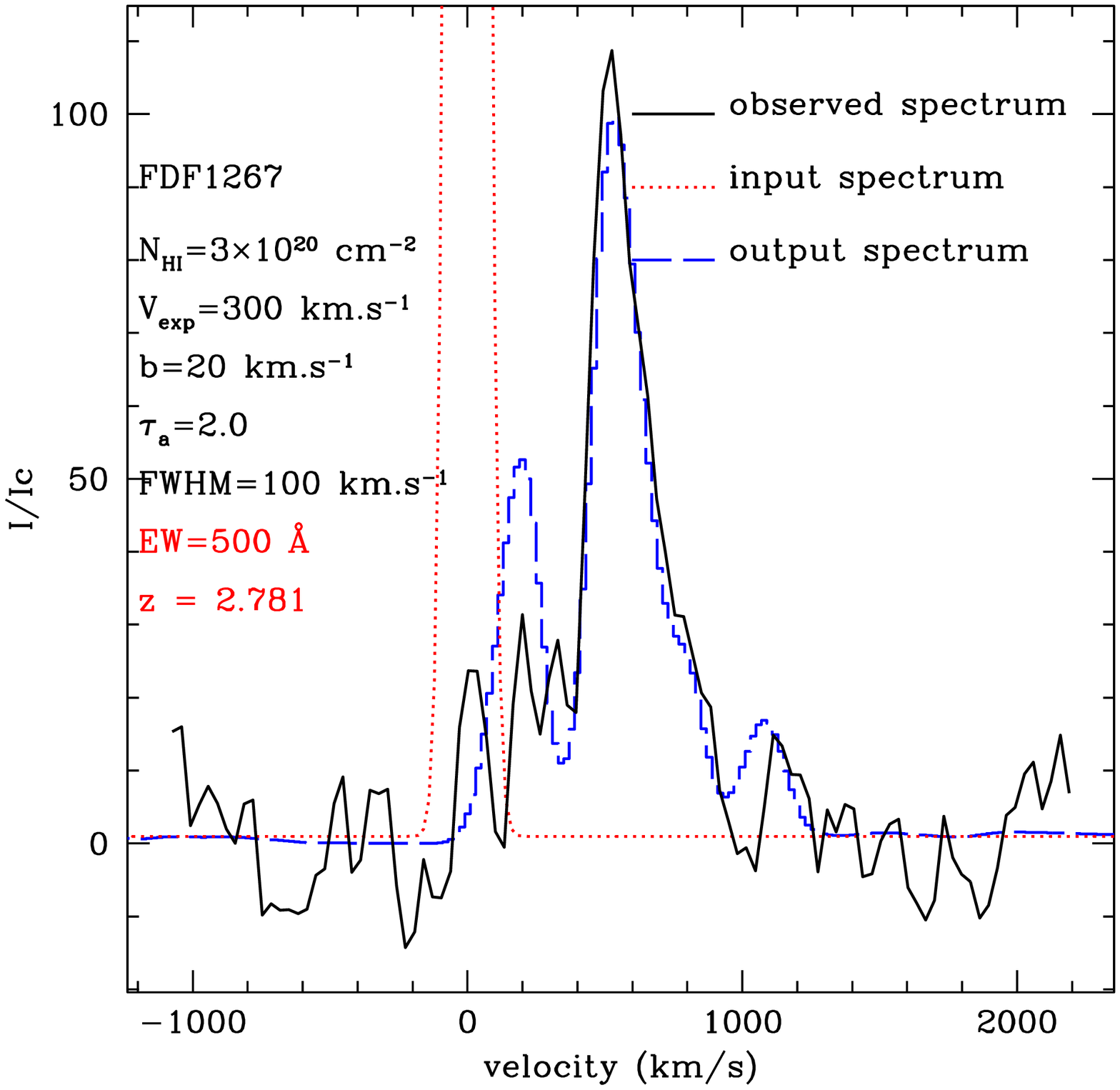}\\
\end{tabular}
\caption{Fitting FDF1267 with two different scenarii : the bump is
  either considered as a blue peak in an almost static shell (left), or as
  the first red peak of \lya\ emission (right).}
\label{FDF1267}
\end{figure*}

\begin{figure*}
\begin{tabular}{cc}
\includegraphics[height=6cm, width=8cm]{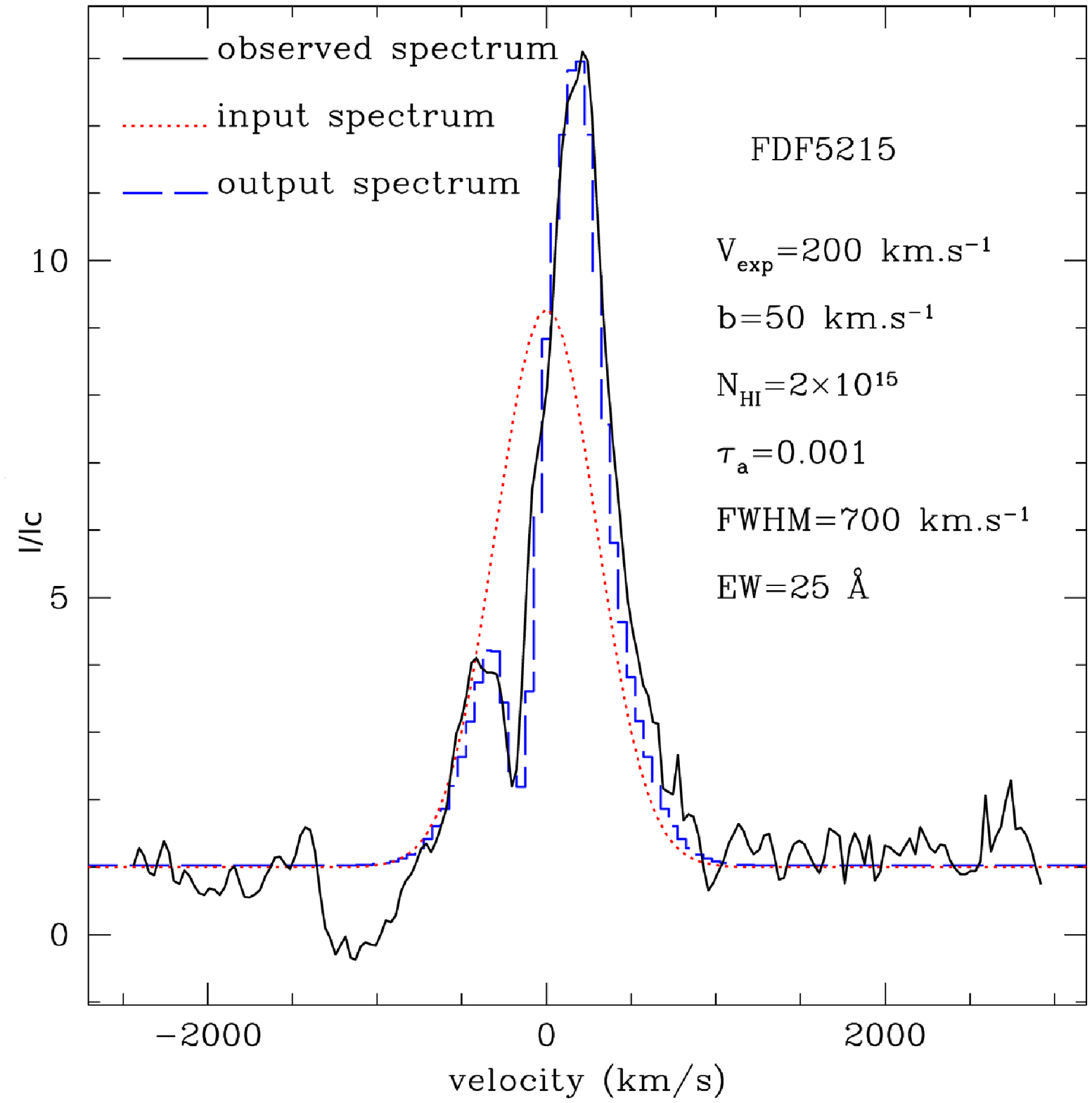} &
\includegraphics[height=6cm, width=8cm]{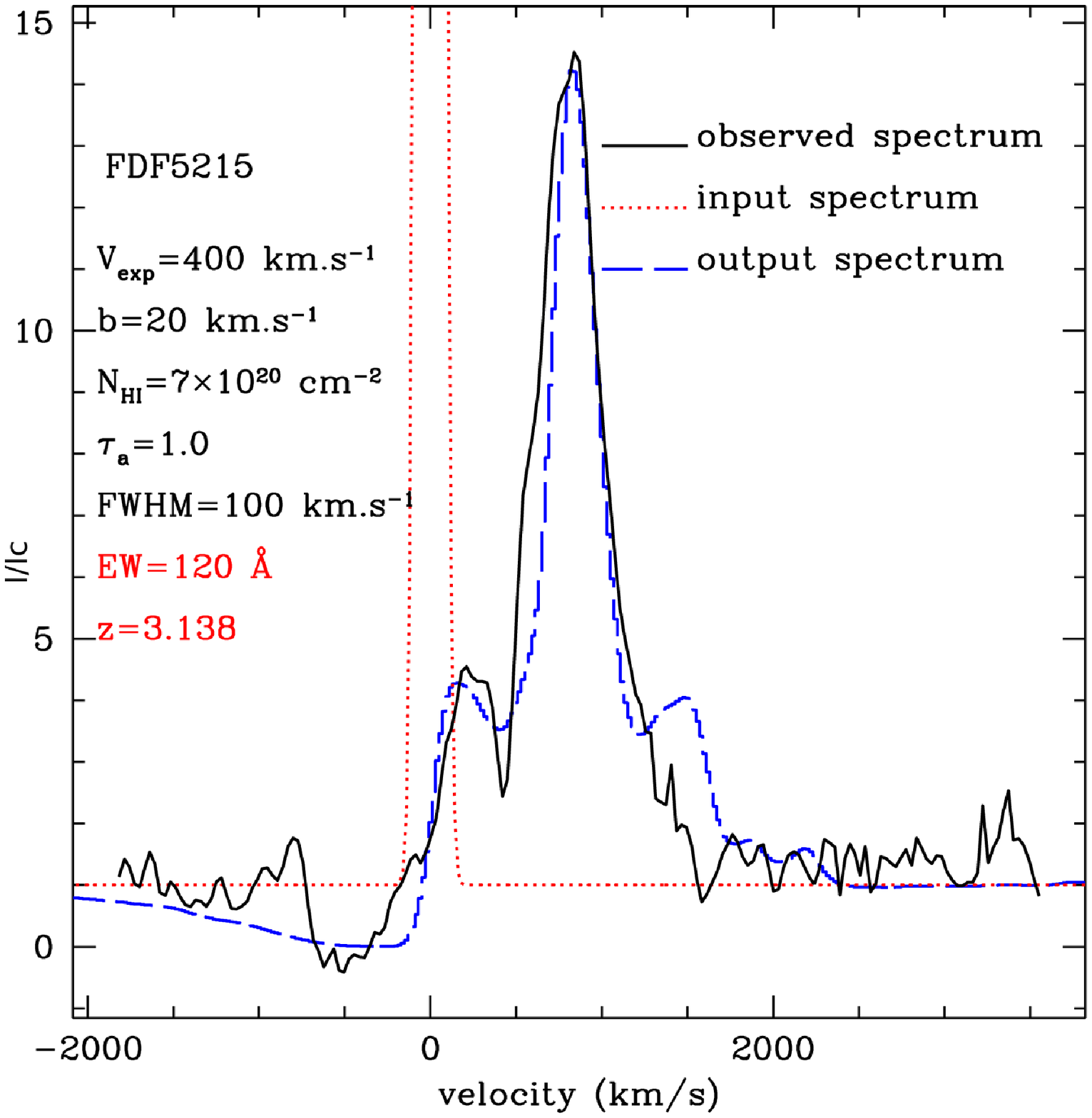} \\
\end{tabular}
\caption{Fitting FDF5215 with two different scenarii : the bump is
  either considered as a blue peak in an almost static shell (left), or as
  the first red peak of \lya\ emission (right).}
\label{FDF5215}
\end{figure*}

\subsubsection{FDF7539} 
For this object the velocity shift between the LIS in absorption and
\lya\ in emission was measured: $\dv=80$ \kms, which
implies a shell velocity $\vexp=80/3\sim27$ \kms, i.e. almost static
as in the case of FDF4691, or at maximum $\la 40$ \kms, in case of a low
column density. Indeed, the spectrum is also double-peaked as for FDF4691.

The large peaks separation ($V_p \pm 500$ \kms, larger than for
FDF4691) implies a high column density. Presumably, the rather low
observed EW(\lya)$_{\rm obs}$ also implies the presence of dust in the shell.
Indeed, the best fit shown in
Fig.\ \ref{FDF4691_7539} has a high column density ($\nh=5\times
10^{20}$ cm$^{-2}$), and dust ($\tau_a=0.3$). It is compatible with the
canonical value for the intrinsic \lya\ spectrum. The resulting escape
fraction is $f_e=0.28$. 

For comparison,
\citet{Tapk07} proposed a fit of similar quality for this object, but
the velocity of the shell they derive from their
modelling is high ($\vexp=190$ \kms), which is in contradiction with
the observed small velocity shift between \lya\ and LIS ($\dv=80$
\kms), and surprising for a double-peaked profile. As their
investigation is restricted to low column densities, the only solution they
have to produce separated and broad peaks is with high values of
$\vexp$ and $b$, and a huge intrinsic FWHM ($\sim 1900$ \kms).

\subsection{Other \lya\ profiles (group C)}

\subsubsection{FDF1267}
This object presents an asymmetric emission peak plus a
bump on the blue side of this peak with a strong observed
EW(\lya)$_{\rm obs}=129$ \AA. The \lya\ profile, shown in  Fig.\
\ref{FDF1267}, can be fitted by different scenarii.

On the left panel of Fig.\ \ref{FDF1267}, the bump is considered as a
blue peak in an almost 
static ($\vexp=50$ \kms) shell with a small column density
($\nh=2\times 10^{19}$ cm$^{-2}$) and a small amount of dust
($\tau_a=0.1$). Values of $b$ higher than $b=20$ \kms lead to too
separated peaks. The escape fraction is $f_e=0.6352$.
On the right panel, we fit the profile with a fast moving (\vexp=300
\kms), dense ($\nh=3\times10^{20}$ cm$^{-2}$) and dusty($\tau_a=2.0$)
shell, leading to $fe=0.02$, and a very large intrinsic EW(\lya)$_{\rm int}=500$ \AA.
Dust is needed in this configuration to reproduce narrow and well
separated peaks. Indeed, dust destroys more efficiently photons which
undergo the highest number of scatterings, i.e.\ all but the backscattered
photons, which leads ``isolates'' and ``slims down'' the peaks. 
Note that this solution requires an adjustment of the source redshift
to $z\sim2.781$ instead of $z\sim2.788\pm0.001$ derived by \citet{Tapk07}.
However, this object is the only one for which the redshift
determination is only based on \lya, so this poses so far no difficulty.

Our favoured solution is the ``quasi-static shell'' picture. 
Indeed, the high EW$_{\rm int}$ inferred in the second fit seems
unlikely. Furthermore, the large observed EW$_{\rm obs}$ of 1267 would
imply a rather low 
column density as discussed in Sect.~\ref{s_obsEW}. Finally, the  SFR values
derived from uncorrected UV and \lya\ fluxes ($SFR(\lya)>SFR(UV)$) indicate a
low dust content.   
An accurate redshift measurement of FDF1267, independent from \lya, should
allow to distinguish between these two solutions.

\subsubsection{FDF5215}
FDF5215, shown in Fig.\ \ref{FDF5215} presents the same spectral shape
as FDF1267: a small bump on the blue side of the asymmetric strong
emission, but the noise level is much lower, and this small bump has
to be taken into account. Again, two different scenarii can reproduce
the spectral shape. The bump is either considered as a blue peak in an
almost static shell, or as the first red peak of \lya\ emission.
The two solutions, differing by more than 5 orders of magnitude in
\nh, are listed in Table \ref{t_summary}. In passing we note that
FDF5215 was also modeled by \citep{Tapk07}; their set of parameters is
similar to our solution at low \nh, except for a higher value of $b$.

None of our fits are really
satisfying. The solution at low-\nh\ (left panel) reproduces well the
observed profile, but the derived column density is very low: at
least 4 orders of magnitude lower than the rest of the
sample. As a consequence, the FWHM of the intrinsic
\lya\ emission is huge (FWHM=700 \kms) to reproduce a broad profile
without efficient broadening due to radiation transfer in an almost transparent
medium. Finally, this solution does not reproduce the absorption at
$V=-1200$ \kms\ discussed below. On the other hand, the solution with high \nh\ and a
standard intrinsic FWHM fits less well.
Furthermore, the redshift derived from this fit is out of the
error bars ($z=3.138$ instead of $3.148\pm0.004$).  

The black absorption component, found at $V \sim - 1200$ \kms\
in this object (see Fig.\ \ref{FDF5215} left), is unaccounted for 
in the fit with low \nh. How likely it is that this represents a chance
alignment of an \hi\ absorber? 
If fitted separately with a Voigt profile the absorption is well described
by $b \sim 70$ \kms, and $\nh({\rm abs}) \sim 2 \times 10^{17}$ cm$^{-2}$.
Using the column density distribution from \citet{Misawa07}
we find that the probability to find an absorber with $>\nh({\rm abs})$ in a 
velocity interval of say 4000 \kms\ is $\sim$ 5 \%.
It seems thus more likely that this feature is related to the galaxy.
We conclude that none of our solutions is clearly favoured,
and further observations are needed to help constraining the models
for FDF5215.

\subsection{Uncertainties and degeneracies in \lya\ fits}

\subsubsection{Uncertainties on each parameter}
The parameters listed in Table~\ref{t_summary} correspond to best-fits
determined by eye, without resorting to minimisation techniques.
We now attempt to indicate the approximate uncertainties of the
derived parameters.

The characteristic uncertainty on the expansion velocity is estimated
to be $\sim 50$ \kms, which is the step in velocity in our grid of
models. The sampling in $\tau_a$, the dust absorption optical depth,
is not linear (we have assumed $\tau_a$ values of 0, 0.1, 0.5, 1., 2.,
3., and 4. for our grid), but it was refined when necessary. 
From the line fits the
characteristic uncertainty on $\tau_a$ and on the neutral column density $\nh$
are estimated as $\pm 50\%$.
We refer to Fig.~\ref{mini_grid} to give an idea of the
uncertainties on \nh\ and $\tau_a$.
Other comparisons, e.g.\ with SED fits and by imposing a consistency
between different SFR indicators, indicate that the uncertainty
on the extinction may be somewhat larger, up to a factor $\sim$ 2 in some
cases (see Sect.\ \ref{s_sfr}).

The Doppler parameter is set to the default value $b=20$ \kms, which
allows to reproduce the relative narrow peaks observed , and secondary features on the
extended red wing of the profiles. Models were computed for other
values of $b$ (10, 20, 40, 80, and 200 \kms), but did not lead to better
fits. We estimate the uncertainty on this parameter around 50\%.      
 
The uncertainty on the intrinsic equivalent width can be fairly
large, and it mostly depends on the determination of the continuum.
Indeed, for strong \lya\ emitters, like FDF5812, the
continuum is so weak that it is poorly constrained, and the uncertainty
on the continuum level is around 20\%.
As already mentioned, all the spectra
presented above have been normalized to the same level as determined by
\citet{Tapk07} to derive observed \lya\ EW$_{\rm obs}$.
However, choosing the continuum level by eye we may also obtain acceptable
solutions with lower intrinsic EW(\lya)$_{\rm int}\sim 80 \AA$ for the
three stronger \lya\ emitters (FDF2384, FDF4454 and FDF5812),
in better agreement with a scenario of a constant star formation over 
$\sim100$ Myr.
Only for FDF1267 do all solutions seem to imply a fairly high intrinsic
\lya\ equivalent width.

The intrinsic FWHM was set to FWHM=100 \kms as a default value, and good 
fits were obtained for almost all spectra;
exceptions are the very low \nh\ solution for FDF5215 and the very broad 
double-peaked profile of FDF4691 (except if we consider a shell with two components
as described before).

\subsubsection{Degeneracies}
\label{s_degen}
Although degeneracies affect in principle our profile fits, it
turns out that asymmetric line profiles provide fairly undegenerate
solutions. We briefly describe how the influence of the main
parameters ($b$, \nh, \vexp, FWHM, and $\tau_a$) can be discerned.

\begin{figure}
\includegraphics[height=6cm, width=8cm]{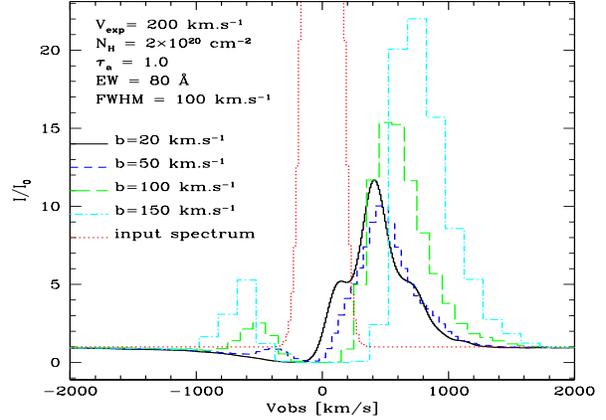} 
\caption{Dependence of the \lya\ line profile on $b$ for typical shell
  parameters. Note the shift of the red peak with increasing $b$, and
  the emergence of a blue counterpart. The escape fraction also
  increases with $b$.}
\label{vary_b}
\end{figure}

{\bf Doppler parameter $b$:} After several attempts to fit the data
with large values $b$ ($>50$ \kms), we adopted a typical
value $b=20$ \kms. The Doppler parameter has a complex influence on
the \lya\ profile: small $b$ lead to asymmetric emission lines on which
the potential multi-peaks due to a high expansion velocity would be
visible (see Fig.\ \ref{mini_grid}). Large values of $b$ ($>50$ \kms) lead
to a smoothed red peak (the multi-peaks are not visible any more),
whose location is redshifted.
Furthermore a blue emission component appears whose strength increases
with $b$,  
reproducing a kind of ``double-peaked profile'' like
in static media, but with asymmetric peaks --- the two sides of each
peak don't have the same slope (see Fig.~\ref{vary_b})--- even in
shells with  high velocities ($\vexp > 200$ \kms). 

In our sample, the spectra from group A (asymmetric profiles) do not
show blue 
components, and faint multi-peaks (better said bumps on the extended
wing) may be visible. Therefore small values of $b$ ($=20$ \kms) are
required to fit our spectra. On the contrary, other spectral types (B
and C) may be fitted with larger values of $b$.

{\bf Column density \nh:} With increasing neutral hydrogen column
density, the blue edge of the \lya\ emission is progressively
redshifted with respect to the systemic galaxy redshift.  Thus, if $z$
is known accurately enough, the neutral column density in the shell is
well constrained.  The full width of the line
increases with increasing \nh\ too. Indeed, it is impossible to fit
narrow lines ($FWHM < 500$ \kms) with high column densities ($\nh >
10^{20}$ cm$^{-2}$), and extended lines ($FWHM > 500$ \kms) with low
column densities ($\nh < 10^{20}$ cm$^{-2}$), if the expanding shell
model applies (cf.\ Sect.\ \ref{s_fwhm} and Fig.\ \ref{fwhm_nh}).


{\bf Expansion velocity:} 
The overall shape of the line profile --- presence of secondary
peaks or not --- constrains the velocity of the shell: fast moving
shells ($\vexp > 200$ \kms) lead to multi-peaks in the \lya\ profile,
whereas static or almost static shells lead to double-peaked profiles
with symmetrical peaks --- the two sides of each peak have the same slope. 
        
{\bf Dust content:} Finally, the dust content is adjusted to fit the
peak width and the relative height of the bumps -- if any -- compared
to the main peak.   

In conclusion, few degeneracies appear in the modeling of asymmetric \lya\
line profiles (group A).
Thanks to the location of the blue edge and to the full width
of the line \nh\ is well constrained for asymmetric spectra. 
Furthermore, the global line shape (one single peak)
implies low values of $b$ ($<50$ \kms) and \vexp\ ($<250$ \kms). 
On the contrary, the 4 spectra with more complex profiles (1267, 5215, 4691 and 7539)
present degeneracies. \citet{Tapk07} proposed \lya\ fits for 3 of them (4691,
5215, 7539). We can reproduce their fits, but propose fits with other
sets of parameters, as our code allows for higher column densities
than the code \citet{Tapk07} used. 
Automated fitting methods and a thorough examination of the
uncertainties and possible degeneracies in \lya\ line profile fits will be 
useful in the near future, also when larger samples of spectra of sufficient
S/N and resolution become available.

\subsubsection{Possible limitations of the model}
As described and motivated in Sect.\ \ref{s_model}, our modeling makes
some simplifying assumptions, including in particular geometry and the
homogeneity of the shell. How far these assumptions would alter our results
is presently unclear and remains to be explored in the future.

Inhomogeneous/clumpy geometries have e.g.\ been explored by \citet{Hans06};
the line profiles obtained from such models do not seem to change
significantly. However, how much our model parameters would be modified
remains to be examined.
For the time being it seems clear that few if any cases are know, where
clumpy geometries would favour \lya\ transmission with respect to the 
continuum \citep[cf.][]{Neuf91,Hans06}. This can e.g.\ be concluded from
the comparison of \ha\ and \lya\ in local starbursts \citep{Atek08},
and from the comparison of UV and \lya\ SFR indicators.
Indications for one possible case of such a \lya\ boosting have been
found among 4 objects analysed by \citet{Finkelstein07b}.
Other geometries have e.g.\ been considered in paper II,
where deviations from the constant velocity shell have been necessary 
for the analysis of cB58.

The effect of the intergalactic medium (IGM) has been neglected in our approach. 
Even if the IGM is almost fully ionised at $z \sim 3$, the redshift of the bulk
of our objects, the effect of the intervening \lya\ forest corresponds
statistically to a transmission of $\sim$ 70 and 40 \%  ($\tau_{\rm eff} \sim$ 0.3--1)
between $z \sim$ 3 and 4 \citep{Faucher07}.
In our modeling we find no need to account for such an IGM reduction within
$\sim$ 1000-2000 \kms\ of the \lya\ line; not even in the two highest redshift ($z \sim$ 4.7--5) objects.
No individual
\lya\ forest absorption components are found in this interval; furthermore
for most objects, except possibly FDF 1337, 5550, 7539, and maybe also 3389,
the predicted continuum flux blueward of \lya\ agrees within the uncertainties 
with the observed continuum. 
In any case, less importance has been given to the line fits on the blue side of \lya.
Also, in a detailed analysis of the \lya\ forest along the line of sight of
the $z \sim 2.7$ LBG cB58, \citet{Sava02} found no indication for neutral gas within
$\sim$ 4000 \kms\ of the systemic velocity of the galaxy.
From these considerations we conclude that our \lya\ line fits are probably
unaffected by additional matter beyond the expanding shell included in our models.


\section{Discussion and implications from our model fits}
\label{s_discuss}

We now discuss the values of the parameters determined for the 11 objects,
possible correlations among them, and we compare them with other
measurements from the literature. 

\subsection{The neutral column density}
\label{s_nh}

The neutral column density we derive from the fitting of 11 \lya\
spectra from the FORS Deep Field is radial --- along a line of sight,
from the center of the shell to the end of the simulation volume ---
and ranges over more than an order of magnitude, from $\nh=2.\times
10^{19}$ to $\nh=5.\times 10^{20}$ cm$^{-2}$. However, the majority of
objects have a small column density (8/11 have $\nh < 1.\times
10^{20}$ cm$^{-2}$).

\begin{figure}[tb]
\includegraphics[width=8.8cm]{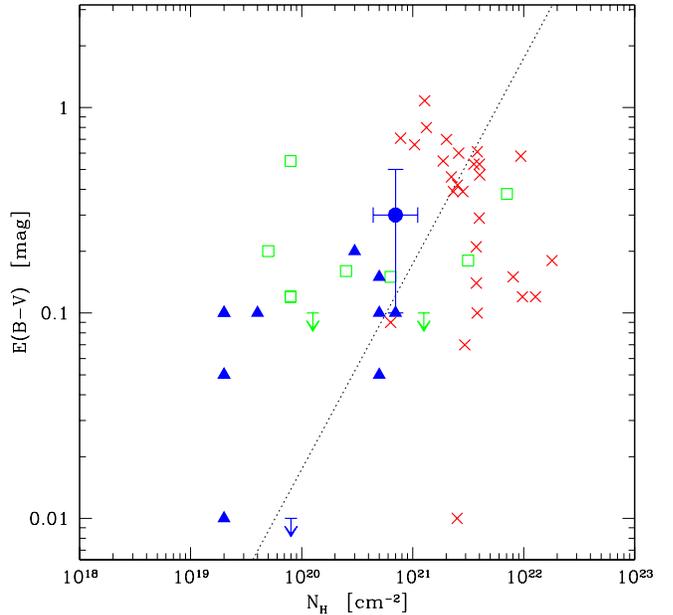}
\caption{Comparison of gas extinction, $E(B-V)$, and \hi\ column density
for the FDF objects (blue triangles), cB58 (blue error bar), local 
starbursts from Calzetti (2007, private communication, red crosses),
and measurements from the nearby starbursts analysed by \citet{Kunth98}
(green squares). The multiple solutions for the FDF objects are also
plotted except for low \nh\ solution of FDF5215; for FDF4691 an arbitrary 
upper limit of $E(B-V)<0.01$ is adopted in the plot. 
The mean Galactic relation $N_H/E(B-V)=5.8\times 10^{21}$
cm$^{-2}$ mag$^{-1}$ from \citet{Bohl78} is indicated by the dotted line.
}
\label{f_ebv_nh}
\end{figure}

How do our \nh\ determinations from \lya\ fitting compare with other \nh\
determinations in starbursts?
Carrying out such a comparison is difficult
for LBGs, since \hi\ column densities are usually not measured.
Even for nearby starbursts the available data is scarce.
Calzetti (2007, private communication) has kindly determined \nh\ for us
from published 21cm RC3 radio observations
and assuming sizes given by $D_{25}$
\footnote{The radii obtained in this way reach from $\sim$ 5 to 50 kpc,
with a median around 17 kpc.}.
For comparison we have also compiled \nh\ and $E(B-V)$ measurements 
from the small sample of nearby starbursts observed in the \lya\ 
region by \citet{Kunth98}. These comparison samples are plotted
in Fig.\ \ref{f_ebv_nh}.
For SBS 0335-052 we have added a second point, adopting \nh\
from \citet{Thuan97} and the extinction from \citet{Atek08}.
Similarly two points are shown for IRAS08339+6517 using the extinction
compiled by \citet{Kunth98} and the one measured by Atek et al.

Despite some overlap, \nh\ is  lower in our objects than  the
column density observed by Calzetti and collaborators, and more similar
to the small sample of Kunth et al.
An attempt to explain
this can be the different ways of determining \nh: for the 
Calzetti sample the determination of the neutral column
density was achieved by radio observations of the whole galaxy, whereas
in the case of Kunth et al., \nh\ is derived from Voigt fitting of the
\lya\ profile, so it only takes into account the neutral gas which
influences \lya\ radiation transfer, even if a Voigt fitting may lead
to an underestimate of \nh\ \citep[ see Sect.\ 4]{Verh06}. This may
explain why the determination from Kunth et al. is closer to our
values than those of Calzetti et al.
In any case, to compare our column densities with those of Calzetti
one needs to increase our \nh\ values typically by a factor $\sim 2$ to 
convert the radial shell column density to a total one.

The range of \nh\ found for the FDF objects is also compatible with
our confirmation of the neutral column density of the gravitationally
lensed $z\sim3$ LBG MS1215-cB58 (cB58, shown as the blue cross) that
we fitted previously (paper II). Indeed, $\nh(cB58)\sim 7.\times 10^{20}$ cm$^{-2}$
is slightly higher than \nh\ of the FDF \lya\ emitters, as expected for
a \lya\ spectrum in absorption.

\subsection{\lya\ equivalent widths}

\subsubsection{Observed EWs}
\label{s_obsEW}
\begin{figure}[tb]
\includegraphics[width=8.8cm]{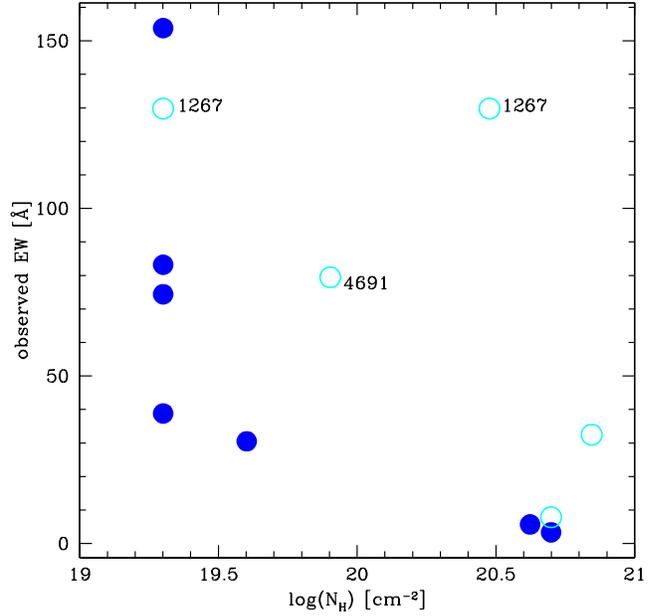}
\caption{Observed \lya\ EW$_{\rm obs}$ versus derived radial \hi\
  column density. 
The three objects with the lowest EW$_{\rm obs}$ are artificially displaced by
a small amount for illustration purposes.
  We may see an anticorrelation between the observed \lya\ EW$_{\rm obs}$
  and the neutral column density in the shell, for the 11 objects from
  the FDF, if we eliminate the high-\nh\ solutions for 1267, and the
  very low-\nh\ solution for 5215. The filled circles are objects presenting
  asymmetric profiles (type A), the open circles are the others (type B, C). 
	The quasi-static object 4691 with a double-peaked profile
	stands out from this trend for unknown reasons.}
\label{f_ew_nh}
\end{figure}

The observed \lya\ EWs 
range from $6$ to $150$ \AA\ in the rest frame. 
Does this range reflect intrinsic differences, or is it somehow
related to the physical conditions of the ISM in which \lya\ radiation
transfer takes place ? 
We examined how EW$_{\rm obs}$ correlates with other
parameters, but no clear correlation is seen. 
We found a trend in EW$_{\rm obs}$ with respect to the
neutral column density in the shell (see Fig.\ \ref{f_ew_nh}): 
EW$_{\rm obs}$ seems to decrease with \nh, at least for the asymmetric
profiles (filled circles). In fact, the objects with a low EW$_{\rm
  obs}$ ($< 10$ \AA, 
i.e.\ FDF1337, 5550, and 7539) can only be fitted with a high value of \nh, since
their profiles are very broad. On the other hand, narrow lines with
large EW$_{\rm obs}$ 
can only fitted with small values of \nh. 
There may be three exceptions to this trend, FDF4691, 5215, and 1267. 
The double-peaked profile of 4691 seems peculiar, as it is static and
dust-free, as suggested by the SFR(UV) and SFR(\lya) which are almost
identical. We imagine that this trend breaks down in dust-free media,
or in media with a very small amount of dust, because dust is needed
to absorb radiation and decrease the intrinsic EW$_{\rm int}$ value.
%
FDF1267 and 5215 are peculiar/degenerate for the reasons discussed above 
(Sect.\ \ref{s_tapken}).

An anti-correlation of EW$_{\rm obs}$ vs \nh\ can be understood by
radiation transfer 
effects if the intrinsic EW$_{\rm int}$ is approximately
constant. When \nh\ increases,  
the path length of \lya\ photons increases, and so does their chance
to be absorbed by dust: the \lya\ escape fraction decreases with
increasing \nh, as mentioned above. Furthermore, if we assume a
constant dust-to-gas ratio, an increase in \nh\ naturally leads
to an increase of the dust quantity (the optical depth). These two
effects explain the decrease of the observed \lya\ EW with increasing
\nh\ from a theoretical point of view.

\subsubsection{Intrinsic EWs}
Three objects (2384, 5812, 1267) have clearly very large intrinsic EW(\lya)$_{\rm int} >100$ \AA,
for one (5215) the two solutions give quite disparate results,
and the remaining 7 objects have all intrinsic EW$_{\rm int}$ of
$\sim$ 50--100 \AA. 
Taking the uncertainties in the continuum placement into account
(cf.\ above) we consider that this latter group (7 of 11 objects) 
have intrinsic EW(\lya)$_{\rm int}$ compatible with expectations for star-forming galaxies with a
constant star formation history over periods $\ga$ 10-100 Myr, as seen
in Fig.~\ref{f_s04_sfr}. 
The strength of \lya\ in three high EW$_{\rm obs}$ objects requires
younger ages, 
irrespectively of their star formation history.

\subsection{Dust extinction}
For the first time we have derived here constraints on the dust 
content of galaxies using the \lya\ profile only. It is therefore
of interest to examine how this determination compares with
other methods.

\begin{figure}[tb]
\includegraphics[width=8.8cm]{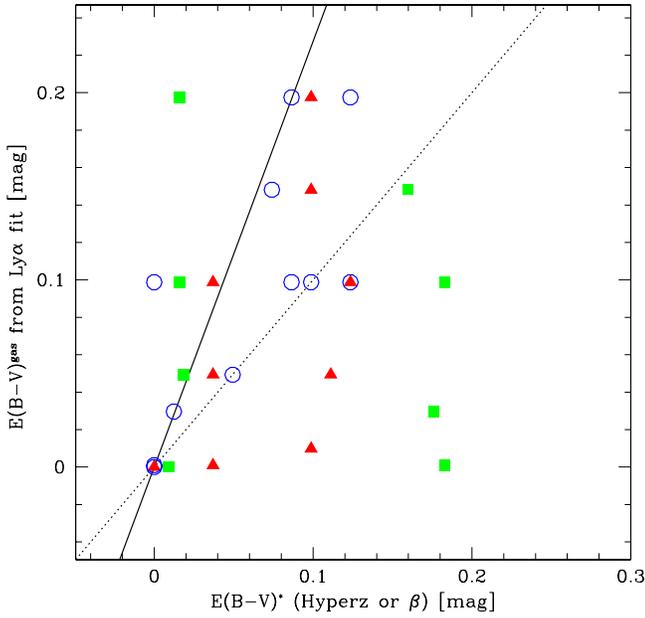}
\caption{Comparison of the extinction $E(B-V)$ determined from the 
\lya\ profile fits versus other methods for objects with sufficient
photometry and/or measured $\beta$ slopes. 
Multiple solutions from \lya\ fits are included.
\hyperz\ SED fits using constant SFR models (red filled triangles)
or arbitrary SF histories (blue open circles). 
Green squares show values of $E(B-V)^{\rm gas}$ determined from the 
$\beta$-slope. 
The dotted line shows the one-to-one relation, the solid line the 
relation $E(B-V)^\star = 0.44 E(B-V)^{\rm gas}$ found empirically 
\protect\citep[cf.][]{Calz00}.}
\label{f_av_tau}
\end{figure}

Using a version of the SED fit and photometric redshift code \hyperz\
described in \citet{SP05} and the UBgRIJKs photometry published by
\citet{Heidt03}, we have modeled the SED of our objects, assuming the 
Calzetti attenuation law \citep{Calz00}. 
Three objects,
5812, 3389, and 6557, have insufficient data (photometry in 3 or less
bands) which does not allow meaningful SED fits.
2384 is also excluded, since it appears to be a multiple source,
where the \lya\ emission is clearly displaced from the continuum.
Results for the remaining objects are shown in Fig.\ \ref{f_av_tau},
where we compare the extinction derived assuming models with
constant star formation (red filled triangles) or exponentially
decreasing SF histories (blue open circles) described by the Bruzual
\& Charlot templates with $E(B-V)$ derived from our \lya\ line fits.
Note that these values, denoted here as $E(B-V)^\star({\rm Hyperz})$, 
measure the  extinction suffered by the stars, whereas $E(B-V)(\lya)$ 
measures that of the \lya\ emitting gas. The two may differ,
as e.g.\ known to hold empirically for local starbursts between
the stellar extinction and the one measured from the Balmer
decrement \citep{Calz00}.

Figure \ref{f_av_tau} shows a good correlation between the different
extinction measures, especially when different SF histories are
allowed for. Indeed, for the bulk of the objects the extinction
derived from \lya\ profile fitting is between the gas extinction
expected from Calzetti's empirical relation, $E(B-V)^{gas} = 1/0.44
E(B-V)^\star$, and a somewhat lower value of $E(B-V)^{gas}$.
%

We can also estimate the extinction from the UV slope.
Excluding again the multiple source 2384, the observed UV slopes 
$\beta$ show basically two groups, whose extinction, estimated
following \citet{Calz00}, is $E(B-V)^\star \sim$ 0.--0.02 and 
0.16--0.18 shown by the green squares in  Fig.\ \ref{f_av_tau}.
These values cover a similar range as $E(B-V)^\star({\rm Hyperz})$,
although with a poor correlation. This could be due to the
statistical nature of the underlying correlation between
attenuation and $\beta$.
We conclude that overall \lya\ line profile fits allow us to obtain
quite consistent extinction values compared to broad band photometry
fits of the individual objects. Our derived extinction values,
corresponding to $E(B-V) \sim$ 0--0.2 are also in good agreement with
the values found for LBGs by \citet{Shap03} and others 
\citep[e.g.][]{Papovich01}.

\subsection{Gas to dust ratio}
How do the gas-to-dust ratios obtained from our line profile fits
compare with other values observed in starbursts? To address this
we turn again to our comparison samples shown in Fig.\ \ref{f_ebv_nh}.

Compared to the Galactic average of $\nh/E(B-V)=5.8 \times 10^{21}$
cm$^{-2}$ mag$^{-1}$ and its $\sim 30\%$ scatter \citep[cf.][]{Bohl78},  
our results show somewhat lower gas-to-dust ratios and apparently a 
larger scatter. The scatter is, however, similar to the one found
among the local starbursts also shown in this Fig.
\citet{Pett00} have already noted this difference in gas-to-dust ratio
for cB58, which they suggest could indicate that a significant fraction
of the gas is not in atomic form, i.e.\ is either ionised and/or in
molecular hydrogen.

The Calzetti objects have a median of $\nh/E(B-V)=7.9 \times 10^{21}$
similar to the ``classical'' Galactic value\, but showing a wide
dispersion.   
Approximately half of our objects show $\nh/E(B-V)$
values lower than those of Calzetti's local starbursts.
For the other half of the sample, and for the LBG cB58 modeled in
paper II, the values of the  gas-to-dust ratio overlap
with those for the Calzetti sample.
In any case the comparison may be hampered for the reasons 
affecting also the \nh\ comparison (cf.\ Sect.\ \ref{s_nh}).
Overall we conclude that the gas-to-dust ratios determined purely from
\lya\ line profile fitting yield values reasonably consistent with
local starbursts or somewhat lower.
For comparison \citet{Vladilo07} find higher gas-to-dust ratios
in DLAs compared to the Galactic value.

If some LBGs show truly lower gas-to-dust ratios, this may be due to a
higher degree of metal enrichment in the outflowing \hi\ gas, and/or an
overall smaller fraction of neutral gas being ``polluted'' in nearby
starbursts, and/or a smaller fraction of atomic hydrogen. 
However, the column densities found in LBGs are also
consistently lower than in local starbursts (cf.\ Fig.\ \ref{f_ebv_nh}),
which may indicate a different ``regime''.
Other, independent determinations of the gas-to-dust ratio in LBGs and
other starbursts and more detailed examinations would be required to
clarify these issues and to understand the possible physical causes for
these apparent differences.

\subsection{Velocity of the outflow}

Assuming a galaxy-scale outflow surrounding our 11 objects (the
relevance of this model is discussed in Sect.\ \ref{s_model}), we have derived
the expansion velocity \vexp\ of the expanding shell from
the \lya\ spectral shape for 8 objects, and deduced it for the 3 other
objects (1337, 5550, 7539) from the measurement of a shift between the
LIS absorption lines and the \lya\ emission.
Overall 9 objects out of 11 have shell velocities around 150-200
\kms, and 2 objects present almost static shells.
No correlation between the expansion velocity and other parameters
is found. We now briefly discuss the high velocity outflows and
the few nearly static cases.
Beforehand we can already mention that we
have no explanation for the causes leading to low ISM velocities
or allowing to distinguishing these objects from the more common
cases showing outflows.

\subsubsection{High velocity objects}
All objects presenting an asymmetric emission line are
reproduced with $\vexp\sim150-200$ \kms.
This velocity range is very similar
to the determination of \citet{Shap03}, from the blueshift of LIS
absorption lines with respect to stellar lines in a sample of
$\sim800$ LBGs at redshift $z\sim 3$, as well as in cB58, where the
outflow velocity is estimated to $\vexp=255$ \kms\ \citep{Pett02}. 

Two objects with a more peculiar spectral shape, FDF 1267 and 5215,
can also be fitted with higher expansion velocities ($\vexp\sim400$
\kms), a high \nh, and high dust content.
However, 1267 has such a large observed equivalent width
($EW_{\rm obs}=129$ \AA) that it may rather have a small column density to
fit better in the plot showing a correlation between the observed
\lya\ EW$_{\rm obs}$ and \nh\ (see Fig.\ \ref{f_ew_nh}). 
In passing we note that this peculiar spectral shape seems also to be
found in observations of an LBG at redshift $z\sim3.7$ by
\citet{Vanz08} in the GOODS-South field.

\subsubsection{Low velocity objects}
The two double-peaked spectra (4691 and 7539) are characterised by a
static (or almost static, $\vexp < 25$ \kms) surrounding shell. In the
case of 7539, the velocity shift between the LIS absorption lines and
\lya\ is even measured ($\dv=80$ \kms), so the shell
velocity is here an observational constraint ($\vexp \sim
\dv/3$). This peculiar spectral shape (double peaks) is
predicted by theory, arising from \lya\ resonant scattering through
static \hi\ media \citep{Neuf90}, but
surprisingly, observed double-peaked profiles are in general not
interpreted as a signature of \lya\ radiation transfer through static
media \citep{Fosb03,Chri04,Wilm05,Vene05,Vanz08}, even if they appear much
less common than the asymmetric emission observed in all high-z LAEs.
In the static case the separation of the peaks $\Delta \lambda$ is not
only related to the thermal velocity of the \hi\ gas, but depends also
on the \hi\ column towards the \lya\ source, \nh. For a homogeneous
slab one has:
\begin{eqnarray}
	\Delta \lambda^{\rm rest} & = & 2 \lambda_0 (-x_p b/c) \nonumber \\
	& = & 2.49\times10^{-7}
	\left(\frac{b}{12.85 \,{\rm km \, s^{-1}}}\right)^{1/3} 
	\left(\frac{\nh}{\rm cm^{-2}}\right)^{1/3} {\rm \AA}
\label{eq_shift}
\end{eqnarray}
where $c$ is the light speed, $\lambda_0$ is the restframe wavelength
of \lya, $b=\sqrt{V_{th}^2 + V_{turb}^2}$ is the Doppler parameter,
and $x_p=0.88(a\tau_0)^{1/3}$ is the location of the peaks in
units of the Doppler width \citep{Neuf90,Dijk06}.  For example, for
$b=12.85$ \kms and $\nh=6.4\times 10^{19}$ cm$^{-2}$ one obtains
$\Delta \lambda^{\rm rest} \sim 1$ \AA.
If applied to the $z \sim 3.65$ double peaked \lya\ object of
\citet{Vanz08}, the observed velocity shift of 13 \AA\ would indicate
$\nh \sim 1.4 \times 10^{21}$ cm$^{-2}$ for $b=12.85$ \kms.
Establishing accurate enough galaxy redshifts for these objects
is important to be able to assert if one is truly dealing
with a nearly static case (in which case zero velocity is between the 
two \lya\ peaks), outflows (with both \lya\ peaks redshifted),
or other situations.

\subsection{Observed \lya\ FWHM}
\label{s_fwhm}

\begin{figure}[tb]
\includegraphics[width=8.8cm]{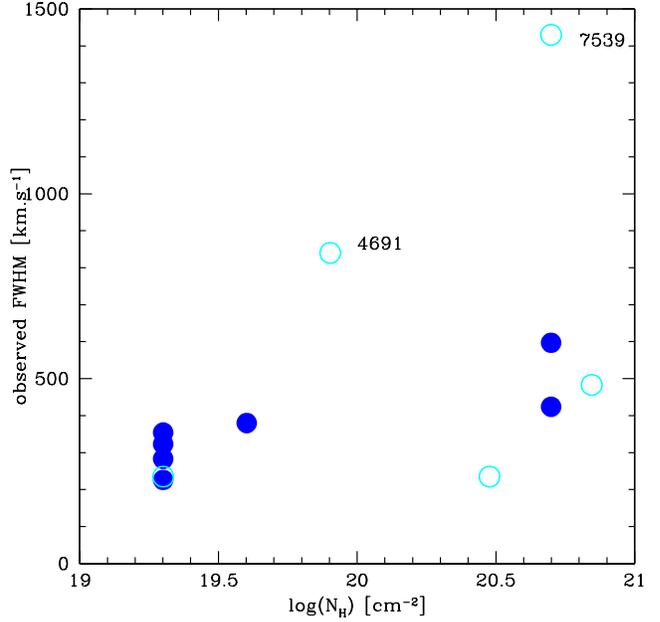}
\caption{Observed \lya\ FWHM versus neutral column density.
    showing a tentative correlation between the observed \lya\ FWHM
  and the \hi\ column density in the shell, for the FDF objects.
  The filled circles are objects with
  asymmetric \lya\ profiles, the open circles are the others. The objects 
  with double-peaked profiles, FDF 4691 and FDF 7539, are clearly distinct,
  showing the highest FWHM.}
\label{fwhm_nh}
\end{figure}

A possible correlation may be found between the observed FWHM and
the neutral column density of the expanding shell, as shown in Fig.~\ref{fwhm_nh}.
If real, it may be used to estimate the \hi\ column density
in starbursts from a simple measurement of FWHM(\lya).
Such a correlation can easily be explained by radiation transfer
effects: the \lya\ optical depth increases with \nh\, so that \lya\
photons have to diffuse further in the wings to be able to escape the
medium, which naturally broadens the \lya\ red wing.

Another point worth noticing about FWHM is that we were able to fit
all observed \lya\ profiles with an intrinsic FWHM of $\sim 100$ \kms, except
for 4691, for which we have to start with an already very extended
intrinsic \lya\ line to reproduce the very broad double-peaked
profile. Otherwise, more exotic scenarii, like a shell with a hole, have
to be invoked. 
 
Finally, the two correlations presented above (FWHM$_{\rm obs}$ vs \nh\
and EW$_{\rm obs}$ vs \nh, Figs.\ \ref{f_ew_nh} and \ref{fwhm_nh}) explain the
observed anti-correlation shown by T07 between FWHM$_{\rm obs}$
and EW$_{\rm obs}$.  Indeed, they are both related to the neutral column
density \nh\ which surrounds the starburst. The EW$_{\rm obs}$ increases with
decreasing \nh\, and the FWHM$_{\rm obs}$ increases with increasing
\nh. Therefore, we should not observe objects with a large EW$_{\rm
  obs}$ and very broad lines, what is also confirmed by the observed
FWHM$_{\rm obs}$ of LAEs which are 
always below $500$ \kms\ \citep{Rhoa03,Daws04,Vene04}.  
If the tentative correlation between FWHM$_{\rm obs}$ and \nh\ really holds 
it would imply a maximum column density of $\nh \la (2-4) \times 10^{20}$ 
cm$^{-2}$ in LAEs.

\subsection{Escape fraction}
\label{s_escape}

\begin{figure}[tb]
\includegraphics[width=8.8cm]{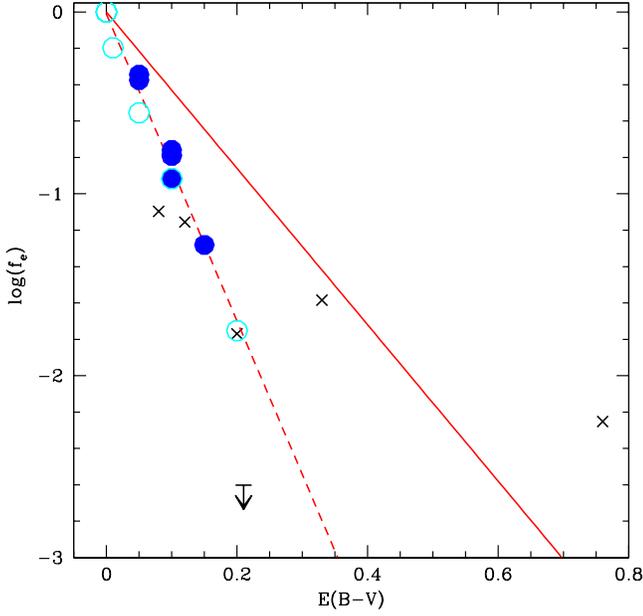}
\caption{\lya\ escape fraction versus dust extinction in the
  gas for the LBG (circles) and local starbursts (crosses).
 We find a clear correlation between the \lya\ escape fraction
  and the dust amount in the shell, for the 11 objects from the
  FDF. The filled circles stand for objects with asymmetric profiles, 
  open circles the remaining ones. The solid line represents the
  continuum attenuation, $fe_{\rm cont} = \exp(-\tau_a) \approx 
\exp(-10\times E(B-V))$, the dashed line the fit proposed in Eq.\
\ref{eq_fe}.
The crosses and the upper limit are the integrated escape fractions
from a sample of 6 local starbursts from \citet{Atek08} plotted as a
function of $E(B-V)$ measured from the Balmer decrement.}
\label{fe_ebv}
\end{figure}

The \lya\ line escape fractions derived for our FDF objects range from
$100\%$ for the dust-free object FDF4691 to $\sim 2\%$ for objects with
the highest extinction. From our modeling we find that the main parameter
determining the escape fraction is the dust amount in the shell.
Furthermore, from all our modeling results
already discussed above, it is quite clear that no single 
value of the \lya\ escape fraction is expected from LBGs and LAEs, in
contrast to simplifying assumptions made in some models
\citep[e.g.][]{Ledelliou06}. 

Our model grids predict the following behaviour for the escape fraction: 
\begin{itemize}
\item $f_e$ increases with increasing \vexp, because the
\lya\ optical depth decreases in a fast moving medium compared to
the static case. Therefore the mean path of \lya\ photons in the medium
decreases, so their chance to be absorbed by dust too.  
\item $f_e$ decreases with increasing \nh, because the \lya\ optical
  depth is proportional to \nh.
\item $f_e$ decreases with increasing $E(B-V)$, obviously since with a larger
number of absorbers in the medium, \lya\ photons increase their chance to
  interact with them.
\end{itemize}
However, the only clear correlation we found in the data
between $f_e$ and other parameters is with $E(B-V)$ 
(see Fig.\ \ref{fe_ebv}).  
No correlation is found in particular with \vexp\ (our objects cover probably
a too small range in velocity range), and with \nh\ 
(the variation in the dust-to-gas ratio is probably more dominant).
On Fig.\ \ref{fe_ebv}, we see also that, as expected, \lya\ photons are more
attenuated by dust than the continuum; indeed all LBG data points
(circles) are 
located below the solid line corresponding to $f_e({\rm
  cont})=\exp(-\tau_a)$.    
The reason is that multiple resonant scattering of \lya\ photons increases 
their path through the medium, and hence their chance to be absorbed by dust, 
compare to the continuum. 
Note that other dependences of $f_e$ on $E(B-V)$ could be expected
with other geometries. For example in clumpy media, the reflection
of \lya\ on the clump surfaces could ease the \lya\ transmission
for the same amount of dust \citep{Neuf91,Hans06}.

We propose a fit to predict the escape
fraction of \lya\ photons knowing the dust extinction (dashed curve on
Fig.\ \ref{fe_ebv}):
\begin{equation}
f_e = 10^{-7.71 \times E(B-V)}.
\label{eq_fe}
\end{equation}
Note that E(B-V) in the precedent formulae is the extinction in the
gas, which may be different from the extinction of the stars
\citep{Calz00}, as already mentioned above.  
Interestingly, the two static objects are also fitted by this formulae, which
illustrates that dust is really the dominant parameter which governs the
\lya\ escape in our objects. One of these (4691) is dust-free, so its escape
fraction is $\sim 1$, but in the other object (7539), 30\% of the
\lya\ photons 
escape the medium. For the same extinction, moving media present an
escape fraction of 40-45\%, which is coherent with the theoretical
prediction that $f_e$ increases with $\vexp$.

Empirical \lya\ escape fractions have recently been measured by
\citet{Atek08} from imaging for a sample of 6 local starbursts.  Their
values are compared to our data for LBGs in Fig.\ \ref{fe_ebv}.  For
$E(B-V) \le 0.2$, our results are in good agreement with three local
objects. SBS 0335-052 with an integrated extinction of $E(B-V)^{\rm
gas} \approx 0.21$ shows no \lya\ emission, it is a net absorber.
For larger $E(B-V)$ values, the
\lya\ escape fraction of two local starbursts (Haro 11 and NGC 6090) are 
higher than $f_e$ predicted by our fit to the LBGs studied here.
Deviations from a simple homogeneous shell geometry are the most
likely explanation for this difference. This will be testable through 
detailed modeling both of the spatially resolved and integrated properties 
of the local objects.

\begin{figure}[tb]
\includegraphics[width=8.8cm]{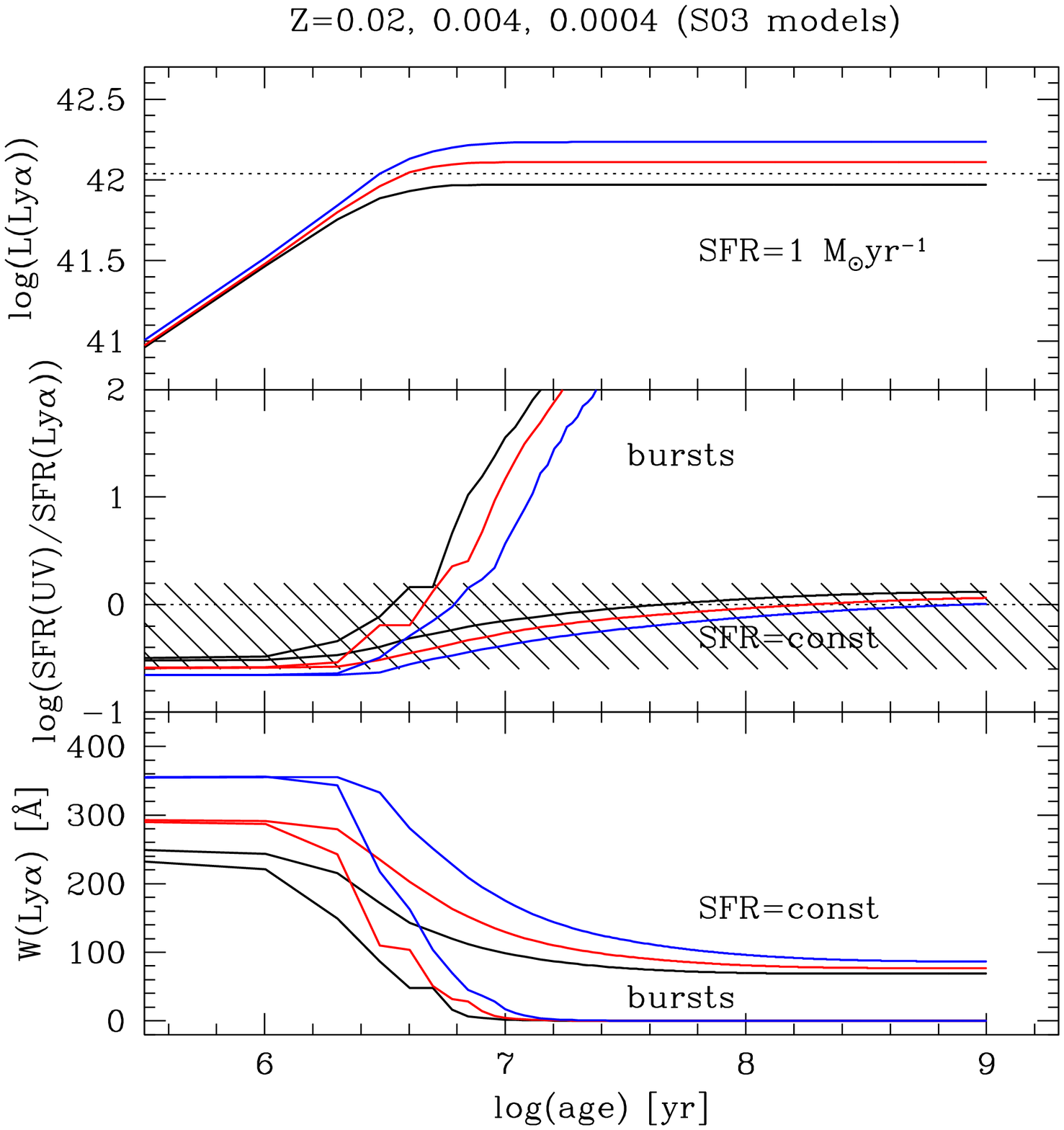}
\caption{Temporal evolution of \lya\ and UV SFR predictions from the 
synthesis models of  S03 for three metallicities 
($Z=0.02=\zsun$ in black, 0.004 in red, and 0.0004 in blue) computed 
for instantaneous bursts and/or constant SF.
{\bf Top:} \lya\ line luminosity in erg s$^{-1}$ emitted per unit 
SF rate, assuming a Salpeter IMF from 0.1 to 100 \msun.
The dotted line shows the ``canonical'' value based on
\citet{Kenn98} and a standard \lya/\ha\ ratio.
{\bf Middle:} logarithm of the UV to \lya\ SFR ratio.
The shaded area shows allowed range allowed for 
constant SF models with metallicities between 1/50 \zsun\ and solar.
{\bf Bottom:} \lya\ equivalent width.
}
\label{f_s04_sfr}
\end{figure}

Since the \lya\ line flux is more strongly reduced (due to multiple scattering
effects) than the adjacent continuum, the \lya\ equivalent width
depends on the extinction. This phenomenon is added to the one
already known to result from the extinction difference between the gas 
and the stellar continuum. In principle, a measurement of EW(\lya)$_{\rm obs}$
could thus be used to determine the extinction, provided the
intrinsic equivalent width  EW$_{\rm int}$ is known. 
Concretely, the fit-relation between $f_e$ and $E(B-V)^{\rm gas}$
proposed above (Eq.\ \ref{eq_fe}) translates to the following behaviour of
the \lya\ equivalent width with extinction:
\begin{eqnarray}
\log\left(\frac{{\rm EW}^{\rm rest}_{\rm obs}}{{\rm EW}_{\rm int}}\right) & = & 
	- E(B-V)^{\rm gas} \left(7.71 - 0.4 k_\lambda \, r \right) 
	\nonumber \\
 & \approx & -5.6 \,\, E(B-V)^{\rm gas}
\label{eq_ext}
\end{eqnarray}
where $k_{\lambda=1216\AA} \approx 12$ and $r=$E(B-V)$^\star$/E(B-V)$^{\rm gas}=0.44$ 
according to \citet{Calz01}.
Adopting reasonable values for EW$_{\rm int}$ (e.g.\ from Fig.\ \ref{f_s04_sfr}),
this formula may be used to obtain a crude estimate of the extinction
in LAEs based on a pure equivalent width measurement.
An extinction corrected SFR(\lya) value can then be obtained from the 
\lya\ luminosity using an appropriate SFR calibration from Fig.\ \ref{f_s04_sfr}, 
consistently with the assumed value of EW$_{\rm int}$.

\begin{figure}[tb]
\includegraphics[width=8.8cm]{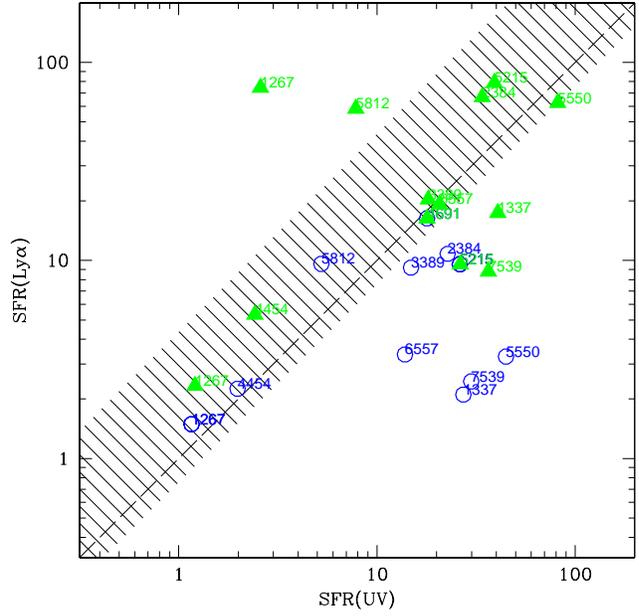}
\caption{Comparison of the SFR values determined from the UV continuum
and from \lya\ for our objects. Blue circles indicated the 
``observed'', uncorrected SFR values from Table \ref{t_overview}.
Green symbols show the ``true'' values corrected for dust and transfer 
effects according to Eqs.\ \ref{eq_lyasfr} and \ref{eq_uvsfr}.
The shaded area shows the range allowed by synthesis models 
for the combination of the ``true'' SFR(\lya) and SFR(UV) values
taking age effects of constant SF models into account and allowing
for metallicities between 1/50 \zsun\ and solar
(cf.\ Fig.\ \ref{f_s04_sfr}). 
The dashed line indicates the one-to-one relation.
}
\label{f_uv_lya}
\end{figure}

\subsection{SFR indicators}
\label{s_sfr}
Given our quantitative analysis of \lya\ radiation transfer, the determination
of the \lya\ escape fraction and of the extinction, we are now able to
examine to what extent \lya\ and the UV continuum provide consistent
measures of the star formation rate.
The main results of this exercise are shown in Fig.\ \ref{f_uv_lya}

First we note that three of our objects (1267, 4454, 5812) show 
observed, i.e.\ uncorrected SFR values corresponding to SFR$(\lya) >
$SFR$(UV)$. 
Such a result need not be inconsistent; this behaviour is indeed expected
for young bursts or objects where constant star formation has not yet
proceeded over long enough timescales, i.e.\ for timescales $\la$ 10--100 Myr
as shown in Fig.\ \ref{f_s04_sfr}. In this case values up to
SFR$(\lya) \sim 4 \times $SFR$(UV)$ can be obtained; this allowed
range of values for SFR$(\lya)/$SFR$(UV)$
for constant SF is shown by the shaded region on
Fig.\ \ref{f_uv_lya}. Indeed  these objects
also show among the largest EW(\lya)$_{\rm obs}$, as expected from Fig.\
\ref{f_s04_sfr} for relatively young, but on average constantly star-forming, 
objects.
Second, 4691 shows SFR(\lya) $\approx$ SFR(UV) (observed), indicative
of little or no dust, and confirmed by our modeling.
Finally the remaining 7 objects show $SFR(UV) > SFR(\lya)$ with UV
star formation rates up to $\sim$ 14 times larger than \lya, a result
found for the majority of LBGs and LAEs \citep[e.g.][]{Yamada05,Gron07}. 

We now correct these SFR indicators for the effects of dust and
radiation transfer. One has 
\begin{equation}
{\rm SFR}(\lya)^{\rm true}={\rm SFR}(\lya) / f_e, 
\label{eq_lyasfr}
\end{equation}
with the \lya\ escape fraction $f_e$. The UV SFR is corrected assuming
the Calzetti  law and the extinction $E(B-V)$ derived from the \lya\
line fit, i.e. 
\begin{equation}
{\rm SFR(UV)}^{\rm true} = {\rm SFR(UV)} \times 10^{+0.4 E(B-V)^\star k_{\rm UV}},
\label{eq_uvsfr}
\end{equation}
with $E(B-V)^\star = 0.44 E(B-V)$. For our objects $k_{\rm UV} =
k_{1600} \approx 10$ is appropriate.  Using $E(B-V)$ derived from our
\lya\ fits, the resulting ``true'' SFR values for all our objects are
plotted in Fig.\ \ref{f_uv_lya} (green triangles). 
For the majority of the objects we find that their corrected SFR
values show much less dispersion between \lya\ and UV based
measurements than observed, uncorrected values. 
Three objects (1337, 5215, 7539) seem to require even lower \lya\
escape fractions (i.e.\ larger extinction) to reduce the differences
between their SFR indicators further. Adopting Eq.\ \ref{eq_fe}
and imposing SFR(\lya)$=$SFR(UV) would imply an extinction
$E(B-V)$ of (0.19, 0.07, 0.18) compared to our estimate of
(0.1, 0.1, 0.05) mag from line fitting.
Similarly the extinction seems to be overestimated in FDF 5812 and
for the high \nh\ solution of 1267.
The maximum extinction allowed to obtain SFR consistencies for these objects 
within the shaded region is $E(B-V)=$ 0.09 and 0.06 for 1267 and 5812
respectively. 
Most of these ``adjustments'' are within a factor of 2 uncertainty.
Overall we therefore conclude that the results based on our \lya\ line
fits using radiation transfer models improve the consistency between
UV continuum and \lya\ line based SFR indicators.

\subsection{Other correlations}
\label{s_correlations}
Several correlations have been found earlier between \lya\
and other properties of LBGs from analysis of large spectroscopic
samples. When grouping their $\sim$ 1000 LBG spectra according to \lya\
equivalent width, \citet{Shap03} found the following main correlations
with decreasing EW(\lya)$_{\rm obs}$ (i.e.\ from emission to absorption):
{\em 1)} The extinction increases, 
{\em 2)} the strength of the low-ionisation interstellar lines increases, 
{\em 3)} the velocity shift between these interstellar lines and the \lya\ peak 
increases $\dv$, and
{\em 4)} the dust-corrected SFR increases.
Correlations 1, 2, and 4 have also been found in other samples 
\citep[e.g.][]{Noll04,Tapk07}.
We now examine if these correlations also hold for our small sample, and how 
our model may or may not explain them.

\subsubsection{$E(B-V)$ versus \lya\ EW}
A decrease of EW(\lya)$_{\rm obs}$ with increasing extinction is naturally
expected from our radiation transfer models, e.g.\ when the intrinsic
equivalent width is constant (see Eq.\ \ref{eq_ext}).  This seems the
most simple/natural explanation for the trend observed between the
average LBG spectra, also given the relative long SF timescales
observed generally in LBGs, which should imply a fairly constant
intrinsic EW(\lya)$_{\rm int}$.

The 11 individual LBGs analysed here show no clear trend
between EW(\lya)$_{\rm obs}$ and $E(B-V)$. Given the relatively large scatter
in the derived gas-to-dust ratio (see Fig.\ \ref{f_ebv_nh}) it appears
that the clearest observational trend found is between
EW$_{\rm obs}$ and \nh\ instead (Fig.\ \ref{f_ew_nh}). Of course, if on average
the  gas-to-dust ratio is constant, any correlation between
EW$_{\rm obs}$ and \nh\ would automatically imply a correlation between 
$E(B-V)$ and EW(\lya)$_{\rm obs}$, as observed by \citet{Shap03}.

\subsubsection{$EW({\rm LIS})$ versus $EW(\lya)$}
Our model does not make direct predictions for the strengths
of interstellar absorption lines. For this reason and since most of 
the observed low ionisation interstellar lines are saturated 
\citep{Shap03} we are not able to examine quantitatively this 
correlation.

\citet{Ferrara06} propose to explain the behaviour of $EW({\rm LIS})$
by cold debris whose covering factor decreases as a function of
time in a dynamical outflow model. $EW({\rm LIS})$ is then mostly
related with the wind velocity (constrained by \dv)
which may vary with  EW(\lya)$_{\rm obs}$ according to these authors.
Verifying observationally whether and how wind velocity truly varies
and quantifying the covering factor of the cold outflowing gas is therefore
important to test this scenario.

\subsubsection{\dv\ versus $EW(\lya)$}

\begin{figure}[tb]
\includegraphics[width=8.8cm]{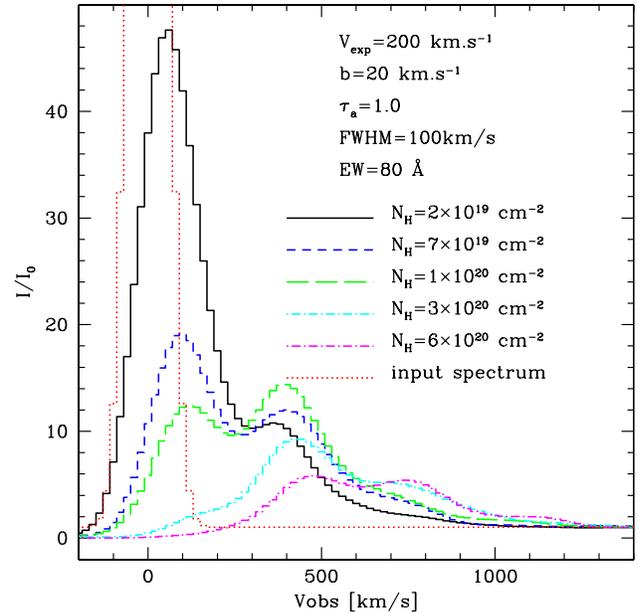}
\caption{Dependence of the \lya\ line profile on \nh\ for typical shell parameters.
Note the clear shift of the peak of the emission profile from low velocity ($\sim 100$
\kms\ $\approx 0.5 \times \vexp)$ to twice the shell velocity with increasing \nh, typical for
$\nh \protect\ga 10^{20}$ cm$^{-2}$.}
\label{vary_nh}
\end{figure}

\begin{figure}[tb]
\includegraphics[width=8.8cm]{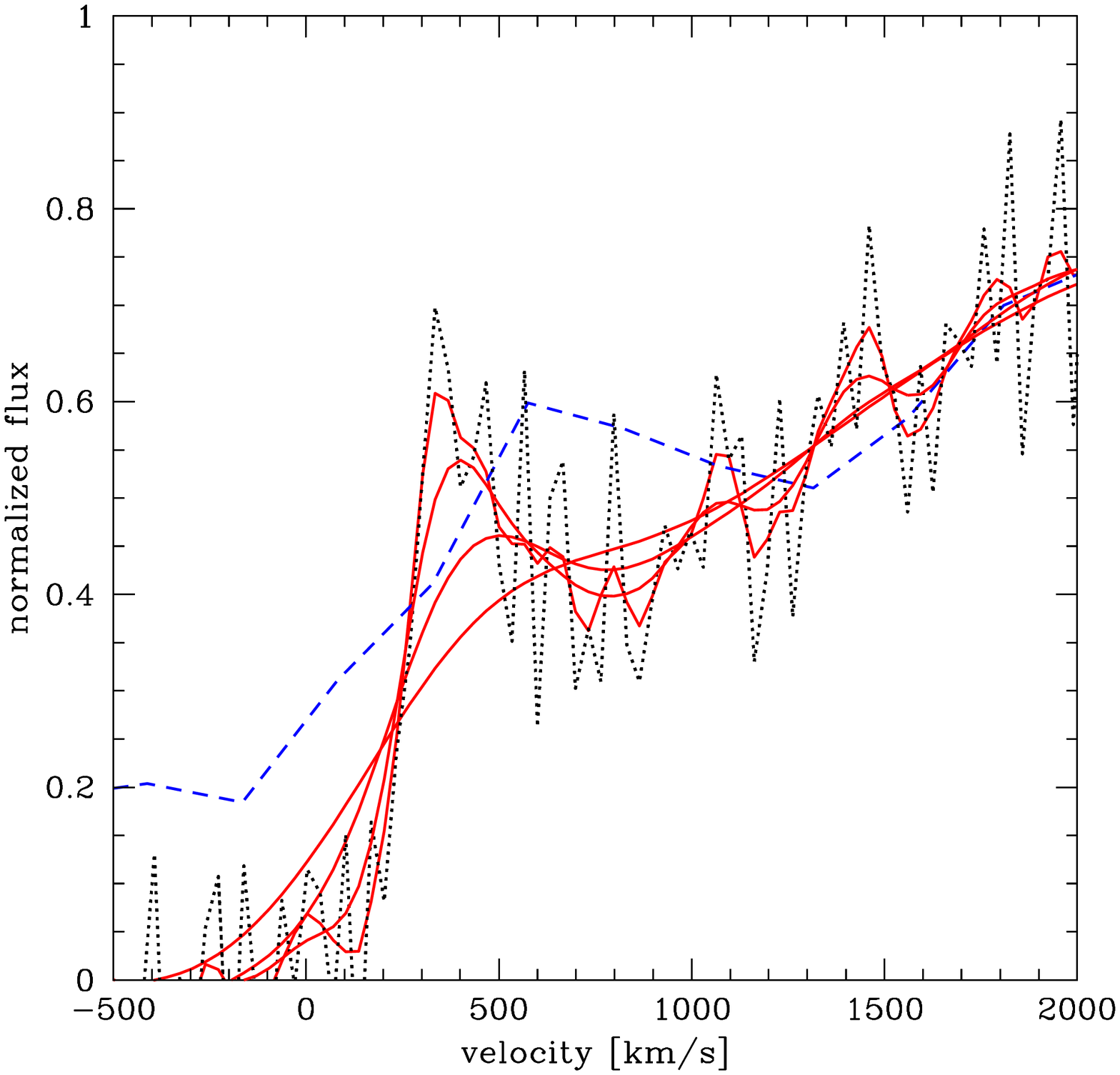}
\caption{Effect of spectral resolution on the \lya\ line shift measurement.
    The dotted line is the observed spectrum of cB58 from \citet{Pett02}
	with a resolution of 58 \kms.
    Red solid lines show the same spectrum convolved to resolutions
    of  100, 200, 400, and 700 \kms. The blue dashed
    line is the spectrum of the first group of \citet{Shap03} normalised
	to the one of cB58 at +1700 \kms.
	Note the progressive redshift of the \lya\ peak (observed at
	$+355$ \kms\ in the original cB58 spectrum) by up to $\sim$ 200 \kms\
	for a resolution of 400 \kms, typical of the composite spectra
	of \protect\citet{Shap03}.
}
\label{f_delta_v}
\end{figure}

\citet{Shap03} measure velocity shifts between $475 \pm 25$ and $795 \pm 3$ 
\kms\ for their four groups (quartiles) of LBGs with decreasing EW(\lya)$_{\rm obs}$.
To verify this trend we have remeasured \dv\ and
other quantities from the composite spectra of \citet{Shap03}; overall
our measurements are within $1 \sigma$, except for group 1 (with the
strongest \lya\ absorption) where we obtain $\dv =
703$ \kms.  Clearly the uncertainty cited for this group must be
underestimated.
Does this observed trend imply a systematic variation of the average
outflow velocity between the spectral groups?  For example, if true, this
could imply an increase of the average wind velocity from $\vexp
\approx 1/3 \times \dv \sim$ 160 to 235
\kms\ (or to 265 \kms\ using Shapley's value for group 1) if
spherical shell models apply as argued earlier.

As already mentioned above, the expansion velocities derived for
our objects and those measured for 2 of them as well as the typical
reddening values, are in good agreement with the data of
\citet{Shap03}. However, within our small sample we do not find any
systematic variation of \vexp, and clearly the reddening variations we
find are large compared to the difference of the mean extinction
between groups 3 and 4 of Shapley (where our objects would lie) and
compared to the dispersion of $E(B-V)$ within these groups.

Shapley stresses that ``none of the correlations with \dv\
is as significant as the trends among EW(\lya)$_{\rm obs}$, $EW({\rm LIS})$, and $E(B-V)$''.
For example, when grouped according to \dv, the objects of 
\citet{Shap03} show several trends opposite to those trend found between the 
groups constructed according to EW(\lya)$_{\rm obs}$\footnote{E.g.\ Instead of decreasing,
$E(B-V)$ increases with increasing EW(\lya)$_{\rm obs}$ between LBGs grouped according 
to \dv\ (cf.\ Figs.\ 3 and 16 of \citet{Shap03}).}.
The reality and significance of this behaviour may thus be questionable.
In fact we find two effects, a radiation transfer and an ``instrumental''
one, which could artificially lead to such a behaviour.

First, while for expanding shells with column densities $\nh \ga 10^{20}$ cm$^{-2}$ 
the main peak of the \lya\ profiles is indeed redshifted by twice
the expansion velocity \citep[as pointed out by][]{Verh06}, the peak
emerges at lower velocities ($\sim 1 \times \vexp$) for lower column densities.
This transition is clearly illustrated in Fig.\ \ref{vary_nh}.
If Shapley's spectral groups correspond on average to a sequence with
increasing \nh\ (from \lya\ in emission to absorption) and if this transition 
happens somewhere within this sample it would mean that the real 
spread of outflow velocities would considerably be reduced to \vexp\ from $\sim$ 
238 to 265 \kms\ (or 235 with our measurement for group 1).

A second, unavoidable effect, also alters the observed velocity
shift between \lya\ and the interstellar lines.
Indeed, given the increasing strength of broad \lya\ absorption in
groups 2 and 1 compared to the other groups, the
relatively low spectral resolution of the composite spectra affects
the measurement of the \lya\ line fit introducing a systematic
shift. For illustration we show in Fig.\ \ref{f_delta_v} the
high-resolution spectrum of cB58 -- an LBG with \lya\ properties
characteristic of group 1 -- and the effect lower spectral resolution
has on the measurement of the \lya\ peak. Interestingly but not really
surprisingly \citep[cf.][]{Mas-Hesse03}, a limited spectral
resolution, such as the one of $\sim$ 400--700 \kms\ of the composite
spectra of \citet{Shap03}, can lead to an overestimate of the true
\lya\ line redshift in objects with a steeply rising ``underlying''
spectrum. In other words, due to the appearance of the broad \lya\
absorption wing, the measurement of \dv\ may be
affected by a systematic shift which increases with decreasing
EW(\lya)$_{\rm obs}$ i.e.\ towards the group 1 with the strongest \lya\
absorption.  As the exercise in Fig.\ \ref{f_delta_v} shows, this
shift can be of the same order as the apparent velocity differences
between different LBG subgroups.  
For the two reasons just discussed we conclude that the observed 
variations between \lya\ and the IS lines found between the
subgroups of \citet{Shap03} and possible correlations with other
quantities cannot reflect a simple behaviour of the 
real outflow velocity. 
More work is needed to clarify the possible link between outflow
velocities and other properties of LBGs.
 
In any case our radiation transfer calculations show that the
variations of \vexp\ found between the objects fitted here have a
relatively minor influence on properties such as $EW(\lya)$, the \lya\
escape fraction etc. It appears that as long as global outflows with
velocities of several 100 \kms\ are present, other parameters are more
dominant.

\subsubsection{SFR(UV) versus $EW(\lya)$}
\label{s_sfr_lya}

\begin{figure}[tb]
\includegraphics[width=8.8cm]{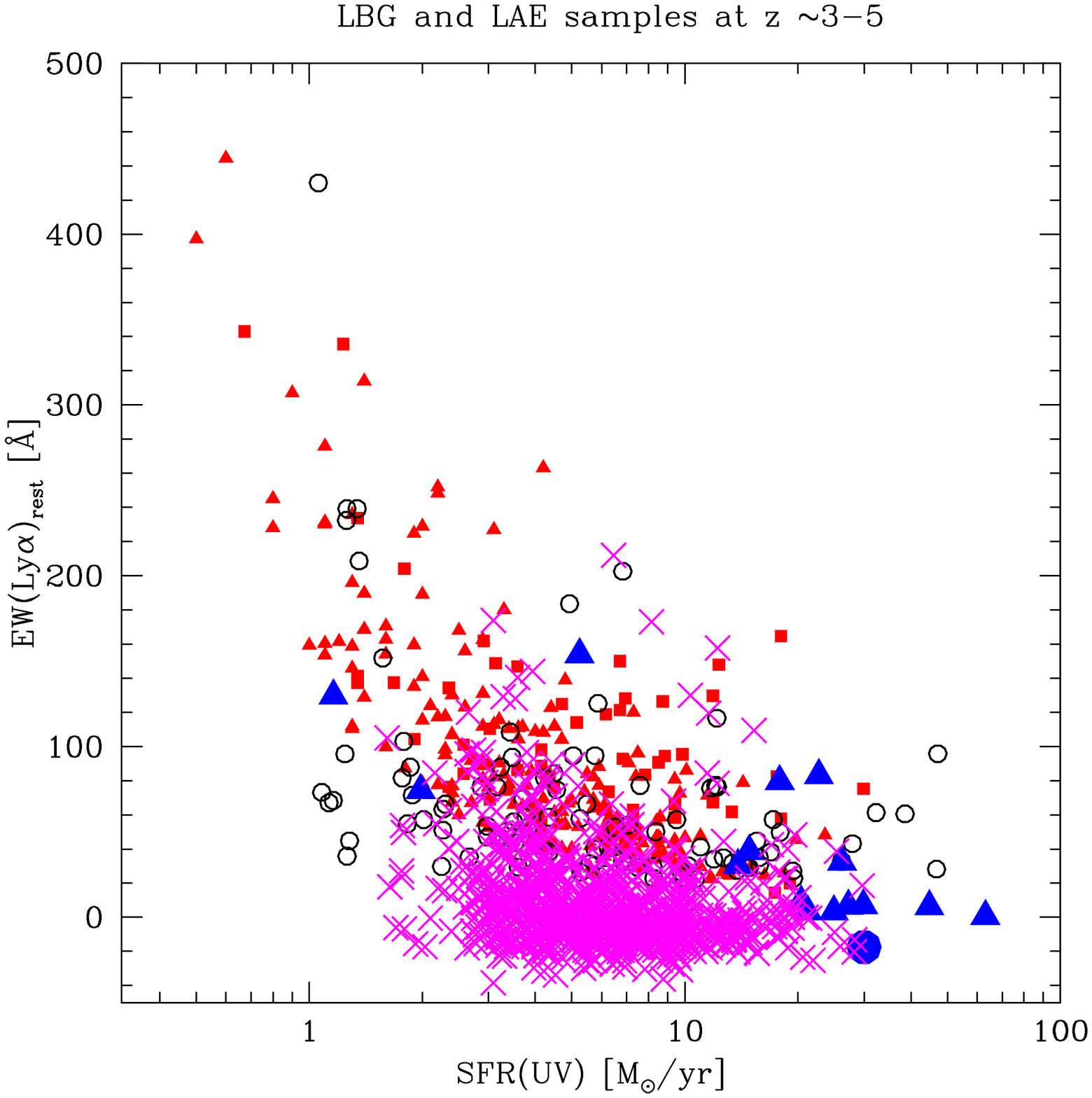}
\caption{Observed restframe \lya\ equivalent width of LBGs and LAEs as a
function of their UV star formation rate (uncorrected for reddening).
The following data is shown: LBGs from \citet{Tapk07} (blue
triangles), cB58 \citet{Pett02} (blue circle), 
and the individual objects from \citet{Shap03} (magenta crosses).
LAEs or LAE-candidates from \citet{Yamada05} (red triangles)
\citet{Ouchi07} (red squares) and from \citet{Gron07} (black circles).
The data clearly shows the absence of high equivalent width objects 
at high SFR, which we explain as being due to dust/radiation transfer
effects. 
For further details see text.}
\label{f_sfruv_ew}
\end{figure}

Trends or correlations between EW(\lya)$_{\rm obs}$ and the star formation rate
have been noted by many groups and are found not only for LBGs, but also
for LAEs \citep[e.g.][]{Shap03,Ando04,Tapk07}.
A compilation of available data is shown in Fig.\ \ref{f_sfruv_ew}, where
we include the following data:
LBGs from \citet{Tapk07}, cB58 \citet{Pett02} (blue symbols),
and the individual objects from \citet{Shap03} (magenta crosses);
LAEs or LAE-candidates from \citet{Yamada05,Ouchi07} (red symbols)
and from \citet{Gron07} (black circles).
Note that Tapken's LBGs include four $z \sim$ 4.5--5 objects, 
Yamada's sample of 198 LAEs spans $z \sim$ 3.3 to 4.8,
Ouchi's 84 spectroscopically confirmed LAEs have $z \sim$ 3.1, 3.7 and 5.7.
All other objects have redshifts have $z \sim$ 2.7--3.2.
Note that none of the data was corrected for reddening.  SFR(UV) is
therefore directly proportional to the observed UV magnitude of the
objects.

The observations summarised in Fig.\ \ref{f_sfruv_ew} show 
{\em 1)} the absence of strong \lya\ emitters (high $EW_{\rm obs}$) at high SFR(UV), 
{\em 2)} the existence of objects with EW(\lya)$_{\rm obs} \ga 100$ \AA\ for low SFR,
{\em 3)} the possible existence of a maximum value of $EW_{\rm max}(\lya)$
as a function of SFR(UV), and
{\em 4)} a wide range of EW(\lya)$_{\rm obs}$ at low star formation rates 
(SFR(UV) $\la 10$ \msunyr).
Although the $z \sim 3$ samples span the largest range in SFR(UV) there does
not seem to be a systematic trend with redshift. We therefore retain
all objects, irrespective of their precise $z$.
We suggest the following explanations for the trends shown in Fig.\
\ref{f_sfruv_ew}. 

{\bf 1)}
The absence of high EW$_{\rm obs}$--high SFR objects (1) is due to radiation transfer
effects and the presence of dust, which unavoidably lead to a stronger
reduction of \lya\ photons compared to the adjacent continuum 
(cf.\ Sect.\ \ref{s_escape}). The probability that high UV-SFR objects
are powered by one or few instantaneous bursts is very small; hence
their intrinsic \lya\ equivalent widths must be close to that expected
for constant SF, i.e.\ EW(\lya)$_{\rm obs}>$ 70--90 \AA\ and up to $\sim$ 250--360
\AA\ depending on metallicity (cf.\ Fig.\ \ref{f_s04_sfr}).
Radiation transfer and the presence of dust reduce EW(\lya)$_{\rm obs}$ from the
intrinsic to the observed values.

{\bf 2)}
Strong \lya\ emitters with EW(\lya)$_{\rm obs} \ga 100$ \AA\ correspond to 
young ($\la 50$ Myr) objects, with the exact age limit depending on the
detailed SF history. Equivalent widths up to $\sim 360$ \AA\
can be explained by stellar populations with ``normal'' Salpeter-like
IMFs and metallicities $Z \ga 1/50$ \zsun. 
The strong emitters probably show little or no reddening, which would otherwise
rapidly reduce the observed equivalent width.
Observationally most of these objects are found as LAEs; few LBGs show
such strong emission \citep[cf.][]{Shap03}.

{\bf 3)} The trend of $EW_{\rm max}(\lya)$ (``upper envelope'') as a
function of SFR(UV) (i.e.\ UV magnitude) is most likely due to an
increase of $E(B-V)$ with SFR(UV). An increasing dust optical depth
will progressively reduce the intrinsic \lya\ equivalent widths providing
a natural continuity between the objects discussed in 2) and 1).
Presumably the \hi\ column density also increases with magnitude
(i.e.\ UV SFR) maintaining a ``reasonable'' spread in the gas-to-dust ratio. 
Ultimately, the main underlying parameter governing the trends
with UV magnitude may be the galaxy mass. We will discuss this further
below.

{\bf 4)} The large spread in EW(\lya)$_{\rm obs}$ at faint magnitudes (SFR(UV)
$\la$ 5--10\msunyr) results most likely from two effects: first the
relatively small amount of dust, which does not eliminate the high
EW$_{\rm obs}$ objects, and second the larger variety of SF
histories/timescales 
-- i.e.\ an enhanced role of ``stochastic  SF events'' -- made
more plausible for objects of smaller absolute scale (mass or total
SFR).

\section{Other derived properties}
\label{s_props}
In other contexts it may be of interest to derive properties
such as the total neutral hydrogen mass in the outflow
and the mass outflow rate.
The following formulae can be used to estimate these quantities.

\subsection{Neutral gas mass}
The \hi\ mass in the shell is related to its column density $\nh$ and
radius $r$ by
\begin{equation}
M_{\rm HI} \approx 10^7 \left(\frac{r}{1 {\rm kpc}}\right)^2 
			\left(\frac{\nh}{10^{20} {\rm cm^{-2}}}\right) \msun 
\end{equation}
Assuming $r=1$ kpc, we find neutral gas masses of the order of $\sim 2
\times 10^{6}$ to $10^{8}$ \msun, with a median of $7 \times 10^{6}$
\msun.  If the \hi\ were found in shells with radii similar to the
optical sizes measured from $D_{25}$ for local starbursts, $r_{25}
\sim$ 5--50 kpc, these estimates have to increased by a factor
25--2500.
A plausible lower limit of $r> 1.6 h^{-1}$ kpc is derived by \citet{Shap03}
from the typical half-light radius of LBGs; similarly the examination
of close LBG pairs puts an upper limit of $r \la 25 h^{-1}$ proper kpc on 
the physical dimension of the gas giving rise to strong interstellar
absorption lines \citep{Adelberger03}.

\subsection{Mass outflow rates}
Following \citet{Pett00} the mass loss rate involved in the outflow
may be estimated as:
\begin{equation}
\dot{M_{\rm HI}} = 6. \left(\frac{r}{1 {\rm kpc}}\right)
			\left(\frac{\nh}{10^{20} {\rm cm^{-2}}}\right) 
			\left(\frac{\vexp}{200 {\rm km s^{-1}}}\right)\msunyr,
\end{equation}
Assuming again $r=1$ kpc,
we derive mass outflow rates between $\sim$ 0.2 and 100 \msunyr.
For larger radii $\dot{M_{\rm HI}}$ has to be increased accordingly.
The estimated outflow rates are thus comparable to the star
formation rate, as already found earlier \citep[e.g.][]{Pett00,Grimes07}.
In cases with the lowest velocities (quasi-static shells, i.e.\ in 2-3
of our objects), there is most likely no true outflow out of the galaxy.  

\section{A unifying scenario for LBGs and LAE, and implications}
\label{s_scenario}

In paper II, based on our modeling results for cB58 and on empirical
data, we have proposed a unifying scenario to explain \lya\ emission and absorption
and the observed trends in LBGs.
We will now discuss this scenario in light of the results from the FDF objects
and of the discussion in the present paper, and we will show how this scenario
should also apply to LAEs.

In paper II
we have suggested that the bulk of the LBGs have intrinsically
EW(\lya)$_{\rm int}$ $\sim$ 60--80 \AA\ or larger, and that the main physical
parameter responsible for the observed variety of \lya\ strengths and
profiles in LBGs are \nh\ and the accompanying variation of the dust
content.  Here, we propose that the same also holds for
most LAEs, and that the larger \lya\ equivalent widths found in (some
of) these objects are due to younger ages of the SF population, quite
independently of their SF histories.  For example, EW(\lya)$_{\rm int}$ up to
$\la$ 300--400 \AA\ can be expected at ages $\la$ 50 Myr and for
metallicities $\la 1/5$ \zsun\ (cf.\ Fig.\ \ref{f_s04_sfr}).  Any
differences between intrinsic and observed \lya\ equivalent widths are
then due to radiation transfer and dust effects.

In paper II, we have shown that these effects transform, e.g.\ for
an extinction of E(B-V) $\sim$ 0.3 as found for cB58, a spectrum with
an intrinsic \lya\ emission with EW(\lya)$_{\rm int}$ $\ga 60$ \AA\ into an
absorption-dominated spectrum of LBGs.  In the objects modeled here,
\lya\ is always found in emission, with observed EW(\lya)$_{\rm obs}$ $\sim$
6--150 \AA. Again, the intrinsic
\lya\ emission determined from our line fits is found to be stronger
with EW(\lya)$_{\rm int}$ $\sim$ 50--280 \AA.
For all these LBG examples, covering the full range of \lya\ 
line strengths observed by \citet{Shap03}, the intrinsic \lya\
properties are thus compatible with expectations of starbursts with
constant star formation.
For LAEs the same scenario can also hold, provided the extinction
and/or \hi\ column density are sufficiently low, and provided the
age of the highest EW(\lya)$_{\rm obs}$ objects is relatively young. 

We now discuss the arguments supporting a continuity and even a strong 
overlap  between LBGs and LAE;
subsequently we will summarise the
main observational evidence supporting our scenario.

\subsection{LBGs and LAE, overlapping population with a sequence driven mostly 
by mass?}
\label{s_overlap}

To be more precise, we suggest the following overlap and distinctions
between LBGs and LAEs at a given redshift:
\begin{enumerate}
\item LAEs brighter than a certain limiting magnitude, $M_{\rm lim}$, in 
the continuum are the same population as LBGs with strong \lya\ emission
(i.e.\ $>$ EW$^{\rm rest}_{\rm lim}$).
\item At magnitudes fainter than $M_{\rm lim}$, LAEs represent less
massive objects than LBGs.
\item The remaining LBGs, i.e.\ those with EW(\lya) $<$ EW$^{\rm rest}_{\rm lim}$
and \lya\ in absorption and hence not selected as LAEs, 
can cover a wide range of galaxy masses, SFR, and SF histories.
\end{enumerate}
At redshift $z \sim 3$ the overlap between LAEs and LBGs corresponds
to approximately $\sim$ 25\% of the LBG population, and $\sim 23$ \% of 
the LAE population, as we will discuss below.
The amount of overlap, hence the value of $M_{\rm lim}$ varies quite
likely with redshift.
Let us now justify these statements in more detail.

\begin{figure}[tb]
\includegraphics[width=8.8cm]{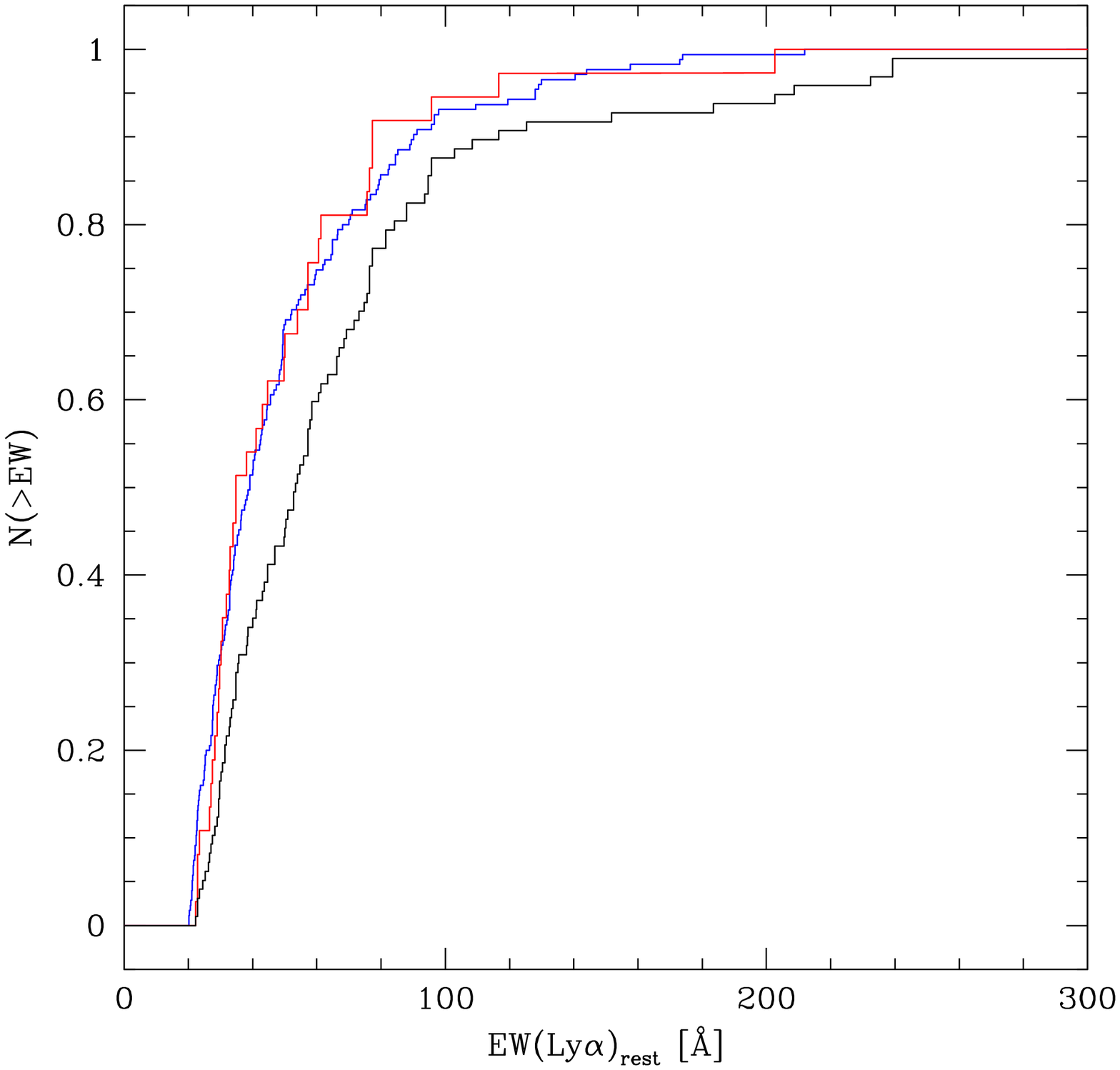}
\caption{Cumulative distribution of the observed \lya\ equivalent widths
of $z \sim 3$ samples of LAEs and LBGs: Shapley's LBGs with $R<25.5$ and
EW(\lya)$^{\rm rest}_{\rm obs}>20$ \AA\ in blue, LAEs from
\citet{Gron07} fulfilling  
the same condition in red, and the full LAE sample of Gronwal et al. in black.
The probability that the blue and red distributions are drawn from the same parent 
distribution is 92 \%, according to the 2 sided Kolmogorov-Smirnov test.}
\label{f_ew_distr}
\end{figure}

That there should be a continuity between LBGs and LAEs and a strong
overlap between the different populations seems quite obvious. 
After all their distinction stems from a ``UV-continuum'' versus 
``emission line'' selection, based, however, on the same spectral
range (restframe UV). Furthermore the Lyman break or drop-out criterion 
is also applied as a selection criterion for many LAE samples 
\citep[e.g.][]{Taniguchi05,Ouchi07}, and in any case LAEs are 
expected to show a Lyman break, even if too faint to be measured from
the current data.
A continuity between LBGs and LAEs is also supported by the overlap of
many of their observed properties, such as magnitudes, colors, and
others \citep[e.g.][]{Gawiser06,Gron07,Ouchi07,Nilsson07,Martin08,Lai08}, 
in addition to the ones already
discussed above (SFR(UV), SFR(\lya), \lya\ equivalent widths).
Furthermore, the spatial distribution of LBGs and LAEs show also very
similar correlation lengths, both at $z \sim 3$ and 4.5 where measurements
are available \citep{Adelberger05,Gawiser07,Kovac07}.
Finally, at $z \sim 3$ the relative number densities of LAE/LBG can also be 
understood: in the magnitude range $R_{\rm AB}<25.5$ covered both by the 
LBG sample of \citet{Shap03} and the LAE sample of \citet{Gron07},
one has a ratio of LAE/LBG $\sim$ 1/3, according to the latter authors. 
This is identical to 
the fraction of $\sim$ 25\% of LBGs showing strong \lya,
EW(\lya)$_{\rm obs}$ $\sim$ 20 \AA, since this limit is basically the same as the
selection limit of the Gronwal sample\footnote{In fact, applying the 
criteria $R_{\rm AB}<25.5$ and
EW(\lya)$^{\rm rest}_{\rm obs}>20$ \AA\ to Shapley's data we find 175 out of 814 LBG,
i.e.\ 21.5 \%. Dropping the restriction on $R$ we obtain 22.5 \%, 
similar to the $\sim$ 25 \% of \citet{Shap03}.}.
Adopting the same magnitude limit ($R<25.5$) for both samples,
we find also no significant difference between the equivalent widths
distributions of the $z \sim 3$ LBG and LAE sample, as shown in
Fig.\ \ref{f_ew_distr}. Indeed, while the EW$_{\rm obs}$ distribution
of Shapley's 
total LBG sample differs from that of Gronwal, the probability that
both distributions are drawn from the same parent distribution is 92 \%
(according to the 2 sided Kolmogorov-Smirnov test), once this same
magnitude limit is applied to both samples.
This overlap with the LBGs concerns 37 out of 160 LAEs 
(23 \%) from Gronwal's statistically complete LAE sample at $z \sim 3.1$.

We therefore conclude that LAEs above a certain magnitude limit 
represent the same population as LBGs with \lya\ emission (point 1 above); 
furthermore the LAE selection method allows one to find strong emission line
objects drawn from a larger range of continuum brightness than the LBG
selection, as also shown in Fig.\ \ref{f_sfruv_ew} (point 2).
The ``remaining'' LBGs, corresponding to three quartiles of Shapley,
cannot be found through the LAE selection technique, and represent thus
a separate class of star forming objects. As many LBG studies have
shown their parameters (mass, SFR etc.) cover a wide range (point 3).

If a continuity and overlap exists between LBGs and LAE, what is then the main parameter(s)
``driving'' the observed \lya\ strength?
From the \lya\ radiation transfer in expanding shells we have seen that
the extinction (here described by $E(B-V)$ or the dust optical depth
respectively) plays a --if not the most important-- role in
determining the strength of \lya 
\footnote{Indeed $E(B-V)$ is proportional
to the dust absorption optical depth, which --associated with the
effect of resonant scatterings-- determines
the absorption probability of \lya\ photons.}.
The next important parameter \nh, and the resulting variation in
dust/gas ratio, as well as possibly deviations from the simple geometry
considered here, likely introduce the required scatter around an 
oversimplified 1-dimensional scaling only.
Instead of the extinction other parameters, such as the stellar mass $M_\star$
of the galaxy, its SFR, or others may be physically more fundamental, as
suggested in paper II. 
In this case  LBGs and LAEs at the same
redshift would represent a continuity with decreasing mass, at
least on average, and correlations between $M_\star$, $E(B-V)$, and \nh\
would naturally explain the observed trends, including those of \lya.

Among LBG samples, \citet{Pentericci07} have e.g.\ found
that LBGs at $z \sim 4$ with \lya\ emission are less massive on average than
LBGs with \lya\ in absorption. The same trend has also been found
by \citet{Erb06} for $z \sim 2$ LBGs.
Comparing LBGs and LAE, 
it is well established that LAEs
show lower UV (i.e.\ continuum-based) star formation rates, and tend to have lower
stellar masses than LBGs at similar redshift
\citep[cf.][]{Reddy06,Pirzkal07,Gron07,Gawiser07,Overzier08,Lai08}.
The measured correlation length and galaxy bias of LBGs and LAEs are also
consistent with a lower mass for the ensemble of LAEs, as discussed
recently by \citet{Gawiser07}.
At $z \sim$ 1--2 clear correlations between the total SFR and stellar
mass $M_\star$ are found for UV star formation rates covering a
similar range as those observed in $z \sim 3$ LAEs and LBGs
\citep{Elbaz07,Daddi07,Noeske07}.
Furthermore we know that extinction scales with the total (IR plus UV)
SFR \citep[e.g.\ at $z \sim 2$ see][]{Reddy06}.
Except for the behaviour of the \hi\ column density with galaxy mass,
which is so far unknown, all quantities show thus the required trend
to be consistent with our scenario.

\subsection{LBGs and LAEs: age and SF history differences?}
\label{s_ages}

As already pointed out in paper II, stellar populations studies of
LBGs consistently find ages of several tens to hundreds of Myr and
favour constant star formation over this timescale \citep[see
e.g.][]{Elli96,Pett00,deMell00,Papovich01,Shap01}. In this case 
intrinsic EW(\lya)$_{\rm int}$ of $\sim$ 60--100 \AA\ are unavoidable 
(cf.\ Fig.\ \ref{f_s04_sfr}).
Although 
less studies have yet been undertaken on the stellar populations of LAEs,
it appears quite clearly that some of them are younger than LBGs.
For example, using the sample of $z=3.1$ LAEs from \citet{Gron07},
\citet{Gawiser07,Lai08} find a large spread of ages (from $\sim$ 20
Myr to $\sim$ 1.6 Gyr) and a decrease of the mean age with decreasing
mass and near-IR luminosity. Similarly, young ages (and low masses)
have been found from studies of LAEs at higher redshift 
\citep[e.g.][]{Pirzkal06,Finkelstein07}.

\begin{figure}[tb]
\includegraphics[width=8.8cm]{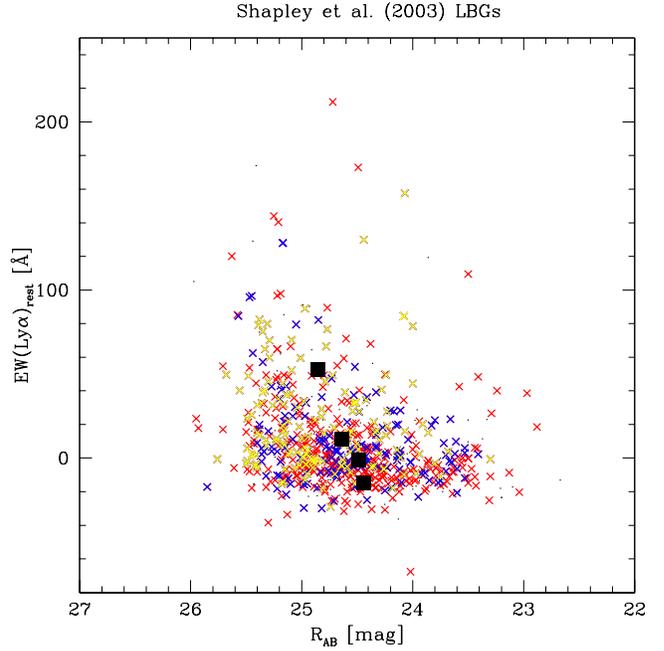}
\caption{Observed restframe \lya\ equivalent width versus $R_{\rm AB}$ magnitude
for the $\sim$ 800 LBGs of \citet{Shap03} (show as crosses).
The total EW$_{\rm obs}$ is plotted here, including both absorption
and emission 
components if present.  Colors indicate different amounts of
extinction $E(B-V)$ estimated from the $G-R$ color as described in 
Shapley et al.: $E(B-V) > 0.2$ in black, red: 0.1--0.2,
blue: 0.05--0.1, and yellow: $E(B-V) < 0.05$.  The corresponding
values of the four LBG groups of Shapley et al.\ are show as filled
squares.Note the apparent existence of low $E(B-V)$, low EW(\lya)$_{\rm obs}$ objects,
which could be dominated by stellar absorption.}
\label{f_shapley}
\end{figure}

No age constraint can be derived from \lya\ alone for objects with
EW(\lya)$_{\rm obs} \la$ 100 \AA, as such equivalent widths can be obtained
for constant SF and taking into account various degrees of \lya\ 
suppression by dust. 
On the other hand, EW(\lya)$_{\rm obs}$ $> 100$ \AA\ requires ages of $\la$ 
10--50 Myr at maximum, depending on the exact SF history (Fig.\ \ref{f_s04_sfr}); 
furthermore in such cases the amount of dust must be relatively low
(e.g.\ $E(B-V) \la$ 0.05--0.1 for $EW^{\rm rest}_{\rm obs}$/EW${\rm int}=0.5$
and $r=$0.44--1, according to Eq.\ \ref{eq_ext}) to maintain
a high \lya\ equivalent width.
In other words, for the vast majority of the LBGs \lya\ provides no age
constraint, whereas LAEs with EW(\lya)$_{\rm obs}$ $\ga 100$ should have ages
of $\la$ 10--50 Myr. These simple predictions appear in good agreement with
the age estimates of LBGs and LAEs discussed above.

What can be inferred about the SF timescale/duration in LBGs and LAEs
and possible differences between the two classes?
As already mentioned, for the LBGs modeled in detail here and in paper
II, our \lya\ modeling results require relatively high intrinsic \lya\
equivalent widths, corresponding to constant star formation over
timescales $\ga$ 20--50 Myr or longer; \lya\ alone does not allow us
to determine an upper limit for the duration of SF.  
For LAEs with high equivalent widths ($\ga$ 100 \AA) there is a
degeneracy between age and SF duration, but long SF timescales are not
excluded from their \lya\ properties. Even less strong LAEs could be
explained with constant SF (and hence unconstrained SF duration).
For example objects with EW(\lya)$_{\rm obs}=20$ \AA\  --- a value close to the
detection 
criterion commonly adopted --- could be explained with
a small amount of extinction, $E(B-V)^{\rm gas} \sim 0.1$,
for an intrinsic equivalent width three times larger, according
to Eq.\ \ref{eq_ext}.
From \lya\ alone, relatively long SF timescales could thus be possible
both for the LBGs studied here and for the typical LAEs.  However,
more ``bursty'', intermittent SF histories may be more plausible for
lower mass objects, such as LAEs.  The suggestion of relatively long
timescales for LAEs is e.g.\ compatible with the ages of $\sim$ 160 Myr
to 1.6 Gyr derived by \citet{Lai08}; 
however these age estimates 
provide probably upper limits, as derived assuming constant SF, and
younger ages are also obtained when multiple components are allowed
for.

Potentially most interesting to constrain the SF timescale are objects 
with a low \lya\ equivalent width (faint emission or even absorption) and 
low extinction. If the super-shell model applies to such objects and if their
velocities are comparable to the typical LBGs, observations of \lya\ in
absorption could indicate a post-starburst phase dominated by stellar 
absorption, as clear from Fig.\ 3 of paper II.
In fact, such objects seem to exist in the $z \sim 3$ LBG sample of 
\citet{Shap03}, as shown Fig.\ \ref{f_shapley}.
Finding objects with shorter SF timescales among the low SFR objects,
which may also be seen in this Fig., appears more plausible given the
stronger dependence on ``stochasticity'' already discussed
earlier. However, whether these objects are truly short duty-cycle,
post-starburst objects with little dust remains to be
verified. Otherwise different geometries and kinematics may need to be
invoked and the radiation transfer models be reexamined for these objects.
Observations of stellar UV features, such as C{\sc iv} and Si~{\sc iv}
and other H recombination lines (e.g.\ \ha) should in principle allow
to test this hypothesis and hence provide interesting constraints on
the duty-cycle of these objects. 
 
\subsection{Open questions and further tests of the scenario}
\label{s_tests}
Although many arguments support our scenario described above,
several important tests should be carried out.
For example, a correlation between the galaxy mass and the \hi\ column
density remains to be established. 
Furthermore, does the dust mass increase with stellar mass?
In other words, how does the dus-to-gas ratio evolve with galaxy mass?
Also, how well do the proposed correlations hold for individual objects?
And what governs the dispersion around such correlations?

Concerning LAEs, direct measurements of their ISM properties (velocity
shifts, ISM covering factors etc.) would be very useful. Although it
can be expected that these sources show outflows, as well established
for the majority of LBGs, such observations remain to be done.  This
could also be important to examine possible differences in the
homogeneity and the covering factor of the ISM in these objects.

An unclear issue raised by our study concerns the fainter LBGs
with low \lya\ equivalent widths or \lya\ in absorption.
Are these, presumably less massive objects, sufficiently dust-rich
to suppress \lya\ emission, are they post starburst objects, or
what other process(es) explain their absence of \lya\ emission?

A larger number of deep high quality spectroscopic observations,
multi-wavelength data, and accompanying modeling should help to
answer these questions in the near future.

\subsection{Implications}
\label{s_implications}
Our scenario proposed for LBGs and LAEs has several notable implications,
most of them already mentioned in paper II.

First, our analysis has clearly shown that LBGs exhibit quite
significant variations in their \lya\ escape fraction, as apparent also
from earlier comparisons of \lya\ and other SFR indicators.
Such variations must be included in models aimed at understanding at \lya\ 
emission from LAEs and LBGs and alike objects.

Second, observed distribution functions of EW(\lya)$_{\rm obs}$ and the
\lya\ luminosity function must be strongly modified by radiation
transfer and dust effects, and the fraction of objects with low EW(\lya)$_{\rm obs}$
($L(\lya)$) must be ``artificially'' overestimated (cf. paper II).
For a proper interpretation this must be taken into account;
simple corrections may be applied using e.g.\ the proposed 
extinction estimate (Eq.\ \ref{eq_ext}) or more sophisticated
methods.

Third, the long/constant star formation rates and ``old'' ages
($\sim$ 20--100 Myr) of LBGs derived from stellar population studies
can be reconciled with the observed \lya\ properties, when radiation
transfer and dust are taken into account.  In these circumstances \lya\
should basically be independent of the stellar population age
(measured since the onset of SF); its strength and line profile are
mostly determined by the extinction and by \nh\ (at least in the case
of a homogeneous expanding shell). Possible trends of \lya\ emission
with age, such as indicated e.g.\ by
\citet{Shap01,Shap03,Ferrara06,Pentericci07} but apparently found to be
conflicting among each other, should then be fortuitous.

Finally, our scenario should naturally explain the increase of the
ratio of LAE/LBG and a higher percentage of LBGs with strong \lya\
emission with redshift. These trends were suggested earlier 
\citep[see e.g.][]{Hu98} and are now quite clearly observed
\citep{Noll04,Shimasaku06,Nagao07,Ouchi07,Dow07,Reddy07}.  Indeed if
extinction decreases with increasing $z$ as expected e.g.\ from the
observed metallicity decrease with redshift
\citep[cf.][]{Tremonti04,Erb06,Ando07,Liu08}, our radiation transfer models
presented here predict stronger \lya\ emission (see Eq.\ \ref{eq_ext}), 
even if the intrinsic \lya\ emission properties remain the same.
Stronger intrinsic emission is additionally expected with
decreasing metallicity, due to changes of the ionising/stellar properties.
Both effects will thus lead to stronger \lya\ emission and
hence to an increase of the LAE/LBG ratio with redshift.

\section{Summary and conclusion}
\label{s_ccl}
Our 3D \lya\ and UV continuum radiation transfer code \citep{Verh06} has
been applied to fit the \lya\ line profiles of 11 LBGs with 
\lya\ emission in the FORS Deep Field with redshifts between 2.8 and 5 
observed with medium spectral resolution ($R \approx 2000$) by
\citet{Tapk07}.
The observed line profiles show a variety of morphologies, including 
redshifted asymmetric profiles commonly observed, two double-peak
profiles, and other more complicated line shapes. 

We have adopted a simple model with a spherically expanding shell, which
consists of neutral hydrogen and mixed in dust, surrounding a 
central starburst emitting a UV continuum plus \lya\ recombination line 
radiation from its associated \hii\ region. With such a model,
described by 4 ``shell parameters'' (the shell velocity \vexp,
its column density \nh, the Doppler parameter $b$, the dust optical depth
$\tau_a$) and 2 ``spectral parameters'' (EW$_{\rm int}$ and FWHM$_{\rm int}$ of the input
\lya\ line), we have been able to successfully reproduce all the 
observed line profiles (see Sect.\ \ref{s_tapken}).
For the majority of the 11 LBGs analysed here the medium resolution
spectra allowed us to constrain these parameters with relatively few
degeneracies. In particular we found that \lya\ line profile fitting alone
allows us to determine the amount of dust extinction.

The parameters we have derived for our sample behave as follows (see Sect.\
\ref{s_discuss}):
\begin{itemize}
\item The expansion velocity is typically between $\vexp \sim$ 150--200 \kms.
Higher velocities (300--400 \kms) may be found in two objects. In two out
of 11 LBGs the neutral medium is most likely quasi-static ($\vexp \sim$ 10--25
\kms); their double-peaked line profiles are clearly distinct from the
majority 
of the spectra. Our velocity determinations are also supported by measurements
of interstellar lines in three cases.

\item Except for one object, FDF4691 with a double-peaked \lya\ profile
with FWHM $\sim$ 840 \kms, all \lya\ line profiles can be reproduced with
a relatively low value of the intrinsic \lya\ line width ($FWHM^{\rm int} \sim 100$ \kms);
radiation transfer effects broaden the line to the observed widths of typically
$FWHM^{\rm obs} \sim$ 200--600 \kms.
We have proposed that the observed FWHM(\lya) is related to the \hi\ column density to first
order.
 
\item The radial \hi\ column density obtained from our \lya\ line fits ranges
from $\nh \sim 2 \times 10^{19}$ to $7 \times 10^{20}$ cm$^{-2}$.
Lower values may also apply in few objects.
We found possible indications for an anti-correlation between the observed \lya\
equivalent width and \nh, and for a correlation between FWHM(\lya) and \nh.

\item Our \lya\ profile fits yield values of $E(B-V) \sim$ 0.05--0.2 for the 
gas extinction. These values are also broadly in agreement with the extinction
derived from broad-band SED fits for our objects.
We found indications for a dust-to-gas ratio higher than the Galactic value,
and for a substantial scatter.

\item The escape fraction of \lya\ photons $f_e$, which follows from our \lya\
profile fits, was found to be determined primarily by the
extinction. We proposes a simple fit between $f_e$ and $E(B-V)$ (Eq.\
\ref{eq_fe}), which for $E(B-V) \la 0.3$ is also in reasonable
agreement    with  the  \lya\  escape    fraction measured  in  nearby
starbursts.  Since \lya\ photons are  more strongly suppressed by dust
than UV continuum photons, a measurement  of EW(\lya)$_{\rm obs}$ can in principle
yield $E(B-V)$ (see Eq.\  \ref{eq_ext}), provided the  intrinsic \lya\
equivalent width is known (or assumed).

\item Although the LBGs modeled here include objects with observed \lya\ 
equivalent widths down to $\sim 6$ \AA\ in the restframe (and EW(\lya)$_{\rm obs}$
up to 130--150 \AA), 10/11 objects can clearly be fit with higher
intrinsic \lya\ equivalent widths (EW(\lya)$_{\rm int}$ $\sim$ 50--100 \AA\ or
larger), which are expected for stellar populations forming constantly
over long periods ($\ga$ 10-100 Myr).  In three cases we found
indications for younger populations.  This result agrees with the one
found in paper II from the modeling of cB58 --- an LBG with \lya\ in
absorption --- and shows that radiation transfer effects and dust can lead to
a great diversity of observed \lya\ strengths (and profile shapes), even
if the intrinsic stellar populations and emission line properties of LBGs
may be fairly similar.

\item Finally, after correction for radiation transfer and dust effects
we obtain a better consistency between the use of the \lya\ and UV continuum 
luminosity  as SFR indicators.

\end{itemize}

Putting together the model results from this paper (LBG with \lya\ in
emission) and those from paper II (modeling of an LBG with \lya\ in
absorption) we have been able to reproduce the main
trends/correlations found in large samples of LBGs and LAEs (see Sect.\
\ref{s_correlations}). In particular we found that the absence of
high $EW_{\rm obs}$--high SFR objects is due to radiation transfer effects and
the presence of dust. We also show that the observed correlation of
velocity shifts between IS lines and \lya\ with \lya\ strength
\citep[cf.][]{Shap03} does not reflect systematic differences of the 
outflow velocities.

We have proposed that most LBGs and LAEs have intrinsically
EW(\lya)$_{\rm int}$ $\sim$ 60--80 \AA\ or larger, and that the main physical
parameter responsible for the observed variety of \lya\ strengths and
profiles in LBGs are \nh\ and the accompanying variation of the dust
content. These quantities, in turn, are thought to scale mostly with
galaxy mass. Such relatively high intrinsic equivalent widths
are compatible with expectations for relatively long star formation
episodes ($\ga$ 10--100 Myr), compatible with typical ages and star 
formation histories derived for LBGs.
Interestingly we found also indications for relatively low 
extinction ($E(B-V) \la 0.1$) LBGs with \lya\ in absorption in the 
sample of Shapley and collaborators. If true, these could be 
short duty-cycle, post-starburst objects, dominated by stellar
\lya\ absorption. 

Analysing the properties of LBGs and LAEs at $z \sim 3$ we have shown 
that there is a clear overlap between these two populations (Sect.\ 
\ref{s_overlap}): $\sim$ 20--25 \% of the LBGs of \citet{Shap03} -- 
those with EW(\lya)$^{\rm rest}_{\rm obs} > 20$ \AA\ -- correspond to
the LAEs brighter than  
$R_{\rm AB} = 25.5$ mag. The remaining $\sim$ 77 \% of the
statistically complete 
LAE population from \citet{Gron07} are fainter in the continuum than
the LBGs.  
Radiation transfer, dust effects, and changes in the stellar/ionising
properties should also naturally explain the increase of the LAE/LBG
ratio, and a higher percentage of LBGs with strong \lya\ emission with
increasing redshift.

Our \lya\ line profile fitting of LBGs with a detailed radiation
transfer code has shown that we are able to understand a diversity of \lya\
line shapes and strengths and thereby to provide interesting quantitative
constraints on their gas and stellar properties.
Most of the observed differences and correlations between \lya\ and other
properties can be understood with a simple outflow model, where
the \hi\ column density and dust content ($E(B-V)$) play the main roles.
The same model allow successfully applies to LAE, who appear to be closely
related to LBGs.

Several additional tests of our scenario have been proposed and 
new questions have been raised, e.g.\ about the star formation
duration of some LBGs with apparently low extinction 
(see Sect.\ \ref{s_tests}). Numerous interesting implications remain
also to be worked, e.g.\ on the interpretation of \lya\ luminosity
functions. It is the hope that radiation transfer models, such 
as those successfully applied here, will help to shed further light
on star formation and galaxy evolution at high redshift.

\acknowledgements
We are happy to thank Daniela Calzetti for unpublished measurements of
\nh\ in local starbursts, and Alice Shapley and Caryl Gronwall for 
communicating data in electronic form. We'd like to thank many
colleagues for stimulating discussions during the last years and at
the \lya\ workshop held recently in Paris. Matthew Hayes, 
Andrea Ferrara, and David Valls-Gabaud, 
also provided useful comments on an earlier version of the manuscript.
This work was supported by the Swiss National Science Foundation.
\bibliographystyle{aa}
\bibliography{references}

\end{document}